\documentclass[fleqn,usenatbib]{mnras}

\usepackage[T1]{fontenc}
\usepackage{ae,aecompl}

\usepackage{graphicx}	
\usepackage{amsmath}	
\usepackage{amssymb}	

\usepackage{newtxtext,newtxmath}

\usepackage[usenames,dvipsnames]{xcolor}
\usepackage{physics}
\usepackage{tikz}
\usetikzlibrary{arrows,babel}
\usepackage[labelfont={bf},labelsep=endash]{caption}
\usetikzlibrary{positioning,arrows}
\usetikzlibrary{decorations.pathmorphing}
\usetikzlibrary{decorations.markings}
\tikzset{
momentum/.style={postaction={decorate},decoration={markings,mark=at position 1 with {\arrow{>}}}},
particle/.style={dashed
    },
photon/.style={decorate, 
    decoration={snake}},
    math/.style={draw, execute at begin node={$\displaystyle}, execute at end node={$}}
 }

\newcommand{{\go}}[1]{\textcolor{RoyalBlue}{{\small[GO: #1]}}}

\usepackage{pifont}  
\usepackage{enumitem}
\usepackage[normalem]{ulem}

\newcommand{\sub}[1]{_{\mathrm{#1}}}

\defcitealias{2023MNRAS.518...93A}{Paper~I}
\defcitealias{Penarrubia2010}{P10}
\defcitealias{2021MNRAS.505...18E}{EN21}
\defcitealias{2024PhRvD.110b3019D}{D24}

\title[DM subhalo tidal tracks via numerical sims]{New insights on low-mass dark matter subhalo tidal tracks via numerical simulations}

\author[A. Aguirre-Santaella et al.]{
Alejandra Aguirre-Santaella,$^{1}$\thanks{e-mail: alejandra.aguirre@uv.es}
Miguel A. S\'anchez-Conde,$^{2, 3}$ 
Go Ogiya$^{4}$ 
\\
$^{1}$ Institute for Computational Cosmology, Department of Physics, Durham University, South Road, Durham DH1 3LE, UK \\
$^{2}$ Instituto de F\'isica Te\'orica UAM-CSIC, Universidad Aut\'onoma de Madrid, C/ Nicol\'as Cabrera, 13-15, 28049 Madrid, Spain\\
$^{3}$ Departamento de F\'isica Te\'orica, M-15, Universidad Aut\'onoma de Madrid, E-28049 Madrid, Spain\\
$^{4}$ Institute for Astronomy, School of Physics, Zhejiang University, Hangzhou 310027, China \\
}

\date{Accepted 2025 November 17. Received YYY; in original form ZZZ}

\pubyear{2025}

\begin{document}
\label{firstpage}
\pagerange{\pageref{firstpage}--\pageref{lastpage}}
\maketitle

\begin{abstract}

Several studies assert that dark matter (DM) subhaloes without a baryonic counterpart and with an inner cusp always survive.
We conduct numerical simulations to analyse the evolution of the circular velocity peaks ($V_\mathrm{max}$, and its radial value $r_\mathrm{max}$) and concentration of low-mass DM subhaloes under tidal stripping. We employ the improved version of the DASH code, introduced in \citet{2023MNRAS.518...93A}.
We follow the tidal evolution of a single DM subhalo orbiting a Milky Way (MW)-size halo modeled with a baryonic disc and a bulge representing the MW's mass distribution, accounting for the time-evolving gravitational potential of the MW.
We simulate subhaloes with unprecedented accuracy, varying their initial concentration, orbital parameters, and inner slope.
Unlike previous literature, we examine the evolution of subhalo structural parameters --{\it tidal tracks}-- not only at orbit apocentres but also at pericentres, finding in the former case both similarities and differences -- particularly pronounced for prompt cusps.
Overall, $r_\mathrm{max}$ shrinks more than $V_\mathrm{max}$, leading to a continuous rise of subhalo concentration with time. The {\it velocity concentration} at present is around two orders of magnitude higher than the one at infall -- about an order of magnitude more compared to the increase found for field haloes -- being comparatively larger for pericentre tidal tracks versus apocentres.
These findings highlight the importance of tidal effects in reshaping low-mass DM subhaloes, providing insights for future research via simulations and observations, such as correctly interpreting data from galaxy satellite populations, subhalo searches with gravitational lensing or stellar streams, and indirect DM searches.



\end{abstract}




\section{Introduction}


A wealth of cosmological and astrophysical evidence leads us to believe that there should exist a form of dark matter (DM) we cannot directly observe in the Universe, accounting for $\sim 85$ per cent of the total matter content \citep{1992ApJ...396L...1S, Bertone:2004pz, 2020A&A...641A...6P}. %
The current most accepted cosmological model states that it is cold (CDM), i.e. non-relativistic. This implies a bottom-up formation scenario where the smallest bound objects, called haloes, form first, with masses similar to the Earth or less, and later merge generating larger structures, up to $\sim 10^{15} \mathrm{M_\odot}$ haloes hosting galaxy clusters \citep{1990eaun.book.....K, 2022LRCA....8....1A}. %
In this case, haloes would naturally be teemed with substructure, called subhaloes. Indeed, millions of subhaloes are expected in a galaxy like the Milky Way (MW) at present time. The most massive ones would host dwarf satellite galaxies, while less massive objects or \textit{dark satellites} would not host any baryons and therefore would not be visible in the optical spectrum. %

Cosmological simulations represent the best tool to study how these small structures form and evolve as %
these so-called \textit{subhaloes} orbit around their host. These simulations have shown that subhaloes are expected to lose a significant amount of their mass due to tidal stripping \citep[see e.g., ][]{2003ApJ...584..541H, 2019MNRAS.490.2091G, 2020MNRAS.491.4591E, 2024arXiv240804470H} and that a relevant fraction of them might eventually end up completely destroyed \citep[see e.g., ][]{2017MNRAS.471.1709G, 2021MNRAS.507.4953G}. Early simulation work did not take into account the baryonic content of the Universe, i.e. they were fully collisionless N-body simulations \citep[e.g.,][]{2008MNRAS.391.1685S, vlii_paper}. Despite providing a fair and realistic approximation of the evolution of the subhalo population, these DM-only simulations have limitations in describing the central region of galaxies, where most baryons reside and are expected to boost up the tidal force at the Galactic centre, leading subhaloes to more significant tidal mass loss. Nowadays, a plethora of cosmological simulations are available where baryonic feedback is also considered \citep[e.g.,][]{2014MNRAS.444.1518V, 2016MNRAS.457.1931S, 2017MNRAS.467..179G, 2018MNRAS.480..800H}. In most cases, these so-called hydrodynamical simulations do not exhibit as much substructure near the centre of galaxies compare to their DM-only counterparts \citep[see e.g.,][]{2014ApJ...786...87B, 2016MNRAS.458.1559Z, 2017MNRAS.471.1709G, Kelley2019, 2024ApJ...964..123J}. 


These DM haloes and subhaloes have been widely studied and characterised. As discussed in \citet{Penarrubia2010}, there exists considerable debate regarding the inner slope of a DM halo density profile in the absence of gas and stars. This region, often referred to as the "inner cusp," is defined by a steep slope in the inner DM density profile. Its absolute value ranges from 1 to 1.5 or greater \citep[see e.g.,][]{1999MNRAS.310.1147M, 2004MNRAS.353..624D, 2008MNRAS.391.1685S, 2018MNRAS.473.4339O}, suggesting that it might not be truly universal.  The term {\it prompt cusp} has been coined for haloes with a slope equal to -1.5 or even more negative \citep{2017MNRAS.471.4687A,2018MNRAS.473.4339O,2023MNRAS.518.3509D}.

A number of studies \citep[see e.g.,][and references therein]{2023MNRAS.518...93A} assert that subhaloes without a baryonic counterpart and with an inner cusp %
will always survive no matter the strength of the tidal force they undergo. Nevertheless, we witness great subhalo disruption in cosmological simulations \citep[e.g.,][]{Kelley2019,2021MNRAS.501.3558G}, most likely due to artificial behaviour derived from lack of resolution.

Even though the inner cusp of a subhalo should always remain, its structural parameters, such as the maximum circular velocity of the particles within the subhalo and its corresponding radial distance, evolve as they are affected by tidal stripping. Several works \citep{Penarrubia2010, 2019MNRAS.490.2091G, 2021MNRAS.505...18E, 2021arXiv211101148A, 2022MNRAS.517.1398B} %
have studied this evolution as the subhalo orbits their host halo, and point to the existence of a \textit{tidal track}, i.e., the evolution of subhalo structural properties is only affected by the mass loss, and is essentially independent of the initial conditions and orbital parameters at accretion.

In \citet{2023MNRAS.518...93A}, hereafter Paper~I, we improved and employed the DASH code \citep{Ogiya2019} to follow the evolution of low-mass ($\lesssim 10^6 M_\odot$) subhaloes around a MW-like potential with superb resolution. In that work, we mainly investigated two relevant quantities, namely the bound mass fraction $f_\mathrm{b}$ and the subhalo annihilation luminosity $L$ \citep[the latter assuming that subhaloes were composed of annihilating DM such as WIMPs, e.g.][]{Bertone10}. In all cases, a Navarro-Frenk-White \citep[NFW,][]{Navarro1997} DM density profile was adopted for subhaloes, and both DM-only and DM+baryons host haloes were considered. %

The aim of this follow-up work is to provide further insight into these questions. We will keep our focus on the evolution of subhalo structural properties, this time analysing a more extensive parameter space of initial conditions and further increasing our particle resolution with respect to our previous work. %
As in our previous study, our subhaloes will be orbiting a DM host halo with both a baryonic disc and a bulge replicating the mass distribution of the MW. We will also follow in great detail the impact of tidal stripping in an individual cuspy DM subhalo, yet this time not only considering an initial NFW density profile but also a more resilient prompt cusp.  
Furthermore, we also broaden our vision with respect to previous literature, by investigating not only the most relevant quantities at the orbit's apocentres but at the pericentres as well.

Knowing with accuracy the internal structure and fate of orbiting subhaloes is of utter importance for various purposes. Indeed, the subhalo population represents a powerful test of the underlying cosmological model -- e.g., subhaloes would not exist in warm DM cosmologies below a certain mass scale (\citet{2014MNRAS.439..300L} and references therein) or would have very different structural properties in self-interacting DM scenarios \citep{2018PhR...730....1T} %
-- and indirect DM searches, which look for the radiation allegedly generated by the annihilation or decay of DM particles in the form of gamma rays, neutrinos or antimatter, e.g., \citet{2011ARA&A..49..155P}. In other words, achieving a better knowledge of the properties of halo substructure within the MW -- in particular at present time -- may be key to both, definitely support or not the standard cosmological model, and to unveil the nature of DM.

The work is organised as follows. In Section~\ref{sec:simssurv}, we describe the technical details behind our simulations and introduce the main quantities that will be relevant for our study.  
We utilise our brand new suite of generated simulations to derive tidal track relations in Section \ref{sec:tidaltrack}, both for NFW profiles and prompt cusps. A similar exercise is performed in Section \ref{sec:modelcv} to obtain a relation between maximum circular velocity and subhalo concentration. %
Our conclusions are given in Section \ref{sec:conclu}.

\section{Simulations} \label{sec:simssurv}
We perform simulations of a single    
DM subhalo, modelled as an $N$-body system, and study its dynamical evolution %
as it orbits within an analytically defined, temporally varying potential resembling that of the MW, 
using the DASH code \citep{Ogiya2019}. The mass and concentration of the host halo grow up, respectively, to $10^{12} \mathrm{M_\odot}$ and nearly $10$ at present time, and the scale radius of the stellar disk at $z=0$ is $2.5$ kpc.  Full details can be found in Section 2 and Figure 1 of \citetalias{2023MNRAS.518...93A}. While in that work we considered both MW DM-only hosts and MW potentials including baryonic components -- stars, gas and bulge --, this study 
 focuses solely on the latter, more realistic scenario. 
We recall that we do not incorporate actual baryonic feedback in our simulations but time-evolving disc-like and bulge-like mass distributions in addition to the DM halo. 
Our subhaloes are always completely dark, i.e. devoid of baryons, as their masses are below the minimum expected for them to host baryons inside \citep{2015MNRAS.448.2941S, 2016MNRAS.456...85S, 2025ApJ...983L..23N}. 
More precisely, our default subhalo mass is a million solar masses, although our results can be easily extrapolated to smaller masses. 
Indeed, we decide not to vary the initial subhalo mass in the present work, since (normalised) results are independent of $m_\mathrm{sub}$ when the ratio with respect to the mass of the host is small enough (i.e. $<1/1000$ if the host is DM-only, and $<1/10000$ if baryons are considered) for both self-friction and dynamical friction to become negligible, as it is the case~\citep[][\citetalias{2023MNRAS.518...93A}]{vandenBosch2018_analytic,Ogiya2019, 2020MNRAS.495.4496M}. %
This eases making predictions for subhalo masses spanning many orders of magnitude, down to the smallest ones, with Earth masses or lower in CDM. %

Our initial subhalo density profile is defined by the generalised NFW parametrisation \citep{1996MNRAS.278..488Z, Navarro1997, 2006ApJ...641..647K},

\begin{equation}
    \rho(r) = 4\rho\sub{s} (r/r\sub{s})^{-\gamma} [1+(r/r\sub{s})^\alpha]^{-(\beta-\gamma)/\alpha},
        \label{eq:nfw}
\end{equation}
where  $ \rho_\mathrm{s}$ is the scale density, $ r_\mathrm{s}$ is the scale radius of the halo,  $\alpha = 1$, $\beta = 3$, and $\gamma$, the inner slope, is either $1$ for an NFW or $1.5$ for a prompt cusp. The DM within our host halo follows an NFW DM profile as well, with a total mass of $10^{12} \mathrm{M_\odot}$. 

As in \citetalias{2023MNRAS.518...93A}, the parameters we vary are the following: 
initial subhalo concentration $c = c_{200} = R_\mathrm{sub,vir} / r_\mathrm{s}$ with $R_\mathrm{sub,vir}$ the initial subhalo virial radius;   
subhalo accretion redshift $z_\mathrm{acc}$; orbital energy parameter $ x_\mathrm{c} \equiv r_\mathrm{c}(E)/r_\mathrm{200,host}(z_\mathrm{acc}),$ where $r_\mathrm{c}(E)$ and $r_\mathrm{200,host}(z_\mathrm{acc})$ are the radius of a circular orbit with orbital energy $E$, and the virial radius of the host halo at the subhalo accretion redshift, $z_\mathrm{acc}$, respectively; orbit circularity $\eta = L/L_\mathrm{c}(E)$ where $L$ and $L_\mathrm{c}(E)$ are the actual angular momentum of the subhalo orbit, and the angular momentum of the circular orbit with energy $E$, both at accretion time; and orbit inclination angle with respect to the baryonic disc, $\theta$, with 0 corresponding to a parallel orbit. Our fiducial setting remains the same as well: $m_\mathrm{sub} = 10^6~\mathrm{M_\odot}$, $c = 10$, $z_\mathrm{acc} = 2$, $x_\mathrm{c}=1.2$, $\eta=0.3$ and $\theta = 45$~deg. In the present work, we vary an additional parameter: the slope of the inner DM density profile, $\gamma$, which is equal to $1$ in our fiducial case, i.e. following a standard NFW. The simulation setup and parameters are summarised in Table~\ref{tab:fiduset}.

Here, we are mainly focusing\footnote{While we cannot determine the exact value of $f_\mathrm{b}$ at $z=0$ for a given set of initial parameters before running the simulation, we aim to explore the region of parameter space where $f_\mathrm{b} < 0.01$. See Appendix~\ref{sec:apsurvive} for details. } 
on subhaloes which have lost a significant amount of their initial mass, at least 99\%, at present time. Thus, our initial concentration values do not reach as high values as in \citetalias{2023MNRAS.518...93A}, so as to avoid too resilient subhaloes. This has the caveat of needing very high particle resolutions in order to obtain robust, converged results. Specifically, results may suffer from lack of resolution when the number of particles drops below several thousands \citep[\mbox{$\sim3000$} inside $r_\mathrm{max}$, according to][]{2021MNRAS.505...18E}. We use a total number of particles $N=2^{25}$ in most cases, thus untrustable results may appear when $\log_{10} f_\mathrm{b} \lesssim -3.5 $. 
We find this limit very reasonable though in order to achieve relevant results and predictions. We have explored our survival criteria and initial parameters space region further in Appendix~\ref{sec:apsurvive}.

\begin{table}
	\centering
	\caption{Set of parameters used in this work, described in Section~\ref{sec:simssurv}. The `fiducial' column refers to fiducial values adopted for each of the parameters. The last column depicts the full range of values studied in the full suite. %
	}
	\label{tab:fiduset}
	\begin{tabular}{c|cc} 
		\hline
		  & fiducial & suite  \\
		  \hline
		  $m_\mathrm{sub} [\mathrm{M}_\odot]$ & $10^6$ & $10^6$ \\
		  $z_\mathrm{acc}$ & $2$ & $[1,4]$ \\
		  $c$ & $10$ & $[5,30]$ \\
		  $ \eta$ & $0.3$ & $[0.1,0.8]$ \\
		  $x_\mathrm{c}$ & $1.2$ & $[0.8,1.6]$ \\
		  $\theta $ [deg] & 45 & $[0,90]$ \\
            $\gamma$ & $1$ & $1, 1.5$ \\
		\hline 
	\end{tabular}
\end{table}



From here, we will elaborate on the quantities relevant for this work. In addition to $f_\mathrm{b}$, we will focus on the evolution of particle circular velocities, namely their maximum value or maximum circular velocity, $V_\mathrm{max}$, and the radius at which this occurs or $r_\mathrm{max}$. We will also investigate the subhalo concentration derived from these two quantities, referred as \textit{velocity concentration}, $c_\mathrm{V}$, from now on. Recall that these parameters are more robustly obtained %
in cosmological simulations compared to the subhalo mass, %
as both $V_\mathrm{max}$ and $c_V$ are less prone to tidal forces than subhalo masses and derived quantities. Furthermore, as they do not depend on the assumed functional form for the subhalo DM density profile, a virial radius, which would be virtual after the stripping, is not even needed \citep{Moline17}.

\subsection{Circular velocities}

Circular velocity profiles can be extracted from  subhalo mass profiles using the following expression:

\begin{equation}
    V_\mathrm{circ} (r) = \sqrt{\frac{G M(<r)}{r}},
\end{equation}
 where the cumulative mass $M(< r)$ includes all particles within the radius $r$, and $G$ is the gravitational constant.

For our purposes, we need to characterize in great detail the subhalo density and velocity profiles. We have chosen to divide our radial interval, $x = r / R_\mathrm{sub,vir} \in [10^{-3}, 10]$, 
in 200 evenly spaced bins on a logarithmic scale. We also need to employ a large enough number of particles to run the simulation so as to avoid noise in the innermost region. This number depends on how much the subhalo material is stripped, but it will typically range between $N \in [2^{22}, 2^{25}]$, always being a power of 2.

\begin{figure}
\begin{center}
\includegraphics[width=\columnwidth]{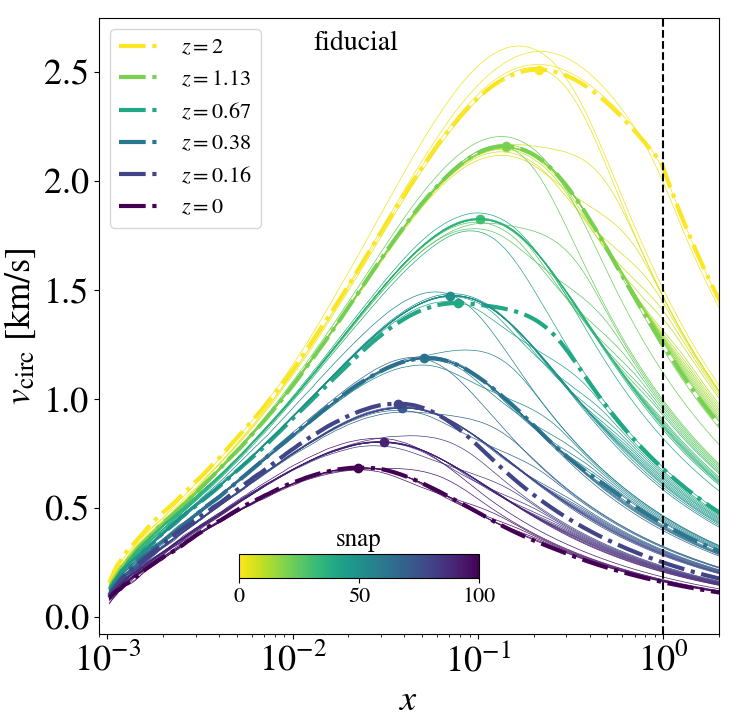}
\caption{Circular velocities for each snapshot in a simulation with our fiducial parameters reported in Table~\ref{tab:fiduset}. The $x$ axis is the radius normalised to the subhalo virial radius at accretion. 
Colour represents time, namely snapshot number, being yellow the beginning of the run, down to the purple at the bottom corresponding to $z=0$. Circles are drawn for the $V_{\mathrm{max}}$ found after every 10 snapshots, while lines every 20 snapshots are labeled in the legend and highlighted in dash-dotted and thicker layout. The dashed vertical line corresponds to $R_\mathrm{sub,vir}$.
} 
\label{fig:fig1}
\end{center}
\end{figure}

Fig.~\ref{fig:fig1} shows the evolution of the subhalo circular velocities for our fiducial set of parameters in Table~\ref{tab:fiduset}. %
From these profiles, we extract values of $V_\mathrm{max}$ 
and $r_\mathrm{max}$ 
 for each snapshot, interpolating our binned results via cubic splines to obtain more reliable values.
 We find that $r_\mathrm{max}$ is reduced, and $V_\mathrm{max}$ also decreases with time, agreeing with results in previous studies \citep{2003ApJ...584..541H, 2008ApJ...673..226P}. In this particular case, $r_\mathrm{max,i} / r_{\mathrm{max}, z=0} = 8.3 $ and $V_\mathrm{max,i} / V_{\mathrm{max}, z=0} = 4.1 $, where the subscript $i$ indicates the initial value of that quantity. 

\subsection{Velocity concentrations}\label{sec:vconc}

The concentration of a subhalo is a structure parameter which gives an intuition of the density of its inner region. This parameter is well defined and `stable' for field haloes, since once formed they are not expected to alter their internal structure significantly. Assuming they follow an NFW density profile (Equation~\ref{eq:nfw}), %
we can define a ``virial'' concentration for them \citep{2001MNRAS.321..559B, 2002ApJ...568...52W, 2008MNRAS.391.1940M, mascprada14}, $c_{200} = R_\mathrm{vir} / r_\mathrm{s}$, where $R_\mathrm{vir}$ is the virial radius and $r_\mathrm{s}$ is the scale radius of the halo, i.e., the radius where the slope of the density profile is equal to $-2$. Therefore, it is reasonable to use this parameter when we initialise our subhalo, since it was a field halo until the start of the simulation, i.e., the accretion time.

Subhaloes are tidally stripped, i.e., the mass of their outskirts is eventually removed, thus concentration values relying on halo virial radius are not possible, as the latter does not exist anymore \citep[see e.g.][]{1998MNRAS.300..146G}. 
 In the absence of a virial radius, subhalo concentrations adopting a scale radius and a tidal radius have been studied in previous works though \citep[see, e.g., ][]{Moline17}. However, there is an alternative, preferable definition of the concentration for subhaloes, which involves both $V_\mathrm{max}$ and $r_\mathrm{max}$, and that is more reliable, as it does not depend on the adoption of a particular density profile \citep{vlii_paper, Moline17, 2023MNRAS.518..157M}:

\begin{equation}
\label{eq:cv-def}
c_{\rm V} = 2\left( \frac{V_{\rm{max}}}{H(z) \, R_{\rm{max}}}\right)^{2} \,,
\end{equation}

where $H(z)$ is the Hubble parameter, $H(z)=H_{0}\,\sqrt{\Omega_{\rm{m},0}(1+z)^3+\Omega_{\Lambda,0}}\equiv H_{0}\,h(z)$. Here, $\Omega_{\rm{m},0} = 0.308$ and $\Omega_{\Lambda,0} = 0.692$ are, respectively, the DM and dark energy density parameters at present time in our standard cosmological model, and $H_0 = 67.8\, \mathrm{km}\, \mathrm{s}^{-1}\, \mathrm{Mpc}^{-1}$ is the Hubble constant. 

\begin{figure}
\begin{center}
\includegraphics[width=\columnwidth]{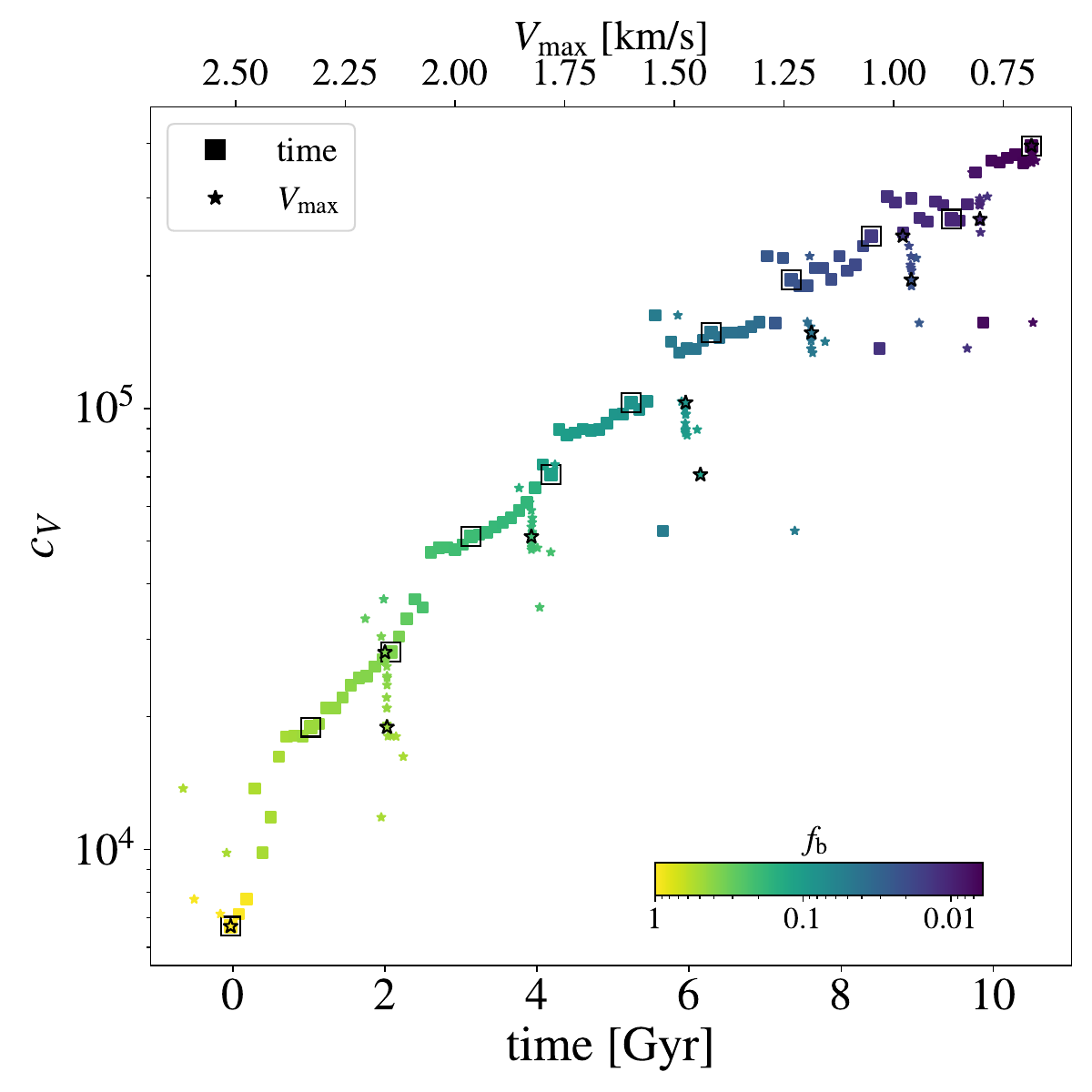} 
\caption{ Evolution of $c_\mathrm{V}$ as a function of $V_\mathrm{max}$ (filled stars, top $x$ axis) and time (squares, bottom $x$ axis) for our fiducial run (Table~\ref{tab:fiduset}). Colour indicates the bound mass fraction $f_\mathrm{b}$, with yellow denoting the beginning of the simulation. Black hollow markers are drawn after every 10 snapshots. }
\label{fig:fig3}
\end{center}
\end{figure}

Figure~\ref{fig:fig3} shows the evolution of subhalo structural properties across several orbits for our fiducial setup. We find subhalo velocity concentrations to increase with time, primarily because $r_\mathrm{max}$ decreases more rapidly than $V_\mathrm{max}$ (see Fig.~\ref{fig:fig1}). This effect is most pronounced for the first couple of orbits, when the largest amount of mass is stripped. The concentration nearly triples after the first orbit, and rises by $\sim1.75$ orders of magnitude over the course of the simulation (in contrast to the $\sim$ order of magnitude increase for field haloes), while there is a 75\% reduction in $V_\mathrm{max}$. Configurations other than fiducial will be discussed later in the paper, where we explore the impact of varying the initial concentration and orbital parameters.

\section{Tidal track relation }\label{sec:tidaltrack}

Tidal stripping does not affect only the subhalo outskirts; its inner region is impacted as well. More specifically, the subhalo  is dynamically heated up and expands inside its tidal radius \citep{2021MNRAS.505...18E}. This has an influence on the subhalo 
circular velocity profile. When following the subhalo evolution, we find that, on average, both $V_\mathrm{max}$ and $r_\mathrm{max}$ decrease with time. We have witnessed this behaviour for our fiducial setting in the previous Section. From the literature \citep[see e.g.,][]{2008ApJ...673..226P, Penarrubia2010, 2019MNRAS.490.2091G, 2021MNRAS.505...18E}, %
we know this holds in general. Actually, this evolution of both quantities has been proposed to be essentially independent of the initial subhalo parameters, depending just on the amount of mass lost or, in other words, the value of $f_\mathrm{b}$. This is the so-called \textit{tidal track} and has been widely studied \citep[and references above]{2022MNRAS.517.1398B, 2024PhRvD.110b3019D} using the apocentre values, where the subhalo is more stable. In this work, we want to dig up further and explore the behaviour in the pericentres and during the whole orbit.

Including baryonic mass distributions within the host is indeed crucial, as it provides a more comprehensive understanding of the dynamics involved and subhaloes mass stripping is much more noticeable. Many previous studies have often ignored this aspect, focusing primarily on DM-only host potentials. Stars and gas distributions, which are disky in opposition to the spherical DM halo, significantly contribute to the total gravitational potential and influence the overall dynamics of orbiting subhaloes  \citep[\citetalias{2023MNRAS.518...93A}]{2022arXiv220700604S}. %

Although \citet{Penarrubia2010} made comparisons between simulations including baryonic mass distributions within the host and DM-only runs, most of the works \citep{2019MNRAS.490.2091G, 2021MNRAS.505...18E, 2022MNRAS.517.1398B, 2024PhRvD.110b3019D} %
overlook the former setting, which is in fact the most realistic one. Moreover, none of them consider the time evolution of the host. %
From now on, we will refer to several of these works as: P10 \citep{Penarrubia2010}, EN21 \citep{2021MNRAS.505...18E}, and D24 \citep{2024PhRvD.110b3019D}.

Section \ref{sec:tidaltrack1} is dedicated to a thorough study of the orbital evolution of our fiducial setting, paying particular attention to how $V_\mathrm{max}$ and $r_\mathrm{max}$ change throughout a single orbit. Section \ref{sec:tidaltrackNFW} analyses the tidal track for standard NFW profiles. Lastly, Section \ref{sec:tidaltrackpc} deals with subhaloes exhibiting a prompt cusp to build the tidal track.

\subsection{Delving into our fiducial setting}\label{sec:tidaltrack1}

Here we want to explore in depth the evolution of $V_\mathrm{max}$ and $r_\mathrm{max}$ as well as take a closer look at one single orbit at a time. To do this, we have run a simulation with our fiducial set of parameters depicted in Section~\ref{sec:simssurv}, $N = 2^{25}$ particles, and 500 snapshots. 

\begin{figure}
\begin{center}
\includegraphics[width=\columnwidth]{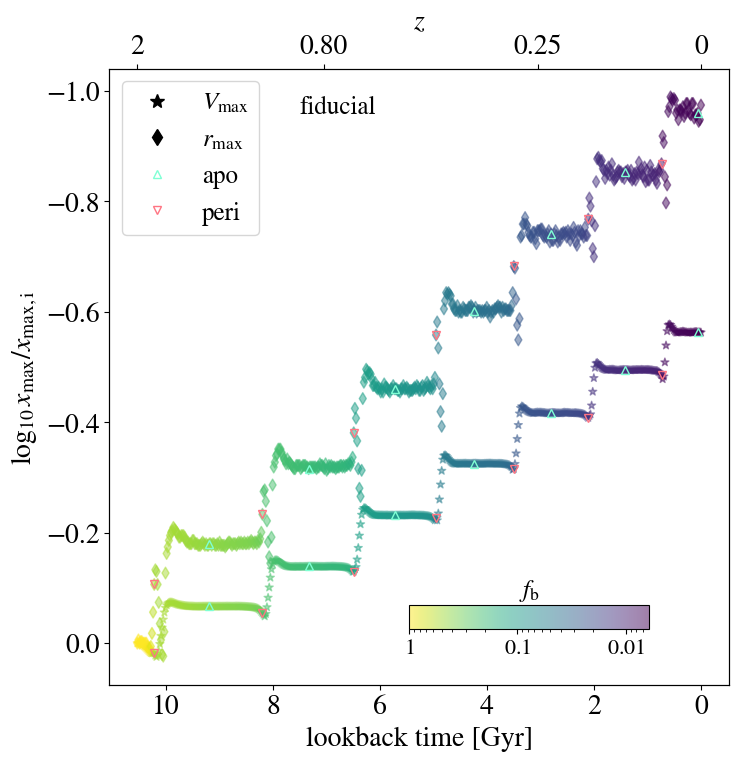} 
\includegraphics[width=\columnwidth]{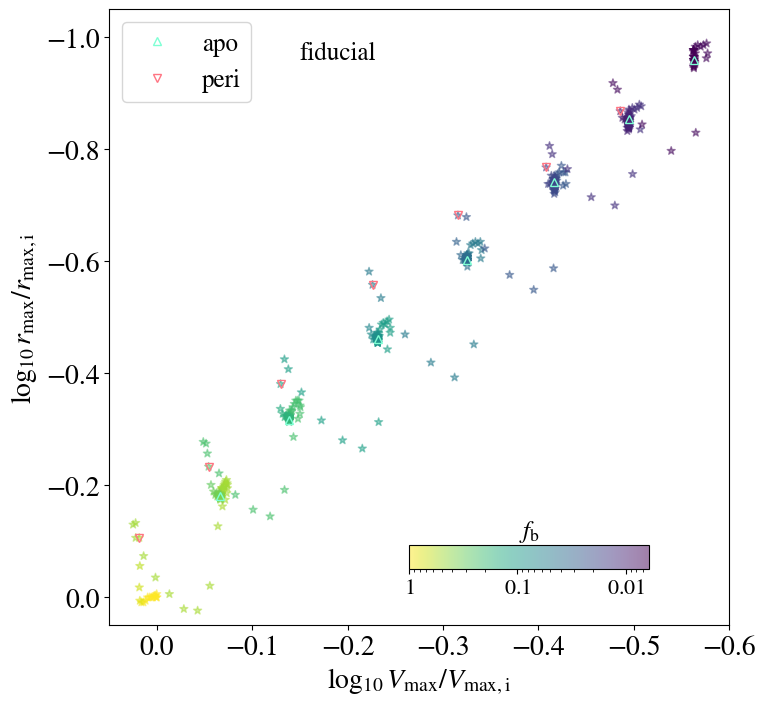} 
\caption{Evolution of $V_\mathrm{max}$ and $r_\mathrm{max}$ normalised to their initial values throughout the whole life of a subhalo since its accretion (yellow) until present day (purple). Top panel: Each quantity against the time (lower axis) or redshift (upper axis). Bottom panel: One quantity against the other.
Apocentres are highlighted as aquamarine hollow triangles while pericentres appear as red and inverted. 
Changes occur near the pericentres. $V_\mathrm{max}$ decreases later than $r_\mathrm{max}$. Besides, $r_\mathrm{max}$ decreases more. See main text for details. 
}
\label{fig:figttorbit1}
\end{center}
\end{figure}

In the top panel of Fig.~\ref{fig:figttorbit1}, the $x$ axis displays the time while $V_\mathrm{max}$ (stars) and $r_\mathrm{max}$ (triangles) share the $y$ axis, and the colour represents $f_\mathrm{b}$. In contrast, the bottom panel shows how $r_\mathrm{max}$ changes against $V_\mathrm{max}$. As we saw in Section \ref{sec:simssurv}, the final $V_\mathrm{max}$ after seven pericentric passages is about four times smaller than the initial one, while the respective $r_\mathrm{max}$ at $z=0$ is nearly one order of magnitude smaller, which implies an increase in the subhalo concentration. Both apocentres (teal triangles) and pericentres (red inverted triangles) are highlighted and illustrate how most drastic changes occur around the pericentre, with $V_\mathrm{max}$ getting reduced later than $r_\mathrm{max}$.

\begin{figure}
\begin{center}
\includegraphics[width=\columnwidth]{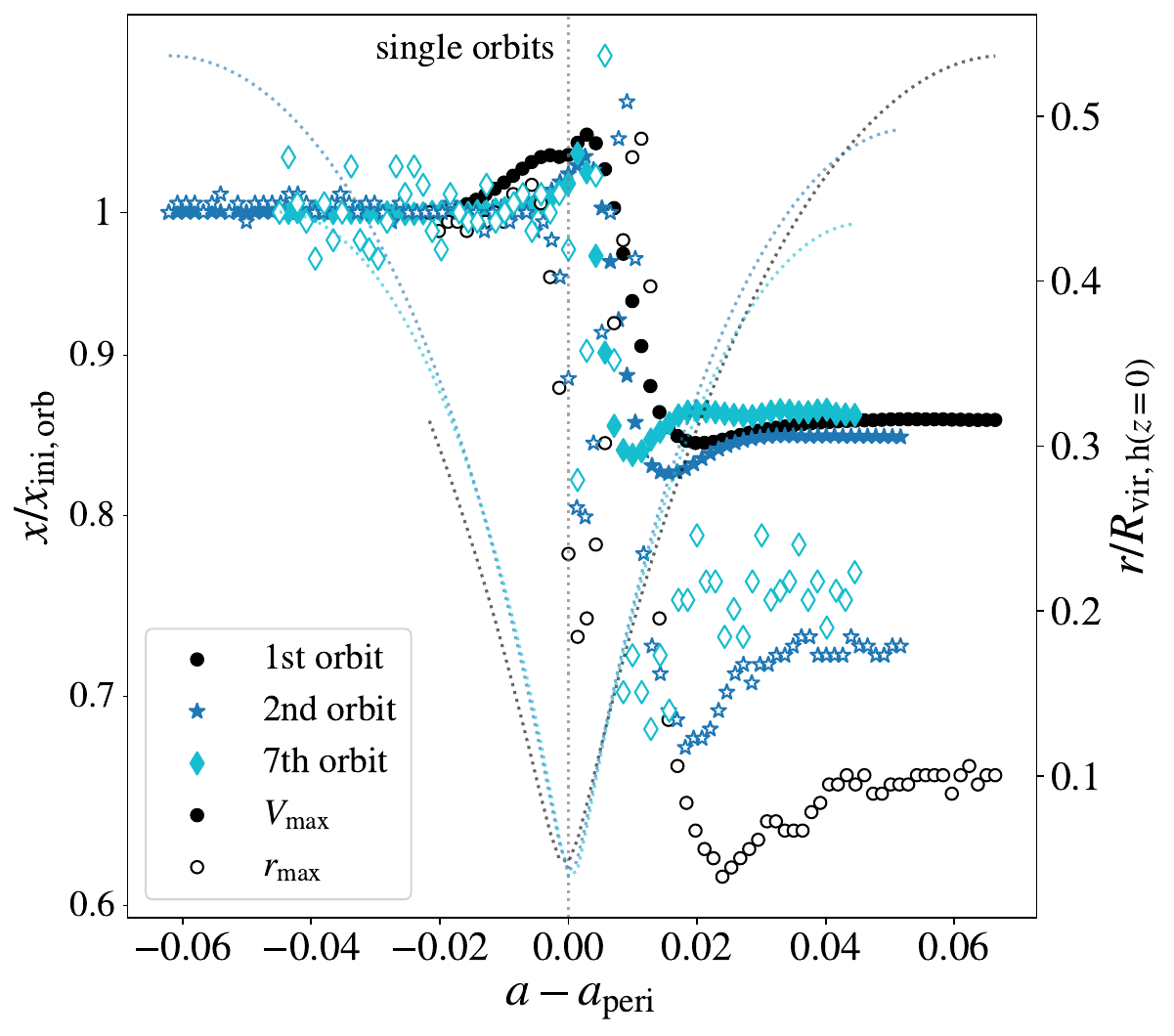} 
\includegraphics[width=\columnwidth]{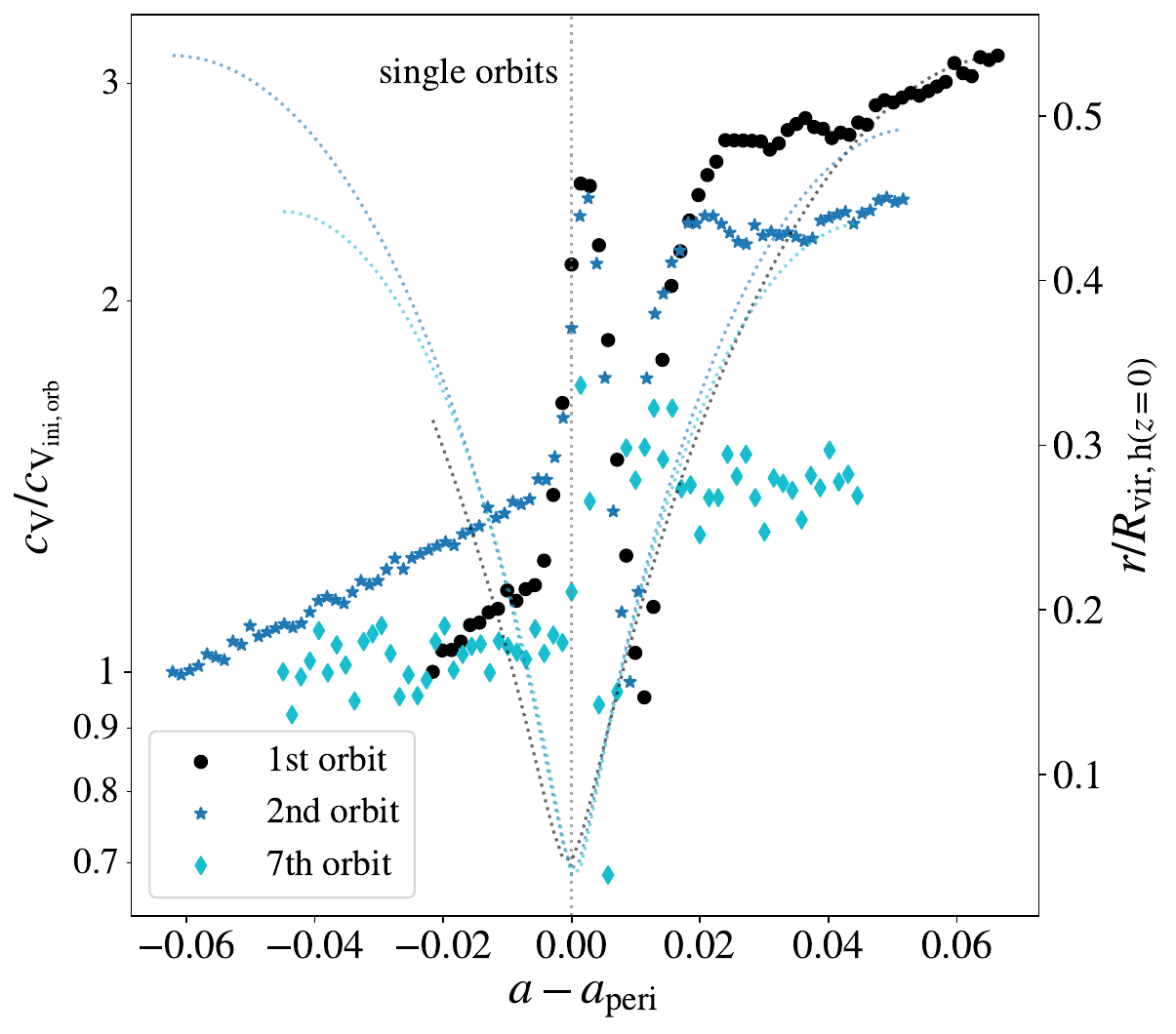}  
\caption{Top panel: Evolution of $V_\mathrm{max}$ (filled markers) and $r_\mathrm{max}$ (hollow markers) as a function of the scale factor $a$, subtracting the value of the respective pericentre, throughout a single orbital period.  Only the two first orbits and the last one are shown, with black circles, blue stars and cyan diamonds, respectively. Values are initialised at the first value of the respective orbit, which is the beginning of the simulation in the former case, and the apocentre in subsequent ones. Dotted curves show the evolution of the distance between the subhalo and the host halo centre for the three mentioned orbits, adopting the same colour scheme and normalised by the virial radius of the host at present.  Bottom panel: Evolution of $c_\mathrm{V}$, showing an increase of the velocity concentration throughout each orbit, which is more significant at the beginning.
} 
\label{fig:figttorbit4}
\end{center}
\end{figure}

Fig.~\ref{fig:figttorbit4} focuses on single orbits: the first, the second, and the last one. The top panel is obtained by dividing the values of $V_\mathrm{max}$ and $r_\mathrm{max}$ at each snapshot by their respective values at the previous apocentre, except for the points before the first apocentre, in which case the values are divided by the 
initial ones (i.e., first snapshot). This way, we can zoom in and study whether the behaviour during each orbit changes. 
This pattern likely arises from tidal compression at pericentre, where strong tidal forces lead to a reduction in $r_\mathrm{max}$. As the subhalo moves away from the host, it becomes more relaxed and $r_\mathrm{max}$ increases again. The continued stripping of material from the subhalo causes both $V_\mathrm{max}$ and $r_\mathrm{max}$ to decrease. By the next apocentre, the system stabilises at its new equilibrium. 
With each completed orbit, the subhalo's $r_\mathrm{max}$ decreases more significantly than its $V_\mathrm{max}$. However, the $V_\mathrm{max}$ reduction is very similar for all orbits, with the ratio relative to its initial value at the beginning of the orbit always being about 0.85. Contrariwise, the final ratio of $r_\mathrm{max}$ after an orbit is smaller for the first passage, approximately 0.65, compared to the subsequent ones, reaching approximately 0.8. Therefore, the increase in velocity concentration is greater -- a factor 3 increment -- at the beginning, as illustrated on the bottom panel of the same figure. Nonetheless, the different $r_\mathrm{max}$ ratio induces a less pronounced change in concentration for successive orbits -- a factor 1.3 after the last.

\subsection{NFW apocentres and pericentres}\label{sec:tidaltrackNFW}

We have seen in Section~\ref{sec:tidaltrack1} that both $V_\mathrm{max}$ and $ r_\mathrm{max}$ vary during a single orbit. 
The tidal track has usually been studied using only the apocentres of the subhalo orbit, because the subhalo is expected to be closer to dynamical equilibrium
as the tidal forces are less strong. We have included the same procedure in our study to be able to compare with other works in the literature.
In addition, we have explored the tidal track of the pericentres, where the opposite happens: the object is undergoing the strongest tidal forces. This is a new path no one has walked before. %

\begin{figure}
\begin{center}
\includegraphics[width=\columnwidth]{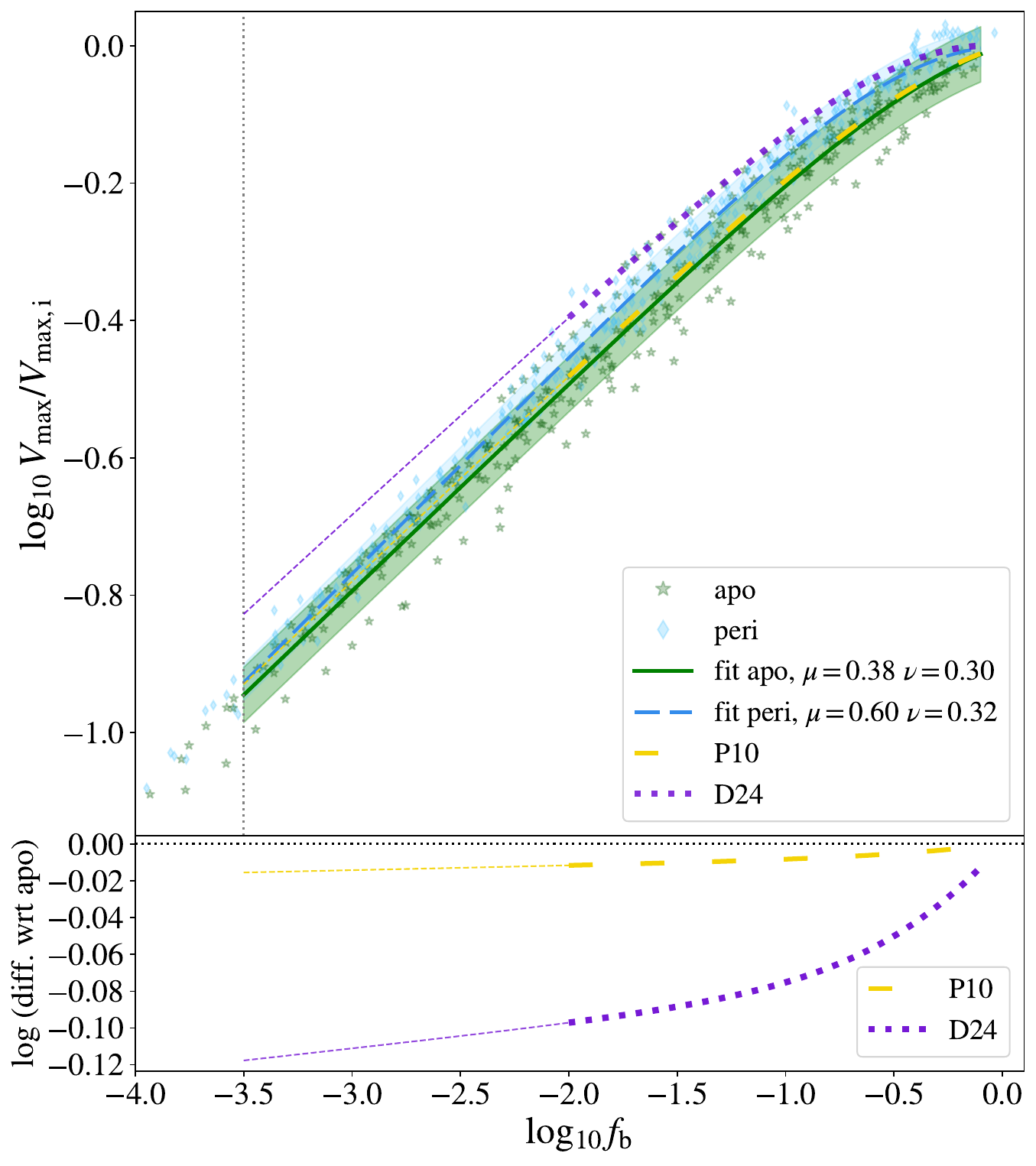}
\caption{Relation between $f_\mathrm{b}$ 
and $V_\mathrm{max}/V_\mathrm{max,i}$, for the 
apocentres (green stars) and the pericentres (sky blue diamonds). Subhaloes have an NFW density profile at accretion. The tidal track found for each subset using Eq.~\ref{eq:tidalP10} is drawn as a solid green line in the former case and a dashed sky blue line in the latter, with shadowed bands for their respective scatter. The fits from \citetalias{Penarrubia2010} (loosely dashed yellow line) and \citetalias{2024PhRvD.110b3019D} (purple dotted line) are included, and thinned when they are extrapolated. Data can be trusted to the right of the vertical dotted line. We have adjoined a lower panel with the difference between our best fit for the apocentres and the literature fits, calculated as $\log_{10}$ [our fit using apocentres] $ - \log_{10}$ [fit in the literature].
} 
\label{fig:ttvmaxfb}
\end{center}
\end{figure}

This tidal track can be analysed in several ways, since there are three variables which depend on one another:  $V_\mathrm{max}, r_\mathrm{max}$ and $f_\mathrm{b}$.
Fig.~\ref{fig:ttvmaxfb} depicts the relation between $V_\mathrm{max}$ and $f_\mathrm{b}$, which is indeed essentially the same for all subhaloes considered in our sample, with some scatter. %
The tidal track between $r_\mathrm{max}$ and $f_\mathrm{b}$ is explored in Appendix~\ref{sec:aptt}. 

The standard fitting function in the literature, given by \citetalias{Penarrubia2010},
\begin{equation}\label{eq:tidalP10}
    g(x) = \frac{2^\mu x^\nu}{(1+x)^\mu} ,
\end{equation}
where $g(x)$ can be either $V_\mathrm{max}$ or $r_\mathrm{max}$ divided by their initial values, and $x = f_\mathrm{b}$, 
has been used to perform the fit, both for $V_\mathrm{max}$ and $r_\mathrm{max}$. %
Our best-fit values are displayed in Table~\ref{tab:ttbestfit} along with values from the literature, and the respective lines are also included in Fig.~\ref{fig:ttvmaxfb}. 
We find a larger scatter for the apocentres than for the pericentres, and a stronger curvature in the latter case for modest mass losses, while the apocentre tidal track behaves as a power law quicker. When the subhalo has been significantly depleted ($\log_{10} f_\mathrm{b} < - 3$), the pericentre tidal track reaches the same values as the apocentre one, since tidal stripping has less effect as time goes by for NFW or cuspier subhaloes %
\citep[\citetalias{2023MNRAS.518...93A}]{2022arXiv220700604S}. Before that, $V_\mathrm{max}$ is larger at the same $r_\mathrm{max}$ for the pericentre values. 

When only the apocentres are considered, our fit 
is very similar to the one given by \citetalias{Penarrubia2010}. If we make use of the pericentre values instead, the fit behaves similarly to \citetalias{2024PhRvD.110b3019D} for larger values of $f_\mathrm{b}$ and $V_\mathrm{max}$, i.e., the first snapshots of the simulations, but then deviates and gets closer to the relation held by our apocentre points. Recall that the tidal track given by \citetalias{2024PhRvD.110b3019D} has been obtained with the apocentres. Moreover, they are using just a couple of different subhalo initial conditions, both of them within a DM-only host potential, so their simulations do not encompass the whole parameter space. Note as well that their definition of the bound mass fraction does not allow particles to become bound again once they have been unbound from the subhalo, leading to smaller bound masses than those found in our simulations -- about a factor of two lower than ours for the same $V_\mathrm{max}$ ratio.

\begin{figure}
\begin{center}
\includegraphics[width=\columnwidth]{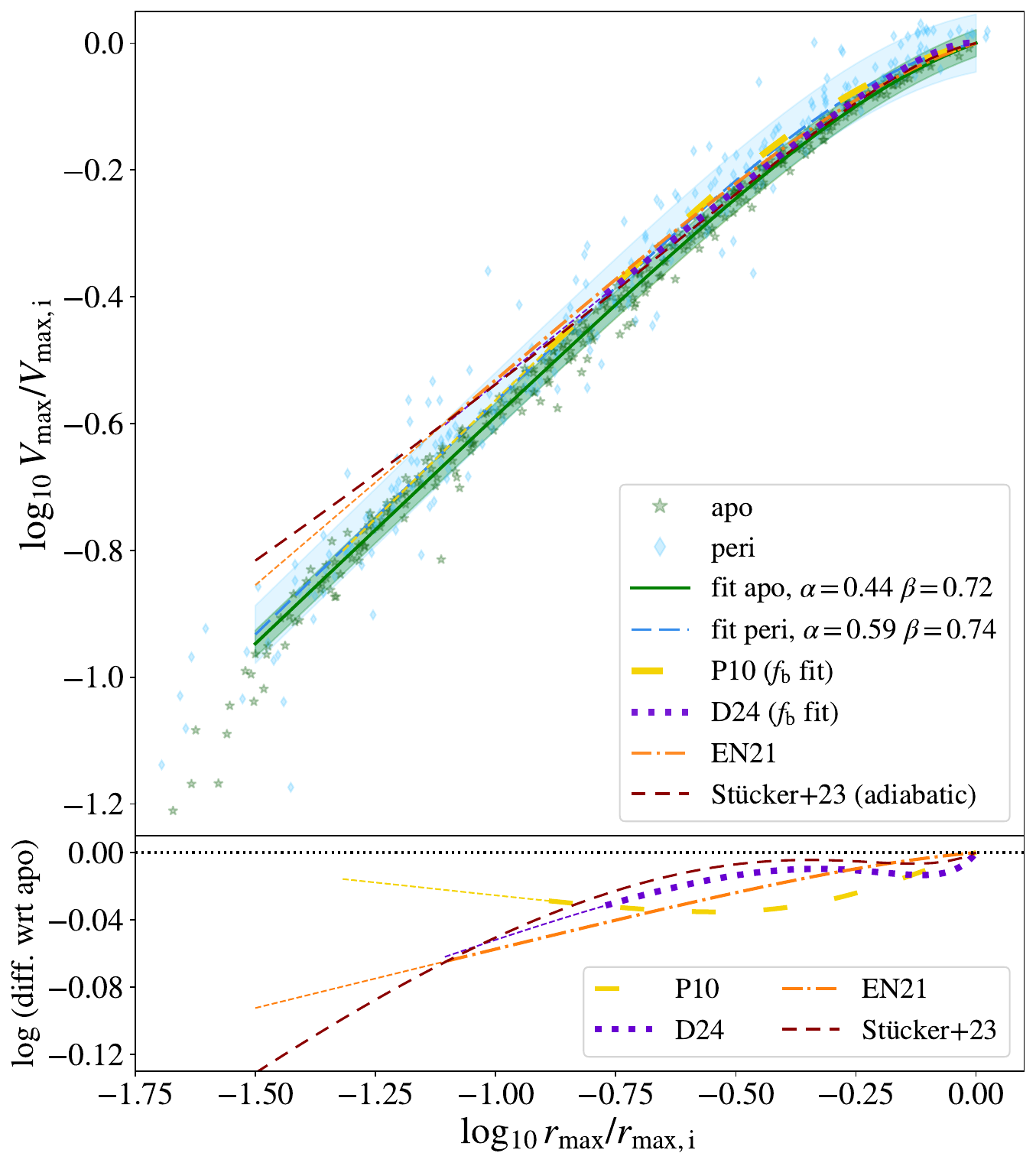} 
\caption{
Relation between $r_\mathrm{max}$ and $V_\mathrm{max}$, divided by their initial values, for the 
apocentres (green stars) and the pericentres (sky blue diamonds). Subhaloes have an NFW density profile at accretion. The tidal track found for each subset using Eq.~\ref{eq:tidalEN21} is drawn as a solid green line in the former case and a dashed sky blue line in the latter, with shadowed bands for their respective scatter. The fits derived from \citetalias{Penarrubia2010} (loosely dashed yellow line) and \citetalias{2024PhRvD.110b3019D} (purple dotted line), the fit from \citetalias{2021MNRAS.505...18E} (dash-dotted orange line), and the result from \citet{2022arXiv220700604S} (dashed dark red line) are included, and thinned when they are extrapolated. We have adjoined a lower panel with the difference between our best fit for the apocentres and the literature fits. 
%
} 
\label{fig:ttvmaxrmax}
\end{center}
\end{figure}

We have also computed the tidal track using the subhalo structural parameters $V_\mathrm{max}$ and $r_\mathrm{max}$. The relation is shown in Fig.~\ref{fig:ttvmaxrmax} and the best-fit parameters are added in Table~\ref{tab:ttbestfit}. In this case, we perform our fits making use of the following equation from \citetalias{2021MNRAS.505...18E}:

\begin{equation}\label{eq:tidalEN21}
    \frac{V_\mathrm{max}}{V_\mathrm{max,i}} = 2^\alpha \left( \frac{r_\mathrm{max}}{r_\mathrm{max,i}} \right)^\beta \left[ 1 + \left(\frac{r_\mathrm{max}}{r_\mathrm{max,i}} \right)^2 \right]^{-\alpha}  , 
\end{equation} 

and we overplot their fitting curve over our findings, as well as the derived ones from \citetalias{Penarrubia2010} and \citetalias{2024PhRvD.110b3019D} -- they do not give a fit of these two parameters, yet the curve can easily be obtained from the other two fits they perform. In this case, the scatter is larger for the fit using the pericentres, which can be explained since the internal structure of the subhalo is hugely impacted. Comparing with earlier work, once more \citetalias{Penarrubia2010} is the closest to our findings for the apocentres (although our pericentre tidal track is more similar to their apocentre tidal track). Nonetheless, \citetalias{2024PhRvD.110b3019D} and \citetalias{2021MNRAS.505...18E} predict higher values of $V_\mathrm{max}$ for the same $r_\mathrm{max}$, especially for lower values. 
We find the tidal track to depend slightly on the pericentre-apocentre ratio. 
The dependence of the tidal track on different initial parameters has been explored in Appendix~\ref{sec:aptt} and can explain both our scatter and our mismatch with \citetalias{2021MNRAS.505...18E}. Note as well that the density profile of their host halo is not modeled as an NFW.

In this case, we have included the semi-analytical result of adiabatically tidally stripped subhaloes given by \citet{2022arXiv220700604S}. Our result behaves similarly for $\log_{10} (r_\mathrm{max} / r_\mathrm{max,i}) \gtrsim -0.5$, while it gives lower $V_\mathrm{max}$ values to the left and a steeper power-law behaviour. Note that in \citet{2022arXiv220700604S} the tidal field is increased infinitely slowly and is isotropic.

When $\log_{10} (r_\mathrm{max} / r_\mathrm{max,i}) \lesssim -1.5$, our runs are affected by artificial two-body relaxation (see Appendix~\ref{sec:apsurvive} for details), and results in this regime cannot be considered reliable. This threshold corresponds to approximately $\log_{10} (V_\mathrm{max} / V_\mathrm{max,i}) \lesssim -0.95$ and $\log_{10} f_\mathrm{b} \lesssim -3.5$. A vertical dotted line is drawn in Fig.~\ref{fig:ttvmaxfb} to indicate this limit. We are aware of the fact that potential resolution issues may be affecting our results even at larger $r_\mathrm{max}$ ratios, $\log_{10} (r_\mathrm{max} / r_\mathrm{max,i}) \lesssim -0.75$, where discrepancies already emerge when comparing to \citet{2022arXiv220700604S}. According to their findings, which are consistent with \citet{2021arXiv211101148A}, heavily stripped subhaloes that retain their inner cusps should follow a tidal track described by a power law with a slope approaching 0.5 for small values of the structural parameters. However, our results indicate larger values of $\beta$ ($\gtrsim0.7$), implying steeper power laws. Artificial two-body relaxation, that transforms the central density cusp of the subhalo to a cored structure \citep{2003MNRAS.338...14P,2008gady.book.....B}, could impact the accuracy of the circular velocity profiles and, in turn, the values of $V_\mathrm{max}$ and $r_\mathrm{max}$ obtained.  
Moreover, truncating the density profiles at $x = 10^{-3}$ can also impact the calculation of $V_\mathrm{max}$. 
Addressing this issue in enough detail is not a trivial task, since a definitive answer is expected to be coupled to an extremely high computational cost. Indeed, further checks in this direction were  performed, with inconclusive results as of today. Yet, we do not consider this a major issue, as the differences are not substantial in the range of parameters tested by our simulations. Further investigation is left for future work. 

We agree with \citetalias{Penarrubia2010} to a certain extent on the fact that, if we assume NFW density profiles, the evolution of the structural parameters of the subhaloes, i.e. $V_\mathrm{max}$ and $r_\mathrm{max}$, essentially depends on how much mass they have lost, and not on how this mass has been stripped. Nonetheless, we also find some scatter in this relation, even if we only consider the apocentres. As shown in Fig. 6 of \citetalias{2021MNRAS.505...18E}, the ratio between the apocentre and the pericentre of the orbit may play a role. Other relevant factors include the concentration of the subhalo \citep{2019MNRAS.490.2091G} and the accretion redshift; see Appendix~\ref{sec:aptt} for details. Moreover, we agree with \citetalias{2021MNRAS.505...18E} on finding a strong curvature of our fitting function when the subhalo has not been highly depleted and a subsequent power-law behaviour for more negative values of the logarithms of the internal structure parameter ratios.

\begin{table*}
	\centering
	\caption{Best-fit parameters for the different tidal tracks considered for subhaloes with initial NFW density profiles. All the fits from the literature refer to the apocentres.
	}
	\label{tab:ttbestfit}
	\begin{tabular}{ccccccccc} %
		\hline
		relation & parameter & apocentres & $1\sigma$ scatter& pericentres & $1\sigma$ scatter &
         \citetalias{Penarrubia2010} & \citetalias{2024PhRvD.110b3019D} & \citetalias{2021MNRAS.505...18E}  \\
		\hline 
		$V_\mathrm{max}, f_\mathrm{b}$  & $\mu$ & $0.38$ & 0.04 & $0.60$ & 0.03 & 0.40 & 0.6175 & - \\ 
		(fig. \ref{fig:ttvmaxfb}, eq. \ref{eq:tidalP10}) & $\nu$ & $0.30 $ &  & $0.32 $ & & 0.30 & 0.2895 & -  
        \\
		\hline 
		$r_\mathrm{max}, f_\mathrm{b}$ & $\mu$ & $-0.04 $ & $0.05 $ & $0.05  $ & $0.06$  & $-0.30$ & $0.5529$ & -  \\   
		 (fig. \ref{fig:figttrmax}, eq. \ref{eq:tidalP10})  & $\nu$ & $0.43  $ &  & $0.43 $ & & $0.40$ & $0.4675$ & - \\
		\hline 
		$V_\mathrm{max}, r_\mathrm{max}$ & $\alpha$ & $0.44 $ & $0.02 $ & $0.59$ & $0.05$ & - & - & $0.4$  \\
		(fig. \ref{fig:ttvmaxrmax}, eq. \ref{eq:tidalEN21}) & $\beta$  & $0.72 $ &  & $0.74$ & & - & - & $0.65$ \\
		\hline
	\end{tabular}
\end{table*}

\subsection{Prompt cusps: apocentres and pericentres}\label{sec:tidaltrackpc}

According to \citet{2010ApJ...723L.195I, 2014ApJ...788...27I, 2017MNRAS.471.4687A, 2023MNRAS.518.3509D}, our cosmological model predicts the rapid formation of very cuspy DM density peaks in the early Universe, with density profiles having an inner slope of $-3/2$ or steeper. These so-called prompt cusps would reside within the smallest subhaloes, with masses similar to or below that of Earth for $\Lambda$CDM, and many are expected to have survived until today %
as they are highly resilient because of the extreme high density. %
These objects are expected to be particularly relevant to searching for DM annihilation signals 
\citep{2010ApJ...723L.195I, 2023JCAP...10..008D}. Current cosmological simulations generally fail to reproduce these prompt cusps due to limited numerical resolution \citep{2014ApJ...788...27I}. 

Our definition of this initial subhalo profile uses Eq.~\ref{eq:nfw} as well, simply adopting $\alpha = 1$, $\beta = 3$, and $\gamma = 1.5$ in this case.\footnote{A gNFW with $\gamma = 1.5$ corresponds to a prompt cusp that did not accrete significant mass after formation. Per \citet{2023MNRAS.518.3509D}, for a typical halo we should expect a full NFW-like profile around the prompt cusp \citep[e.g.,][]{2025ApJ...993...93D}.} This way, we can also define an initial virial concentration for prompt cusps the same way we do for an NFW.

\begin{figure}
\begin{center}
\includegraphics[width=\columnwidth]{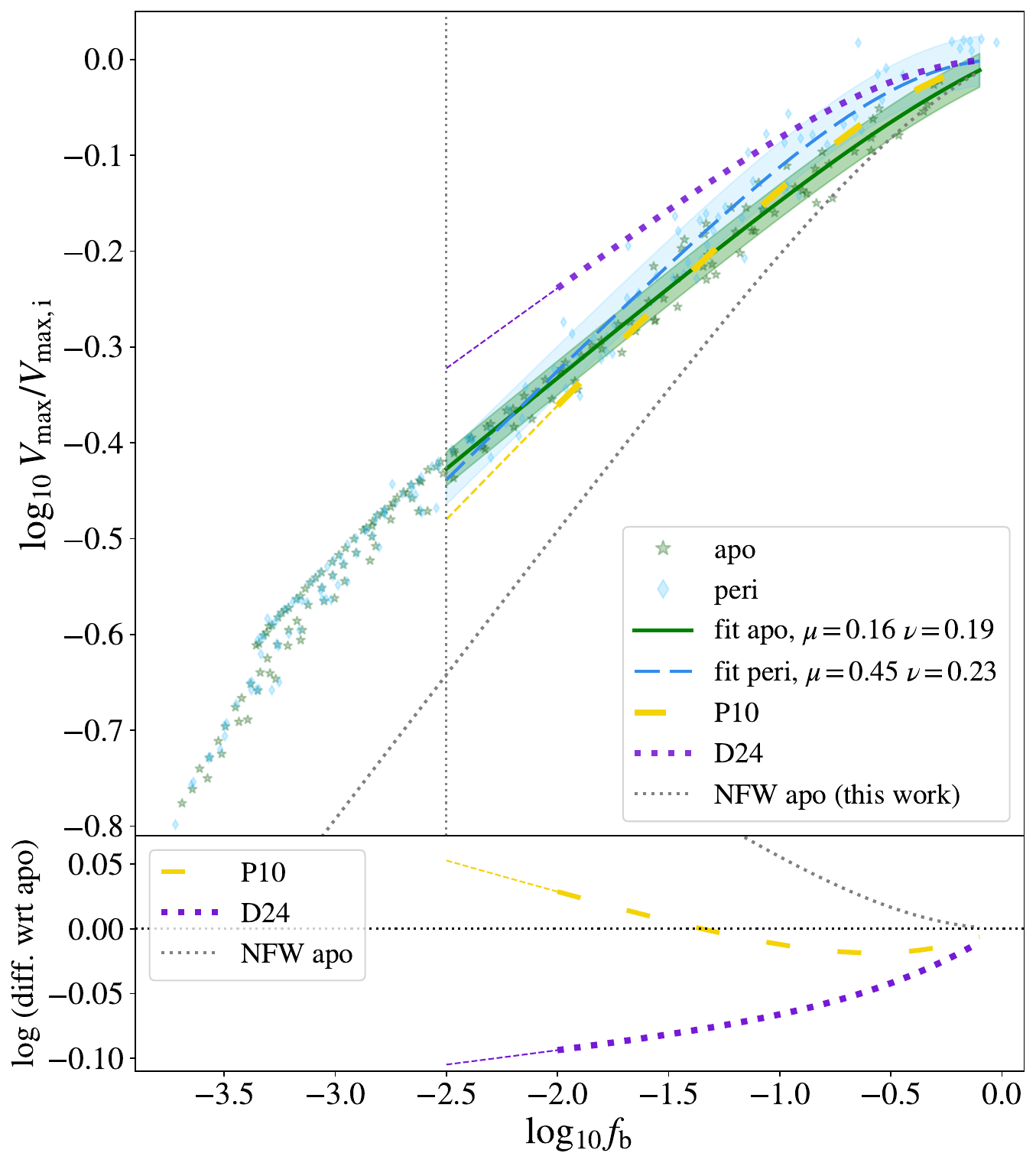} 
\caption{ Relation between $f_\mathrm{b}$ 
 and $V_\mathrm{max}$ divided by the initial values for the apocentres (green stars) and the pericentres (sky blue diamonds). Subhaloes exhibit an inner prompt cusp and an NFW tail at accretion. The tidal track found for each subset using Eq.~\ref{eq:tidalP10} is drawn as a solid green line in the former case and a dashed sky blue line in the latter, with shadowed bands for their respective scatter. Data can be trusted to the right of the vertical dotted line. The fits derived from \citetalias{Penarrubia2010} (loosely dashed yellow line) and \citetalias{2024PhRvD.110b3019D} (purple dotted line) are included, and thinned when they are extrapolated. The lower panel displays the difference between our best fit for the apocentres and the literature fits. }
\label{fig:figpc1}
\end{center}
\end{figure}

Here, we explore the effect of tidal stripping on prompt cusps and its impact on the tidal track, as depicted in Fig.~\ref{fig:figpc1}, where we relate $f_\mathrm{b}$ and $V_\mathrm{max}$. As expected, the change in $V_\mathrm{max}$ is less significant for prompt cusps than for NFW profiles ($\log_{10} (V_\mathrm{max} / V_\mathrm{max,i})$ is $-0.42$ versus $-0.65$ at $\log_{10} f_\mathrm{b} = -2.5$). We also find a smaller scatter for the apocentres, approximately half that of the NFW value. Regardless, note that we needed to simulate subhaloes with very elliptical orbits in order to get $\log_{10} f_\mathrm{b}$ values below $-2$ at present time. Our result lies between \citetalias{2024PhRvD.110b3019D}'s and \citetalias{Penarrubia2010}'s after great disruption, $\log_{10} f_\mathrm{b} < -1.5$, being ours closer to the latter once more. During the first orbits, however, \citetalias{Penarrubia2010}'s curve stands between our fits. This discrepancy may be due to the larger pericentre-to-apocentre ratios in our simulations and different $f_\mathrm{b}$ definitions compared to \citetalias{2024PhRvD.110b3019D}, and to the fact that we initialise our subhaloes with smaller masses than in \citetalias{Penarrubia2010}, so self-friction can be neglected in our case, but not in theirs. 
We have included the relation between $f_\mathrm{b}$ and $r_\mathrm{max}$ in Appendix~\ref{sec:aptt}. Our best-fit values are reported in Table~\ref{tab:ttbestfitpc} along with values from the literature.

\begin{figure}
\begin{center}
\includegraphics[width=\columnwidth]{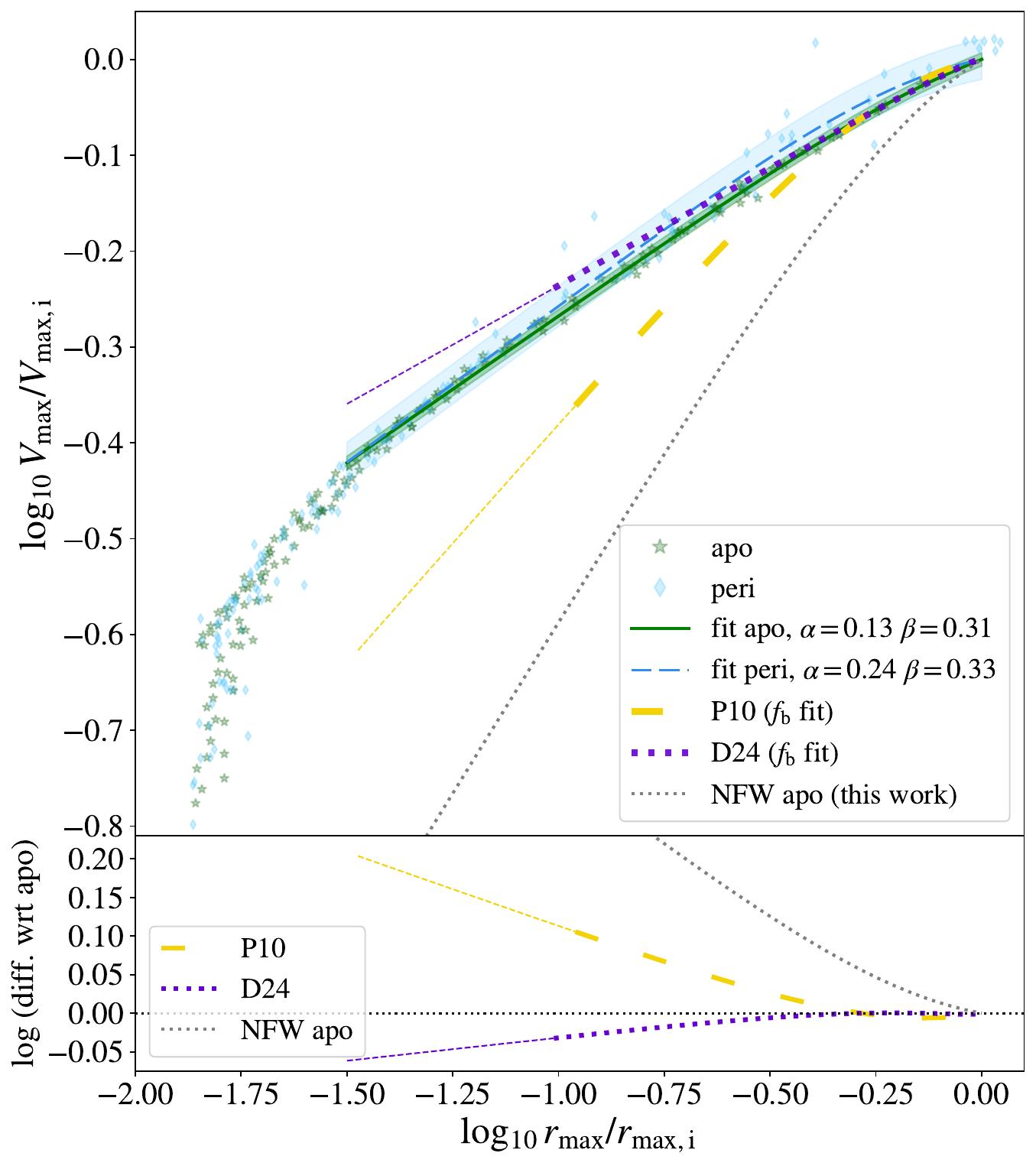}  
\caption{
Relation between $r_\mathrm{max}$ and $V_\mathrm{max}$, divided by their initial values, for the 
apocentres (green stars) and the pericentres (sky blue diamonds). Subhaloes exhibit an inner prompt cusp and an NFW tail at accretion. The tidal track found for each subset using Eq.~\ref{eq:tidalEN21} is drawn as a solid green line in the former case and a dashed sky blue line in the latter, with shadowed bands for their respective scatter. The fits derived from \citetalias{Penarrubia2010} (loosely dashed yellow line) and \citetalias{2024PhRvD.110b3019D} (purple dotted line) are included, and thinned when they are extrapolated. The lower panel displays the difference between our best fit for the apocentres and the literature fits.
} 
\label{fig:figpc3}
\end{center}
\end{figure}

The corresponding tidal track using both structural parameters is displayed in Fig.~\ref{fig:figpc3}. Whilst the reduction in $V_\mathrm{max}$ is lower than for initial NFW profiles, as found before, the corresponding $r_\mathrm{max}$ ratios are very similar, i.e., for the same $r_\mathrm{max}$ reduction, $V_\mathrm{max}$ is higher for prompt cusps. In this case there is no direct comparison we can make with earlier works, yet we can derive the curves ourselves from the fits in \citetalias{2024PhRvD.110b3019D} and \citetalias{Penarrubia2010}. For strong tidal stripping, $\log_{10} (V_\mathrm{max} / V_\mathrm{max,i}) < -0.3$, we lie between both results, with the former above and closer, and the latter below and further away. 
That is, here we are more in agreement with the results from \citetalias{2024PhRvD.110b3019D}.

Again, we partially agree with \citetalias{Penarrubia2010} on finding that, if we assume prompt cusp initial density profiles, the evolution of the internal parameters of the subhaloes essentially depends on how much mass they have lost and not on how this mass has been stripped. Nevertheless, we also find some scatter in this relation. %

When $\log_{10} (r_\mathrm{max} / r_\mathrm{max,i}) \lesssim -1.5$, our runs are affected by artificial two-body relaxation (see Appendix~\ref{sec:apsurvive} for details), and results in this regime cannot be considered reliable. This threshold corresponds to approximately $\log_{10} (V_\mathrm{max} / V_\mathrm{max,i}) \lesssim -0.42$ and $\log_{10} f_\mathrm{b} \lesssim -2.5$. A vertical dotted line is drawn in Fig.~\ref{fig:figpc1} to indicate this limit.

\begin{table*}
	\centering
	\caption{Best-fit parameters for the different tidal tracks considered for subhaloes with initial density profiles exhibiting an inner prompt cusp. All the fits from the literature refer to the apocentres.
	}
	\label{tab:ttbestfitpc}
	\begin{tabular}{cccccccc} %
		\hline
		relation & parameter & apocentres & $1\sigma$ scatter &  pericentres & $1\sigma$ scatter & \citetalias{Penarrubia2010} & \citetalias{2024PhRvD.110b3019D}  \\
		\hline 
		$V_\mathrm{max}, f_\mathrm{b}$ & $\mu$ & $0.16 $ & $0.02$ & $0.45$ & $0.03$ & 0.40 & 0.3358 \\
		(fig. \ref{fig:figpc1}, eq. \ref{eq:tidalP10}) & $\nu$ & $0.19 $ &  & $0.23 $ & & 0.24 & 0.1692\\
		\hline 
		$r_\mathrm{max}, f_\mathrm{b}$  & $\mu$ & $0.04$ & $0.05$ & $0.62  $ & $0.06$ & 0.00 & 1.207 \\
		(fig. \ref{fig:figpc2}, eq. \ref{eq:tidalP10}) & $\nu$ & $0.61$ &  & $0.70$ & & 0.48 & 0.6845 \\
		\hline 
		$V_\mathrm{max}, r_\mathrm{max}$ & $\alpha$ & $0.13 $ & $0.007$ & $0.24$ & $0.02$ & - & - \\
		(fig. \ref{fig:figpc3}, eq. \ref{eq:tidalEN21}) & $\beta$  & $0.31$ & & $0.33$ & & - & - \\
		\hline
	\end{tabular}
\end{table*}

\section{The tidal track of velocity concentrations }\label{sec:modelcv}
One still open question in the field is the precise evolution of subhalo concentrations with time. Previous literature found in simulations that, at present time, subhaloes closer to the Galactic centre exhibit higher concentrations for the same subhalo mass \citep{Moline17, 2023MNRAS.518..157M}. The subhalo concentration at different redshifts has also been explored in \citet{2023MNRAS.518..157M}. Here, we are answering the following questions: i) \textit{how and how much does the velocity concentration increase due to tidal stripping?} and ii) \textit{does the evolution of the velocity concentration depend on the initial subhalo parameters?}

\begin{figure}
\begin{center}
\includegraphics[width=\columnwidth]{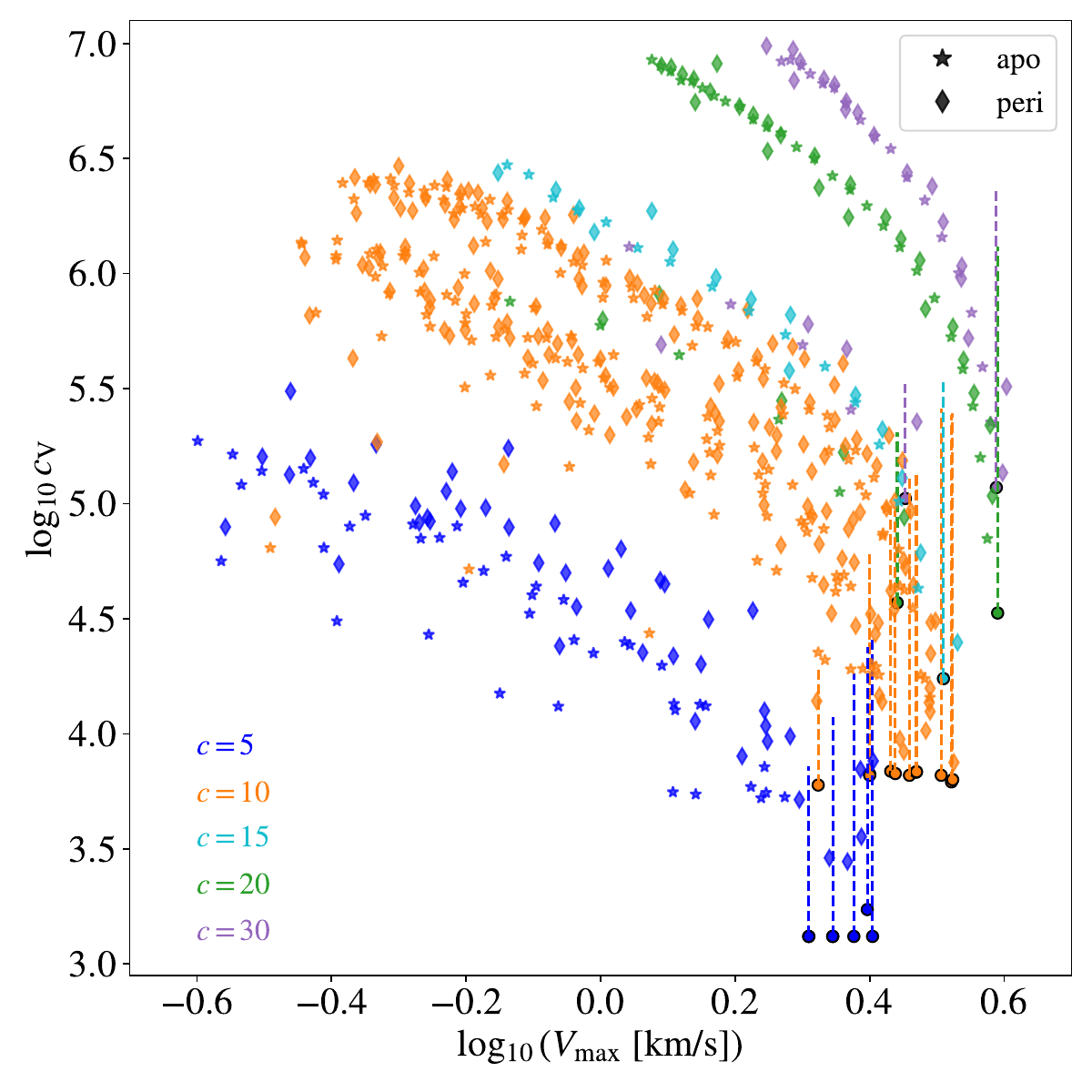} 
\caption{ Evolution of $c_\mathrm{V}$ versus $V_\mathrm{max}$ for different simulations adopting an initial NFW profile for subhaloes. Each colour depicts a specific initial virial concentration. Apocentre values are plotted as stars, while pericentres are diamonds. Circles and their respective dashed vertical lines describe the evolution of $c_\mathrm{V}$ for a field halo with that initial virial concentration, from the different considered accretion redshifts until present. These lines are longer for earlier $z_\mathrm{acc}$, which range from 1 to 4.  
The velocity concentration increases for isolated haloes solely due to its dependency on the Hubble parameter; see Eq.~\ref{eq:cv-def}. 
} 
\label{fig:figvelc1}
\end{center}
\end{figure}

We can define this velocity concentration for NFW profiles in terms of $V_\mathrm{max}$ and $r_\mathrm{max}$ using Eq.~\ref{eq:cv-def}. However, we cannot simply extract a velocity concentration tidal track from the ones obtained previously, since this expression includes $H$, which depends on the redshift. Therefore, we need to derive this relation directly from the data. The $V_\mathrm{max}$ and their corresponding $c_\mathrm{V}$ values for both the apocentres and the pericentres of subhalo orbits are shown in Fig.~\ref{fig:figvelc1}, for different initial concentration values. As can be seen, there is a clear trend that depends on the latter. We find  the velocity concentration to increase around two orders of magnitude from accretion redshift to present. Even for higher initial concentrations, one witnesses approximately the same increase in velocity concentration, yet accompanied by a smaller change in $V_\mathrm{max}$ for moderate mass losses. Note that any variation of $V_\mathrm{max}$ is a direct consequence of tidal stripping, as field haloes would not experience such changes; if we consider field haloes instead, the increase in concentration is the one given by the vertical dashed lines in Fig.~\ref{fig:figvelc1}, which reach different heights depending on $z_\mathrm{acc}$.

We have checked that the initial $c_\mathrm{V}$ for subhaloes with different accretion redshifts is the same if their initial virial concentration is the same, even though $V_\mathrm{max}$ is higher for earlier $z_\mathrm{acc}$ and $r_\mathrm{max}$ is smaller; circles of the same colour in Fig.~\ref{fig:figvelc1} describe this behaviour. %
This is due to the role of the Hubble parameter, which depends on $z_\mathrm{acc}$ and compensates for the larger ratio between the structural parameters. The increase in concentration when the halo is in isolation, from our accretion redshifts to present, is between one and one and a half orders of magnitude, meaning that subhaloes get more concentrated than field haloes. 

The driving parameter responsible for the scatter in velocity concentrations after dividing them by their initial values is the accretion redshift; see Appendix~\ref{sec:aptt} for details. This is partly due to the fact that subhaloes are accreted at the virial radius of the host at that time, which is smaller for earlier $z_\mathrm{acc}$ and thus those orbits are also smaller, leading to a larger number of pericentric passages and stronger tidal forces. 
 Finally, because of the redshift dependence through the Hubble parameter, the velocity concentration increases more for haloes having higher $z_\mathrm{acc}$. The combination of all these factors imprint larger concentration ratios at present time for subhaloes accreted earlier.

\begin{figure}
\begin{center}
\includegraphics[width=\columnwidth]{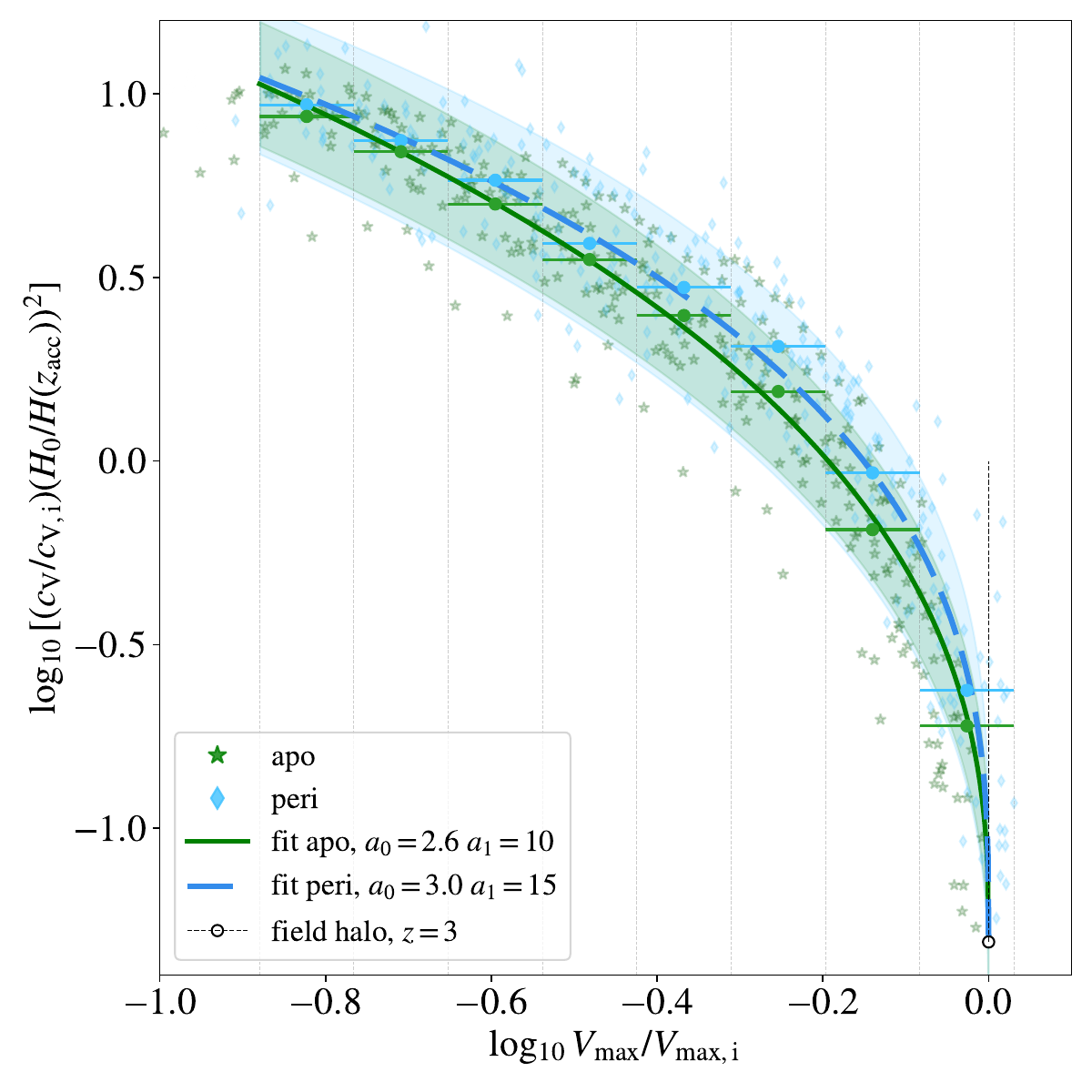} 
\caption{ Evolution of $c_\mathrm{V}$, normalised to its initial value $c_\mathrm{V,i}$ and by the accretion redshift through the Hubble parameter, for different simulations with an initial NFW profile. The $x$ axis is the ratio $V_\mathrm{max} / V_\mathrm{max,i}$, which becomes more negative with time.  Apocentre values are plotted as green stars, while pericentres are sky blue diamonds.  Filled circles show the mean value in each velocity bin, and the line refers to the fit to Eq.~\ref{eq:cvmax}. Best-fit parameters are given in Table~\ref{tab:ttbestfitcv}. 
Shadowed bands indicate the $1\sigma$ scatter. The vertical black line shows the evolution of the velocity concentration, solely due to the Hubble parameter, of a field halo formed at $z=3$ (black hollow circle).  } %
\label{fig:figvelc2}
\end{center}
\end{figure}

With the intention to isolate the increase of $c_\mathrm{V}$ due to tidal stripping from the mentioned redshift-related effects, in Fig.~\ref{fig:figvelc2} we divide both $V_\mathrm{max}$ and $c_\mathrm{V}$ by their initial values, then normalise the latter taking into account the accretion redshift through the Hubble parameter. This way we end up for a single trend for the apocentres and another for the pericentres.  This allows to perform fits to the following relation: 

\begin{equation}\label{eq:cvmax}
    \log_{10} \left[ \frac{c_\mathrm{V}}{c_\mathrm{V,i}} \left(\frac{H_0}{H(z_\mathrm{acc})} \right)^2 \right] =  \left( a_1 \left| \log_{10}\frac{V_\mathrm{max}}{V_\mathrm{max,i}}\right|  \right)^{1/a_0}, 
\end{equation} 

\begin{table}
	\centering
	\caption{Best-fit parameters for the concentration tidal track for subhaloes with initial NFW density profiles.
    }
	\label{tab:ttbestfitcv}
	\begin{tabular}{cccccc} %
		\hline
		relation & param. & apo & $1\sigma$ scat &  peri & $1\sigma$ scat  \\
		\hline 
		$c_\mathrm{V}, V_\mathrm{max}$ & $a_0$ & $2.6 $ & $0.17$ & $3.0$ & $0.19$ \\
		(fig. \ref{fig:figvelc2}, eq. \ref{eq:cvmax}) & $a_1$ & $10 $ &  & $15 $ & \\
		\hline 
	\end{tabular}
\end{table}

 
  Best-fit values and the respective scatters are reported in Table~\ref{tab:ttbestfitcv}. Concentration ratios are generally higher for the pericentres, as reflected by the corresponding fitting curves in Fig.~\ref{fig:figvelc2}, which is in agreement with subhaloes closer to the Galactic centre being more concentrated. %
  In other words, the higher subhalo $c_\mathrm{V}$ values found near the Galactic centre are driven by tidal stripping, as initially reported (but not yet explained) by \citet{Moline17}. Furthermore, Fig.~\ref{fig:figvelc2} also shows that the difference of $c_\mathrm{V}$ best-fit values for both pericentres and apocentres increases at intermediate epochs, while it stabilises when the subhalo has been greatly disrupted, i.e., it asymptotically reaches its maximum mass loss and thus its minimum $V_\mathrm{max}$.

\section{Conclusions}\label{sec:conclu}

Cosmological simulations have proven to be invaluable in understanding the formation and evolution of subhaloes as they orbit their host galaxies. They reveal that subhaloes undergo significant mass loss due to tidal stripping. Despite their effectiveness, cosmological simulations face limitations, particularly in resolving smaller structures due to their computational expense. This cost is further increased when hydrodynamics are taken into account.

Our study addresses these limitations and provides significant insights into the evolution of the structural parameters of small CDM subhaloes, particularly those completely dark and thus exhibiting an inner cusp. Through our suite of very high-resolution numerical simulations using the DASH code, achieving unprecedented accuracy, we have analysed the changes in the maximum circular velocities and their radial positions of low-mass DM subhaloes subjected to tidal stripping. We have found, in agreement with previous literature, that subhaloes follow a tidal track which essentially depends on the amount of mass stripped, rather than the initial subhalo configuration.

We focused on the substructures within a MW-sized halo which includes a baryonic disc and bulge that accurately replicate the MW's mass distribution. By varying the single DM subhalo's initial concentration, accretion redshift, orbital configuration, and inner slope -- considering both NFW and prompt cusps --, and accounting for the time-evolving gravitational potential of the MW, we have broadened our analysis beyond previous studies. Notably, we explored tidal tracks at both apocentres and pericentres, our study representing the first work to address the latter. Our findings can be summarised as follows. 

\begin{itemize}

\item[$\star$] Both $V_\mathrm{max}$ and $r_\mathrm{max}$ shrink after each subhalo orbit (Figs.~\ref{fig:fig1}-\ref{fig:figttorbit1}). For our fiducial case, while $V_\mathrm{max}$ decreases approximately by the same factor after each orbital period, $r_\mathrm{max}$ decreases less drastically in later orbits (top panel of Fig.~\ref{fig:figttorbit4}). Overall, $r_\mathrm{max}$ shrinks more than $V_\mathrm{max}$.

\item[$\star$] This results in a continuous rise in subhalo velocity concentrations over time, with a larger increase during the first orbit compared to subsequent ones (bottom panel of Fig.~\ref{fig:figttorbit4}). At present, when $V_\mathrm{max}$ is typically nearly one order of magnitude smaller than the initial one, the velocity concentration can reach values above two orders of magnitude higher than at infall (Figs.~\ref{fig:fig3} and \ref{fig:figvelc2}).

\item[$\star$] As in previous literature, when focusing on orbit apocentres, we confirm the existence of a very distinct tidal track for the structural parameters and $f_\mathrm{b}$ of initial subhalo NFW profiles (Figs.~\ref{fig:ttvmaxfb} and \ref{fig:ttvmaxrmax}). A significant scatter is found, which arises from differences in pericentre-to-apocentre ratios, the precise value of the pericentre, accretion redshift, circularity, or a combination of these. The concentration and the accretion redshift imprint a scatter on the $V_\mathrm{max} - f_\mathrm{b}$ parameter space, while the circularity does -- weakly -- on the $V_\mathrm{max} - r_\mathrm{max}$ parameter space (Figs. \ref{fig:figttcirc} - \ref{fig:figttc}). These tidal tracks are reproduced using Eqs.~\ref{eq:tidalP10} and \ref{eq:tidalEN21} with the best-fit parameters listed in Table~\ref{tab:ttbestfit}.

\item[$\star$] Similarly, we find the corresponding pericentre values for the same simulations to also follow a tidal track, with larger $V_\mathrm{max}$ for the same $f_\mathrm{b}$ and $r_\mathrm{max}$ compared to the apocentre tidal track (Figs.~\ref{fig:ttvmaxfb} and \ref{fig:ttvmaxrmax}). Again, these tidal tracks can be recovered using Eqs.~\ref{eq:tidalP10} and \ref{eq:tidalEN21} with the best-fit parameters reported in Table~\ref{tab:ttbestfit}.

\item[$\star$] Our results for NFW profiles using $V_\mathrm{max}$ and $f_\mathrm{b}$ are mostly in agreement with \citetalias{Penarrubia2010}, yet we deviate from \citetalias{2024PhRvD.110b3019D} (Fig.~\ref{fig:ttvmaxfb}). When we look at $r_\mathrm{max}$, the agreement on the curvature with \citetalias{Penarrubia2010} is weaker, even though their curve is contained within our scatter, which is larger (Fig. \ref{fig:figttrmax}).

\item[$\star$] We also find a tidal track for profiles exhibiting an initial inner prompt cusp, however our fits deviate significantly with respect to previous works, especially when $r_\mathrm{max}$ is considered, where we consistently obtain lower values for the same $f_\mathrm{b}$ (Figs. \ref{fig:figpc1}, \ref{fig:figpc3}, and \ref{fig:figpc2}). These tidal tracks can be recovered using Eqs.~\ref{eq:tidalP10} and \ref{eq:tidalEN21} with the best-fit parameters reported in Table~\ref{tab:ttbestfitpc}.

\item[$\star$] Subhaloes with initial inner prompt cusps remain more stable than those with an NFW, in the sense that the reduction in $V_\mathrm{max}$ is more pronounced for the latter: at $\log_{10} f_\mathrm{b} = -2.5$, $V_\mathrm{max}$ drops to $\sim 20\%$ of its initial value for NFW, compared to $\sim 40\%$ for prompt cusps. 

\item[$\star$] A tidal track for subhalo velocity concentrations is also derived from the simulations, that shows an increase of two orders of magnitude or more once $V_\mathrm{max}$ has decreased by one order of magnitude. This includes both the increase due to the concentration definition itself, which depends on redshift. Yet, the latter can only account for about one order of magnitude of this enhancement, being the effect of tidal stripping responsible for the rest (Figs. \ref{fig:figvelc1} and \ref{fig:figvelc2}). A significant scatter is found, whose main driver is the accretion redshift (Fig. \ref{fig:figvelc3}).  This tidal track is described by Eq.~\ref{eq:cvmax} with the best-fit parameters provided in Table~\ref{tab:ttbestfitcv}.

\end{itemize}

It is important to note that all subhaloes considered in our simulations have an initial mass of $10^6 \mathrm{M_\odot}$. Consequently, the significant difference in mass compared to that of the host makes the drag forces of dynamical and self-friction negligible. This effect can influence the results for more massive subhaloes, which have often been the focus of earlier studies, such as \citetalias{Penarrubia2010}.

Our work underscores the importance of baryonic effects and the evolving gravitational potential in accurately characterising the tidal evolution of DM subhaloes, especially for very eccentric orbits and/or small pericentre distances. Here, in deriving tidal tracks, we considered the time evolution of both the DM and baryonic potentials for the first time. We are aware, though, that we are not accounting for proper hydrodynamical feedback, which could impact our results. Besides, we conclude that circularity and/or pericentre values have an impact on the tidal track, as well as the accretion redshift, which will be studied in more detail elsewhere.  Furthermore, one must always be cautious about the accuracy of the inner density profile and how resolution effects may influence the results, as discussed in Section~\ref{sec:tidaltrackNFW}. A larger $V_\mathrm{max}$ ratio for the same $r_\mathrm{max}$ ratio would imply an even higher concentration increase. Since higher concentrations enhance, e.g., annihilation signals, our current estimates for the latter may therefore be somewhat conservative. In addition, larger concentrations would make subhalos more resilient to tidal stripping, leading to higher survival rates, which would also have a positive impact on DM subhalo searches.

Overall, these results greatly improve our understanding of the evolution of low-mass DM subhalo structural properties, offering valuable insights for future research via simulations and observations, such as gravitational lensing, stellar stream analyses, and indirect DM searches.     There are several reasons to include the pericentre tidal tracks from the latter point of view.  We are particularly interested in subhaloes in the solar vicinity, i.e., small subhalo distances with respect to the host. Not only these subhaloes are likely expected to be more concentrated due to having been accreted earlier, but also one tends to find more subhaloes close to their pericentre passages at these low radii, therefore being more concentrated, as shown by our pericentre tidal tracks.

\section*{Acknowledgements}


The authors thank Alastair Basden, Peter Draper, Mark Lovell and Paul Walker for their technical help. They also thank Jorge Peñarrubia, Adrian Jenkins and Jens St\"{u}cker for useful discussions, as well as the anonymous referee for their valuable feedback.


AAS acknowledges support from the Science and Technology Facilities Council funding grant ST/X001075/1. 
AAS and MASC were supported by the grants PID2021-125331NB-I00 and CEX2020-001007-S, both funded by MCIN/AEI/10.13039/501100011033 and by ``ERDF A way of making Europe''. MASC also acknowledges the MultiDark Network, ref. RED2022-134411-T. 
GO was supported by the National Key Research and Development Program of China (No. 2022YFA1602903), the National Natural Science Foundation of China (No. 12373004, W2432003), and the Fundamental Research Fund for Chinese Central Universities (No. NZ2020021, 226-2022-00216). 
This work used the DiRAC@Durham facility managed by the
Institute for Computational Cosmology on behalf of the STFC
DiRAC HPC Facility (www.dirac.ac.uk). The equipment was funded
by BEIS capital funding via STFC capital grants ST/K00042X/1, 
ST/P002293/1, ST/R002371/1 and ST/S002502/1, Durham University and STFC operations grant ST/R000832/1. DiRAC is part of the National e-Infrastructure. 
This work was partially supported by cosmology simulation database (CSD) in the National Basic Science Data Center (NBSDC-DB-10).

This research made use of Python, along with community-developed or maintained software packages, including IPython \citep{Ipython_paper}, Matplotlib \citep{Matplotlib_paper}, NumPy \citep{Numpy_paper} and SciPy \citep{2020SciPy-NMeth}.

\section*{Data availability}
The data underlying this article are publicly available on the \nobreak DAMASCO group website at \url{https://projects.ift.uam-csic.es/damasco/?page_id=831}.




\bibliographystyle{mnras}
\bibliography{References}

@article{Moline17,
   title={Characterization of subhalo structural properties and implications for dark matter annihilation signals},
   ISSN={1365-2966},
   url={http://dx.doi.org/10.1093/mnras/stx026},
   DOI={10.1093/mnras/stx026},
   journal={Monthly Notices of the Royal Astronomical Society},
   publisher={Oxford University Press (OUP)},
   author={Moliné, \'Angeles and Sánchez-Conde, Miguel A. and Palomares-Ruiz, Sergio and Prada, Francisco},
   year={2017},
   month={Jan},
   pages={stx026}
}

@ARTICLE{2023MNRAS.518..157M,
       author = {{Molin{\'e}}, {\'A}ngeles and {S{\'a}nchez-Conde}, Miguel A. and {Aguirre-Santaella}, Alejandra and others},
      journal = {\ MNRAS},
         year = 2023,
        month = jan,
       volume = {518},
       number = {1},
        pages = {157-173},
          doi = {10.1093/mnras/stac2930},
archivePrefix = {arXiv},
       eprint = {2110.02097},
 primaryClass = {astro-ph.CO},
       adsurl = {https://ui.adsabs.harvard.edu/abs/2023MNRAS.518..157M}
}

@Article{vlii_paper,
  author       = {J. Diemand and M. Kuhlen and P. Madau and M. Zemp and B. Moore and D. Potter and J. Stadel},
  title        = {{Clumps and streams in the local dark matter distribution}},
  journal      = {Nature},
  year         = {2008},
  date         = {2008-05-08},
  doi          = {10.1038/nature07153},
  eprint       = {0805.1244v2},
  eprintclass  = {astro-ph},
  eprinttype   = {arXiv},
  journaltitle = {Nature 454:735-738,2008},
}

@ARTICLE{2008MNRAS.391.1685S,
       author = {{Springel}, V. and {Wang}, J. and {Vogelsberger}, M. and {Ludlow}, A. and
         {Jenkins}, A. and {Helmi}, A. and {Navarro}, J.~F. and {Frenk}, C.~S. and
         {White}, S.~D.~M.},
        title = "{The Aquarius Project: the subhaloes of galactic haloes}",
      journal = {Monthly Notices of the Royal Astronomical Society},
         year = 2008,
        month = dec,
       volume = {391},
       number = {4},
        pages = {1685-1711},
          doi = {10.1111/j.1365-2966.2008.14066.x},
archivePrefix = {arXiv},
       eprint = {0809.0898},
 primaryClass = {astro-ph},
       adsurl = {https://ui.adsabs.harvard.edu/abs/2008MNRAS.391.1685S}
}

@ARTICLE{2003ApJ...584..541H,
       author = {{Hayashi}, Eric and {Navarro}, Julio F. and {Taylor}, James E. and {Stadel}, Joachim and {Quinn}, Thomas},
        title = "{The Structural Evolution of Substructure}",
      journal = {APJ},
         year = 2003,
        month = feb,
       volume = {584},
       number = {2},
        pages = {541-558},
          doi = {10.1086/345788},
archivePrefix = {arXiv},
       eprint = {astro-ph/0203004},
 primaryClass = {astro-ph},
       adsurl = {https://ui.adsabs.harvard.edu/abs/2003ApJ...584..541H}
}

@ARTICLE{vandenBosch2018_analytic,
       author = {{\swap{Bosch}{van den }}, Frank C. and {Ogiya}, Go and {Hahn}, Oliver and {Burkert}, Andreas},
        title = "{Disruption of dark matter substructure: fact or fiction?}",
      journal = {MNRAS},
         year = 2018,
        month = mar,
       volume = {474},
       number = {3},
        pages = {3043-3066},
          doi = {10.1093/mnras/stx2956},
archivePrefix = {arXiv},
       eprint = {1711.05276},
 primaryClass = {astro-ph.GA},
       adsurl = {https://ui.adsabs.harvard.edu/abs/2018MNRAS.474.3043V}
}

@ARTICLE{2020MNRAS.491.4591E,
       author = {{Errani}, Rapha{\"e}l and {Pe{\~n}arrubia}, Jorge},
        title = "{Can tides disrupt cold dark matter subhaloes?}",
      journal = {MNRAS},
         year = 2020,
        month = feb,
       volume = {491},
       number = {4},
        pages = {4591-4601},
          doi = {10.1093/mnras/stz3349},
archivePrefix = {arXiv},
       eprint = {1906.01642},
 primaryClass = {astro-ph.GA},
       adsurl = {https://ui.adsabs.harvard.edu/abs/2020MNRAS.491.4591E}
}

@ARTICLE{2021arXiv211101148A,
       author = {{Amorisco}, Nicola C.},
        title = "{Cold dark matter subhaloes at arbitrarily low masses}",
      journal = {arXiv e-prints},
         year = 2021,
        month = nov,
          eid = {arXiv:2111.01148},
        pages = {arXiv:2111.01148},
archivePrefix = {arXiv},
       eprint = {2111.01148},
 primaryClass = {astro-ph.CO},
       adsurl = {https://ui.adsabs.harvard.edu/abs/2021arXiv211101148A}
}

@ARTICLE{2024arXiv240804470H,
       author = {{He}, Feihong and {Han}, Jiaxin and {Li}, Zhaozhou},
        title = "{Why artificial disruption is not a concern for current cosmological simulations}",
      journal = {arXiv e-prints},
         year = 2024,
        month = aug,
          eid = {arXiv:2408.04470},
        pages = {arXiv:2408.04470},
          doi = {10.48550/arXiv.2408.04470},
archivePrefix = {arXiv},
       eprint = {2408.04470},
 primaryClass = {astro-ph.GA},
       adsurl = {https://ui.adsabs.harvard.edu/abs/2024arXiv240804470H}
}

@ARTICLE{2017MNRAS.471.1709G,
       author = {{Garrison-Kimmel}, Shea and {Wetzel}, Andrew and {Bullock}, James S. and {Hopkins}, Philip F. and {Boylan-Kolchin}, Michael and {Faucher-Gigu{\`e}re}, Claude-Andr{\'e} and {Kere{\v{s}}}, Du{\v{s}}an and {Quataert}, Eliot and {Sanderson}, Robyn E. and {Graus}, Andrew S. and {Kelley}, Tyler},
        title = "{Not so lumpy after all: modelling the depletion of dark matter subhaloes by Milky Way-like galaxies}",
      journal = {MNRAS},
         year = 2017,
        month = oct,
       volume = {471},
       number = {2},
        pages = {1709-1727},
          doi = {10.1093/mnras/stx1710},
archivePrefix = {arXiv},
       eprint = {1701.03792},
 primaryClass = {astro-ph.GA},
       adsurl = {https://ui.adsabs.harvard.edu/abs/2017MNRAS.471.1709G}
}

@ARTICLE{2021MNRAS.501.3558G,
       author = {{Grand}, Robert J.~J. and {White}, Simon D.~M.},
        title = "{Baryonic effects on the detectability of annihilation radiation from dark matter subhaloes around the Milky Way}",
      journal = {MNRAS},
         year = 2021,
        month = mar,
       volume = {501},
       number = {3},
        pages = {3558-3567},
          doi = {10.1093/mnras/staa3993},
archivePrefix = {arXiv},
       eprint = {2012.07846},
 primaryClass = {astro-ph.GA},
       adsurl = {https://ui.adsabs.harvard.edu/abs/2021MNRAS.501.3558G}
}

@ARTICLE{vandenBosch2018_num_criteria,
       author = {{\swap{Bosch}{van den }}, Frank C. and {Ogiya}, Go},
        title = "{Dark matter substructure in numerical simulations: a tale of discreteness noise, runaway instabilities, and artificial disruption}",
      journal = {MNRAS},
         year = 2018,
        month = apr,
       volume = {475},
       number = {3},
        pages = {4066-4087},
          doi = {10.1093/mnras/sty084},
archivePrefix = {arXiv},
       eprint = {1801.05427},
 primaryClass = {astro-ph.GA},
       adsurl = {https://ui.adsabs.harvard.edu/abs/2018MNRAS.475.4066V}
}

@ARTICLE{Ogiya2019,
       author = {{Ogiya}, Go and {van den Bosch}, Frank C. and {Hahn}, Oliver and {Green}, Sheridan B. and {Miller}, Tim B. and {Burkert}, Andreas},
        title = "{DASH: a library of dynamical subhalo evolution}",
      journal = {MNRAS},
         year = 2019,
        month = may,
       volume = {485},
       number = {1},
        pages = {189-202},
          doi = {10.1093/mnras/stz375},
archivePrefix = {arXiv},
       eprint = {1901.08601},
 primaryClass = {astro-ph.GA},
       adsurl = {https://ui.adsabs.harvard.edu/abs/2019MNRAS.485..189O}
}

@ARTICLE{2021MNRAS.507.4953G,
       author = {{Grand}, Robert J.~J. and {Marinacci}, Federico and {Pakmor}, R{\"u}diger and {Simpson}, Christine M. and {Kelly}, Ashley J. and {G{\'o}mez}, Facundo A. and {Jenkins}, Adrian and {Springel}, Volker and {Frenk}, Carlos S. and {White}, Simon D.~M.},
        title = "{Determining the full satellite population of a Milky Way-mass halo in a highly resolved cosmological hydrodynamic simulation}",
      journal = {\mnras},
         year = 2021,
        month = nov,
       volume = {507},
       number = {4},
        pages = {4953-4967},
          doi = {10.1093/mnras/stab2492},
archivePrefix = {arXiv},
       eprint = {2105.04560},
 primaryClass = {astro-ph.GA},
       adsurl = {https://ui.adsabs.harvard.edu/abs/2021MNRAS.507.4953G}
}

@ARTICLE{Kelley2019,
       author = {{Kelley}, Tyler and {Bullock}, James S. and {Garrison-Kimmel}, Shea and {Boylan-Kolchin}, Michael and {Pawlowski}, Marcel S. and {Graus}, Andrew S.},
        title = "{Phat ELVIS: The inevitable effect of the Milky Way's disc on its dark matter subhaloes}",
      journal = {MNRAS},
         year = 2019,
        month = aug,
       volume = {487},
       number = {3},
        pages = {4409-4423},
          doi = {10.1093/mnras/stz1553},
archivePrefix = {arXiv},
       eprint = {1811.12413},
 primaryClass = {astro-ph.GA},
       adsurl = {https://ui.adsabs.harvard.edu/abs/2019MNRAS.487.4409K}
}

@ARTICLE{2003MNRAS.338...14P,
       author = {{Power}, C. and {Navarro}, J.~F. and {Jenkins}, A. and {Frenk}, C.~S. and {White}, S.~D.~M. and {Springel}, V. and {Stadel}, J. and {Quinn}, T.},
        title = "{The inner structure of {\ensuremath{\Lambda}}CDM haloes - I. A numerical convergence study}",
      journal = {\mnras},
         year = 2003,
        month = jan,
       volume = {338},
       number = {1},
        pages = {14-34},
          doi = {10.1046/j.1365-8711.2003.05925.x},
archivePrefix = {arXiv},
       eprint = {astro-ph/0201544},
 primaryClass = {astro-ph},
       adsurl = {https://ui.adsabs.harvard.edu/abs/2003MNRAS.338...14P}
}

@BOOK{2008gady.book.....B,
       author = {{Binney}, James and {Tremaine}, Scott},
        title = "{Galactic Dynamics: Second Edition}",
         year = 2008,
       adsurl = {https://ui.adsabs.harvard.edu/abs/2008gady.book.....B}
}

@ARTICLE{2015MNRAS.448.2941S,
       author = {{Sawala}, Till and {Frenk}, Carlos S. and {Fattahi}, Azadeh and {Navarro}, Julio F. and {Bower}, Richard G. and {Crain}, Robert A. and {Dalla Vecchia}, Claudio and {Furlong}, Michelle and {Jenkins}, Adrian and {McCarthy}, Ian G. and {Qu}, Yan and {Schaller}, Matthieu and {Schaye}, Joop and {Theuns}, Tom},
        title = "{Bent by baryons: the low-mass galaxy-halo relation}",
      journal = {\mnras},
         year = 2015,
        month = apr,
       volume = {448},
       number = {3},
        pages = {2941-2947},
          doi = {10.1093/mnras/stu2753},
archivePrefix = {arXiv},
       eprint = {1404.3724},
 primaryClass = {astro-ph.GA},
       adsurl = {https://ui.adsabs.harvard.edu/abs/2015MNRAS.448.2941S}
}

@ARTICLE{2016MNRAS.456...85S,
       author = {{Sawala}, Till and {Frenk}, Carlos S. and {Fattahi}, Azadeh and {Navarro}, Julio F. and {Theuns}, Tom and {Bower}, Richard G. and {Crain}, Robert A. and {Furlong}, Michelle and {Jenkins}, Adrian and {Schaller}, Matthieu and {Schaye}, Joop},
        title = "{The chosen few: the low-mass haloes that host faint galaxies}",
      journal = {\mnras},
         year = 2016,
        month = feb,
       volume = {456},
       number = {1},
        pages = {85-97},
          doi = {10.1093/mnras/stv2597},
archivePrefix = {arXiv},
       eprint = {1406.6362},
 primaryClass = {astro-ph.CO},
       adsurl = {https://ui.adsabs.harvard.edu/abs/2016MNRAS.456...85S}
}

@ARTICLE{2025ApJ...983L..23N,
       author = {{Nadler}, Ethan O.},
        title = "{The Impact of Molecular Hydrogen Cooling on the Galaxy Formation Threshold}",
      journal = {\apjl},
         year = 2025,
        month = apr,
       volume = {983},
       number = {1},
          eid = {L23},
        pages = {L23},
          doi = {10.3847/2041-8213/adbc6e},
archivePrefix = {arXiv},
       eprint = {2503.04885},
 primaryClass = {astro-ph.GA},
       adsurl = {https://ui.adsabs.harvard.edu/abs/2025ApJ...983L..23N}
}

@ARTICLE{2019MNRAS.490.2091G,
       author = {{Green}, Sheridan B. and {van den Bosch}, Frank C.},
        title = "{The tidal evolution of dark matter substructure - I. subhalo density profiles}",
      journal = {\mnras},
         year = 2019,
        month = dec,
       volume = {490},
       number = {2},
        pages = {2091-2101},
          doi = {10.1093/mnras/stz2767},
archivePrefix = {arXiv},
       eprint = {1908.08537},
 primaryClass = {astro-ph.GA},
       adsurl = {https://ui.adsabs.harvard.edu/abs/2019MNRAS.490.2091G}
}

@ARTICLE{2020A&A...641A...6P,
       author = {{Aghanim}, N. and {Akrami}, Y. and {Ashdown}, M. and {Aumont}, J. and {Baccigalupi}, C. and {Ballardini}, M. and {Banday}, A.~J. and {Barreiro}, R.~B. and {Bartolo}, N. and {Basak}, S. and {Battye}, R. and {Benabed}, K. and {Bernard}, J. -P. and {Bersanelli}, M. and {Bielewicz}, P. and {Bock}, J.~J. and {Bond}, J.~R. and {Borrill}, J. and {Bouchet}, F.~R. and {Boulanger}, F. and {Bucher}, M. and {Burigana}, C. and {Butler}, R.~C. and {Calabrese}, E. and {Cardoso}, J. -F. and {Carron}, J. and {Challinor}, A. and {Chiang}, H.~C. and {Chluba}, J. and {Colombo}, L.~P.~L. and {Combet}, C. and {Contreras}, D. and {Crill}, B.~P. and {Cuttaia}, F. and {de Bernardis}, P. and {de Zotti}, G. and {Delabrouille}, J. and {Delouis}, J. -M. and {Di Valentino}, E. and {Diego}, J.~M. and {Dor{\'e}}, O. and {Douspis}, M. and {Ducout}, A. and {Dupac}, X. and {Dusini}, S. and {Efstathiou}, G. and {Elsner}, F. and {En{\ss}lin}, T.~A. and {Eriksen}, H.~K. and {Fantaye}, Y. and {Farhang}, M. and {Fergusson}, J. and {Fernandez-Cobos}, R. and {Finelli}, F. and {Forastieri}, F. and {Frailis}, M. and {Fraisse}, A.~A. and {Franceschi}, E. and {Frolov}, A. and {Galeotta}, S. and {Galli}, S. and {Ganga}, K. and {G{\'e}nova-Santos}, R.~T. and {Gerbino}, M. and {Ghosh}, T. and {Gonz{\'a}lez-Nuevo}, J. and {G{\'o}rski}, K.~M. and {Gratton}, S. and {Gruppuso}, A. and {Gudmundsson}, J.~E. and {Hamann}, J. and {Handley}, W. and {Hansen}, F.~K. and {Herranz}, D. and {Hildebrandt}, S.~R. and {Hivon}, E. and {Huang}, Z. and {Jaffe}, A.~H. and {Jones}, W.~C. and {Karakci}, A. and {Keih{\"a}nen}, E. and {Keskitalo}, R. and {Kiiveri}, K. and {Kim}, J. and {Kisner}, T.~S. and {Knox}, L. and {Krachmalnicoff}, N. and {Kunz}, M. and {Kurki-Suonio}, H. and {Lagache}, G. and {Lamarre}, J. -M. and {Lasenby}, A. and {Lattanzi}, M. and {Lawrence}, C.~R. and {Le Jeune}, M. and {Lemos}, P. and {Lesgourgues}, J. and {Levrier}, F. and {Lewis}, A. and {Liguori}, M. and {Lilje}, P.~B. and {Lilley}, M. and {Lindholm}, V. and {L{\'o}pez-Caniego}, M. and {Lubin}, P.~M. and {Ma}, Y. -Z. and {Mac{\'\i}as-P{\'e}rez}, J.~F. and {Maggio}, G. and {Maino}, D. and {Mandolesi}, N. and {Mangilli}, A. and {Marcos-Caballero}, A. and {Maris}, M. and {Martin}, P.~G. and {Martinelli}, M. and {Mart{\'\i}nez-Gonz{\'a}lez}, E. and {Matarrese}, S. and {Mauri}, N. and {McEwen}, J.~D. and {Meinhold}, P.~R. and {Melchiorri}, A. and {Mennella}, A. and {Migliaccio}, M. and {Millea}, M. and {Mitra}, S. and {Miville-Desch{\^e}nes}, M. -A. and {Molinari}, D. and {Montier}, L. and {Morgante}, G. and {Moss}, A. and {Natoli}, P. and {N{\o}rgaard-Nielsen}, H.~U. and {Pagano}, L. and {Paoletti}, D. and {Partridge}, B. and {Patanchon}, G. and {Peiris}, H.~V. and {Perrotta}, F. and {Pettorino}, V. and {Piacentini}, F. and {Polastri}, L. and {Polenta}, G. and {Puget}, J. -L. and {Rachen}, J.~P. and {Reinecke}, M. and {Remazeilles}, M. and {Renzi}, A. and {Rocha}, G. and {Rosset}, C. and {Roudier}, G. and {Rubi{\~n}o-Mart{\'\i}n}, J.~A. and {Ruiz-Granados}, B. and {Salvati}, L. and {Sandri}, M. and {Savelainen}, M. and {Scott}, D. and {Shellard}, E.~P.~S. and {Sirignano}, C. and {Sirri}, G. and {Spencer}, L.~D. and {Sunyaev}, R. and {Suur-Uski}, A. -S. and {Tauber}, J.~A. and {Tavagnacco}, D. and {Tenti}, M. and {Toffolatti}, L. and {Tomasi}, M. and {Trombetti}, T. and {Valenziano}, L. and {Valiviita}, J. and {Van Tent}, B. and {Vibert}, L. and {Vielva}, P. and {Villa}, F. and {Vittorio}, N. and {Wandelt}, B.~D. and {Wehus}, I.~K. and {White}, M. and {White}, S.~D.~M. and {Zacchei}, A. and {Zonca}, A.},
       authortype = "{Planck Collaboration}",
        title = "{Planck 2018 results. VI. Cosmological parameters}",
      journal = {\aap},
         year = 2020,
        month = sep,
       volume = {641},
          eid = {A6},
        pages = {A6},
          doi = {10.1051/0004-6361/201833910},
archivePrefix = {arXiv},
       eprint = {1807.06209},
 primaryClass = {astro-ph.CO},
       adsurl = {https://ui.adsabs.harvard.edu/abs/2020A&A...641A...6P}
}

@ARTICLE{2022LRCA....8....1A,
       author = {{Angulo}, Raul E. and {Hahn}, Oliver},
        title = "{Large-scale dark matter simulations}",
      journal = {Living Reviews in Computational Astrophysics},
         year = 2022,
        month = dec,
       volume = {8},
       number = {1},
          eid = {1},
        pages = {1},
          doi = {10.1007/s41115-021-00013-z},
archivePrefix = {arXiv},
       eprint = {2112.05165},
 primaryClass = {astro-ph.CO},
       adsurl = {https://ui.adsabs.harvard.edu/abs/2022LRCA....8....1A}
}

@ARTICLE{2011ARA&A..49..155P,
       author = {{Porter}, Troy A. and {Johnson}, Robert P. and {Graham}, Peter W.},
        title = "{Dark Matter Searches with Astroparticle Data}",
      journal = {\araa},
         year = "2011",
        month = "Sep",
       volume = {49},
       number = {1},
        pages = {155-194},
          doi = {10.1146/annurev-astro-081710-102528},
archivePrefix = {arXiv},
       eprint = {1104.2836},
 primaryClass = {astro-ph.HE},
       adsurl = {https://ui.adsabs.harvard.edu/abs/2011ARA&A..49..155P}
}

@Article{mascprada14,
      author         = "S\'anchez-Conde, Miguel A. and Prada, Francisco",
      title          = "{The flattening of the concentration–mass relation
                        towards low halo masses and its implications for the
                        annihilation signal boost}",
      journal        = "Mon. Not. Roy. Astron. Soc.",
      volume         = "442",
      year           = "2014",
      number         = "3",
      pages          = "2271-2277",
      doi            = "10.1093/mnras/stu1014",
      eprint         = "1312.1729",
      archivePrefix  = "arXiv",
      primaryClass   = "astro-ph.CO",
      SLACcitation   = "%%CITATION = ARXIV:1312.1729;%%"
}

@article{Bertone:2004pz,
      author         = "Bertone, Gianfranco and Hooper, Dan and Silk, Joseph",
      title          = "{Particle dark matter: Evidence, candidates and
                        constraints}",
      journal        = "Phys. Rept.",
      volume         = "405",
      year           = "2005",
      pages          = "279-390",
      doi            = "10.1016/j.physrep.2004.08.031",
      eprint         = "hep-ph/0404175",
      archivePrefix  = "arXiv",
      primaryClass   = "hep-ph",
      reportNumber   = "FERMILAB-PUB-04-047-A",
      SLACcitation   = "%%CITATION = HEP-PH/0404175;%%"
}

@Article{Bertone10,
  author       = {Gianfranco Bertone},
  title        = {{The moment of truth for WIMP Dark Matter}},
  journal      = {Nature},
  volume       = {468},
  year         = {2010},
  date         = {2010-11-15},
  pages        = {389},
  doi          = {10.1038/nature09509},
  eprint       = {1011.3532v1},
  eprintclass  = {astro-ph.CO},
  eprinttype   = {arXiv},
  journaltitle = {Nature 468:389-393,2010},
}

@Article{Ipython_paper,
  author    = {Fernando Perez and Brian E. Granger},
  title     = {{IPython}: A System for Interactive Scientific Computing},
  journal   = {Comput. Sci. Eng.},
  year      = {2007},
  volume    = {9},
  number    = {3},
  pages     = {21--29},
  doi       = {10.1109/mcse.2007.53},
  publisher = {Institute of Electrical and Electronics Engineers ({IEEE})},
}

@Article{Matplotlib_paper,
  author    = {John D. Hunter},
  title     = {{Matplotlib: A 2D Graphics Environment}},
  journal   = {Comput. Sci. Eng.},
  year      = {2007},
  volume    = {9},
  number    = {3},
  pages     = {90--95},
  doi       = {10.1109/mcse.2007.55},
  publisher = {Institute of Electrical and Electronics Engineers ({IEEE})},
}

@Article{Numpy_paper,
  author    = {{\swap{Walt}{van der }}, St\'efan and {Colbert}, S. Chris and {Varoquaux}, Ga\"el},
  title     = {The {NumPy} {Array: A Structure} for {Efficient Numerical Computation}},
  journal   = {Comput. in Sci. Eng.},
  year      = {2011},
  volume    = {13},
  number    = {2},
  pages     = {22--30},
  doi       = {10.1109/mcse.2011.37},
  publisher = {Institute of Electrical and Electronics Engineers ({IEEE})},
}

@ARTICLE{2020SciPy-NMeth,
       author = {{Virtanen}, Pauli and {Gommers}, Ralf and {Oliphant},
         Travis E. and others},
        title = {SciPy 1.0: {Fundamental Algorithms} for {Scientific
                  Computing} in {Python}},
      journal = {Nat. Methods},
      year = "2020",
      volume={17},
      pages={261--272},
      adsurl = {https://rdcu.be/b08Wh},
      doi = {https://doi.org/10.1038/s41592-019-0686-2},
}

@ARTICLE{Navarro1997,
   author = {{Navarro}, J.~F. and {Frenk}, C.~S. and {White}, S.~D.~M.},
    title = "{A Universal Density Profile from Hierarchical Clustering}",
  journal = {\apj},
   eprint = {astro-ph/9611107},
     year = 1997,
    month = dec,
   volume = 490,
    pages = {493-508},
      doi = {10.1086/304888},
   adsurl = {http://cdsads.u-strasbg.fr/abs/1997ApJ...490..493N}
}

@ARTICLE{1996MNRAS.278..488Z,
       author = {{Zhao}, Hongsheng},
        title = "{Analytical models for galactic nuclei}",
      journal = {\mnras},
         year = 1996,
        month = jan,
       volume = {278},
       number = {2},
        pages = {488-496},
          doi = {10.1093/mnras/278.2.488},
archivePrefix = {arXiv},
       eprint = {astro-ph/9509122},
 primaryClass = {astro-ph},
       adsurl = {https://ui.adsabs.harvard.edu/abs/1996MNRAS.278..488Z}
}

@ARTICLE{2020MNRAS.495.4496M,
       author = {{Miller}, Tim B. and {van den Bosch}, Frank C. and {Green}, Sheridan B. and {Ogiya}, Go},
        title = "{Dynamical self-friction: how mass loss slows you down}",
      journal = {\mnras},
         year = 2020,
        month = jul,
       volume = {495},
       number = {4},
        pages = {4496-4507},
          doi = {10.1093/mnras/staa1450},
archivePrefix = {arXiv},
       eprint = {2001.06489},
 primaryClass = {astro-ph.GA},
       adsurl = {https://ui.adsabs.harvard.edu/abs/2020MNRAS.495.4496M}
}

@ARTICLE{2002ApJ...568...52W,
       author = {{Wechsler}, Risa H. and {Bullock}, James S. and {Primack}, Joel R. and {Kravtsov}, Andrey V. and {Dekel}, Avishai},
        title = "{Concentrations of Dark Halos from Their Assembly Histories}",
      journal = {\apj},
         year = 2002,
        month = mar,
       volume = {568},
       number = {1},
        pages = {52-70},
          doi = {10.1086/338765},
archivePrefix = {arXiv},
       eprint = {astro-ph/0108151},
 primaryClass = {astro-ph},
       adsurl = {https://ui.adsabs.harvard.edu/abs/2002ApJ...568...52W}
}

@ARTICLE{2008MNRAS.391.1940M,
       author = {{Macci{\`o}}, Andrea V. and {Dutton}, Aaron A. and {van den Bosch}, Frank C.},
        title = "{Concentration, spin and shape of dark matter haloes as a function of the cosmological model: WMAP1, WMAP3 and WMAP5 results}",
      journal = {\mnras},
         year = 2008,
        month = dec,
       volume = {391},
       number = {4},
        pages = {1940-1954},
          doi = {10.1111/j.1365-2966.2008.14029.x},
archivePrefix = {arXiv},
       eprint = {0805.1926},
 primaryClass = {astro-ph},
       adsurl = {https://ui.adsabs.harvard.edu/abs/2008MNRAS.391.1940M}
}

@ARTICLE{2014MNRAS.444.1518V,
       author = {{Vogelsberger}, Mark and {Genel}, Shy and {Springel}, Volker and {Torrey}, Paul and {Sijacki}, Debora and {Xu}, Dandan and {Snyder}, Greg and {Nelson}, Dylan and {Hernquist}, Lars},
        title = "{Introducing the Illustris Project: simulating the coevolution of dark and visible matter in the Universe}",
      journal = {\mnras},
         year = 2014,
        month = oct,
       volume = {444},
       number = {2},
        pages = {1518-1547},
          doi = {10.1093/mnras/stu1536},
archivePrefix = {arXiv},
       eprint = {1405.2921},
 primaryClass = {astro-ph.CO},
       adsurl = {https://ui.adsabs.harvard.edu/abs/2014MNRAS.444.1518V}
}

@ARTICLE{2016MNRAS.457.1931S,
       author = {{Sawala}, Till and {Frenk}, Carlos S. and {Fattahi}, Azadeh and {Navarro}, Julio F. and {Bower}, Richard G. and {Crain}, Robert A. and {Dalla Vecchia}, Claudio and {Furlong}, Michelle and {Helly}, John. C. and {Jenkins}, Adrian and {Oman}, Kyle A. and {Schaller}, Matthieu and {Schaye}, Joop and {Theuns}, Tom and {Trayford}, James and {White}, Simon D.~M.},
        title = "{The APOSTLE simulations: solutions to the Local Group's cosmic puzzles}",
      journal = {\mnras},
         year = 2016,
        month = apr,
       volume = {457},
       number = {2},
        pages = {1931-1943},
          doi = {10.1093/mnras/stw145},
archivePrefix = {arXiv},
       eprint = {1511.01098},
 primaryClass = {astro-ph.GA},
       adsurl = {https://ui.adsabs.harvard.edu/abs/2016MNRAS.457.1931S}
}

@ARTICLE{2022arXiv220700604S,
       author = {{St{\"u}cker}, Jens and {Ogiya}, Go and {Angulo}, Raul E. and {Aguirre-Santaella}, Alejandra and {S{\'a}nchez-Conde}, Miguel A.},
        title = "{Tidal stripping in the adiabatic limit}",
      journal = {\mnras},
         year = 2023,
        month = may,
       volume = {521},
       number = {3},
        pages = {4432-4461},
          doi = {10.1093/mnras/stad844},
archivePrefix = {arXiv},
       eprint = {2207.00604},
 primaryClass = {astro-ph.CO},
       adsurl = {https://ui.adsabs.harvard.edu/abs/2023MNRAS.521.4432S}
}

@ARTICLE{2008ApJ...673..226P,
       author = {{Pe{\~n}arrubia}, Jorge and {Navarro}, Julio F. and {McConnachie}, Alan W.},
        title = "{The Tidal Evolution of Local Group Dwarf Spheroidals}",
      journal = {\apj},
         year = 2008,
        month = jan,
       volume = {673},
       number = {1},
        pages = {226-240},
          doi = {10.1086/523686},
archivePrefix = {arXiv},
       eprint = {0708.3087},
 primaryClass = {astro-ph},
       adsurl = {https://ui.adsabs.harvard.edu/abs/2008ApJ...673..226P}
}

@ARTICLE{Penarrubia2010,
       author = {{Pe{\~n}arrubia}, Jorge and {Benson}, Andrew J. and {Walker}, Matthew G. and {Gilmore}, Gerard and {McConnachie}, Alan W. and {Mayer}, Lucio},
        title = "{The impact of dark matter cusps and cores on the satellite galaxy population around spiral galaxies}",
      journal = {\mnras},
         year = 2010,
        month = aug,
       volume = {406},
       number = {2},
        pages = {1290-1305},
          doi = {10.1111/j.1365-2966.2010.16762.x},
archivePrefix = {arXiv},
       eprint = {1002.3376},
 primaryClass = {astro-ph.GA},
       adsurl = {https://ui.adsabs.harvard.edu/abs/2010MNRAS.406.1290P}
}

@ARTICLE{2023MNRAS.518...93A,
       author = {{Aguirre-Santaella}, Alejandra and {S{\'a}nchez-Conde}, Miguel A. and {Ogiya}, Go and {St{\"u}cker}, Jens and {Angulo}, Raul E.},
        title = "{Shedding light on low-mass subhalo survival and annihilation luminosity with numerical simulations}",
      journal = {\mnras},
         year = 2023,
        month = jan,
       volume = {518},
       number = {1},
        pages = {93-110},
          doi = {10.1093/mnras/stac2921},
archivePrefix = {arXiv},
       eprint = {2207.08652},
 primaryClass = {astro-ph.GA},
       adsurl = {https://ui.adsabs.harvard.edu/abs/2023MNRAS.518...93A}
}

@ARTICLE{2018MNRAS.480..800H,
       author = {{Hopkins}, Philip F. and {Wetzel}, Andrew and {Kere{\v{s}}}, Du{\v{s}}an and {Faucher-Gigu{\`e}re}, Claude-Andr{\'e} and {Quataert}, Eliot and {Boylan-Kolchin}, Michael and {Murray}, Norman and {Hayward}, Christopher C. and {Garrison-Kimmel}, Shea and {Hummels}, Cameron and {Feldmann}, Robert and {Torrey}, Paul and {Ma}, Xiangcheng and {Angl{\'e}s-Alc{\'a}zar}, Daniel and {Su}, Kung-Yi and {Orr}, Matthew and {Schmitz}, Denise and {Escala}, Ivanna and {Sanderson}, Robyn and {Grudi{\'c}}, Michael Y. and {Hafen}, Zachary and {Kim}, Ji-Hoon and {Fitts}, Alex and {Bullock}, James S. and {Wheeler}, Coral and {Chan}, T.~K. and {Elbert}, Oliver D. and {Narayanan}, Desika},
        title = "{FIRE-2 simulations: physics versus numerics in galaxy formation}",
      journal = {\mnras},
         year = 2018,
        month = oct,
       volume = {480},
       number = {1},
        pages = {800-863},
          doi = {10.1093/mnras/sty1690},
archivePrefix = {arXiv},
       eprint = {1702.06148},
 primaryClass = {astro-ph.GA},
       adsurl = {https://ui.adsabs.harvard.edu/abs/2018MNRAS.480..800H}
}

@ARTICLE{2017MNRAS.467..179G,
       author = {{Grand}, Robert J.~J. and {G{\'o}mez}, Facundo A. and {Marinacci}, Federico and {Pakmor}, R{\"u}diger and {Springel}, Volker and {Campbell}, David J.~R. and {Frenk}, Carlos S. and {Jenkins}, Adrian and {White}, Simon D.~M.},
        title = "{The Auriga Project: the properties and formation mechanisms of disc galaxies across cosmic time}",
      journal = {\mnras},
         year = 2017,
        month = may,
       volume = {467},
       number = {1},
        pages = {179-207},
          doi = {10.1093/mnras/stx071},
archivePrefix = {arXiv},
       eprint = {1610.01159},
 primaryClass = {astro-ph.GA},
       adsurl = {https://ui.adsabs.harvard.edu/abs/2017MNRAS.467..179G}
}

@ARTICLE{2018MNRAS.473.4339O,
       author = {{Ogiya}, Go and {Hahn}, Oliver},
        title = "{What sets the central structure of dark matter haloes?}",
      journal = {\mnras},
         year = 2018,
        month = feb,
       volume = {473},
       number = {4},
        pages = {4339-4359},
          doi = {10.1093/mnras/stx2639},
archivePrefix = {arXiv},
       eprint = {1707.07693},
 primaryClass = {astro-ph.CO},
       adsurl = {https://ui.adsabs.harvard.edu/abs/2018MNRAS.473.4339O}
}

@ARTICLE{2017MNRAS.471.4687A,
       author = {{Angulo}, Raul E. and {Hahn}, Oliver and {Ludlow}, Aaron D. and {Bonoli}, Silvia},
        title = "{Earth-mass haloes and the emergence of NFW density profiles}",
      journal = {\mnras},
         year = 2017,
        month = nov,
       volume = {471},
       number = {4},
        pages = {4687-4701},
          doi = {10.1093/mnras/stx1658},
archivePrefix = {arXiv},
       eprint = {1604.03131},
 primaryClass = {astro-ph.CO},
       adsurl = {https://ui.adsabs.harvard.edu/abs/2017MNRAS.471.4687A}
}

@ARTICLE{2023MNRAS.518.3509D,
       author = {{Delos}, M. Sten and {White}, Simon D.~M.},
        title = "{Inner cusps of the first dark matter haloes: formation and survival in a cosmological context}",
      journal = {\mnras},
         year = 2023,
        month = jan,
       volume = {518},
       number = {3},
        pages = {3509-3532},
          doi = {10.1093/mnras/stac3373},
archivePrefix = {arXiv},
       eprint = {2207.05082},
 primaryClass = {astro-ph.CO},
       adsurl = {https://ui.adsabs.harvard.edu/abs/2023MNRAS.518.3509D}
}

@ARTICLE{2023JCAP...10..008D,
       author = {{Delos}, M. Sten and {White}, Simon D.~M.},
        title = "{Prompt cusps and the dark matter annihilation signal}",
      journal = {\jcap},
         year = 2023,
        month = oct,
       volume = {2023},
       number = {10},
          eid = {008},
        pages = {008},
          doi = {10.1088/1475-7516/2023/10/008},
archivePrefix = {arXiv},
       eprint = {2209.11237},
 primaryClass = {astro-ph.CO},
       adsurl = {https://ui.adsabs.harvard.edu/abs/2023JCAP...10..008D}
}

@ARTICLE{2025ApJ...993...93D,
       author = {{Delos}, M. Sten},
        title = "{The Cusp─Halo Relation}",
      journal = {\apj},
         year = 2025,
        month = nov,
       volume = {993},
       number = {1},
          eid = {93},
        pages = {93},
          doi = {10.3847/1538-4357/ae0625},
archivePrefix = {arXiv},
       eprint = {2506.03240},
 primaryClass = {astro-ph.CO},
       adsurl = {https://ui.adsabs.harvard.edu/abs/2025ApJ...993...93D}
}

@ARTICLE{2014ApJ...788...27I,
       author = {{Ishiyama}, Tomoaki},
        title = "{Hierarchical Formation of Dark Matter Halos and the Free Streaming Scale}",
      journal = {\apj},
         year = 2014,
        month = jun,
       volume = {788},
       number = {1},
          eid = {27},
        pages = {27},
          doi = {10.1088/0004-637X/788/1/27},
archivePrefix = {arXiv},
       eprint = {1404.1650},
 primaryClass = {astro-ph.CO},
       adsurl = {https://ui.adsabs.harvard.edu/abs/2014ApJ...788...27I}
}

@ARTICLE{2010ApJ...723L.195I,
       author = {{Ishiyama}, Tomoaki and {Makino}, Junichiro and {Ebisuzaki}, Toshikazu},
        title = "{Gamma-ray Signal from Earth-mass Dark Matter Microhalos}",
      journal = {\apjl},
         year = 2010,
        month = nov,
       volume = {723},
       number = {2},
        pages = {L195-L200},
          doi = {10.1088/2041-8205/723/2/L195},
archivePrefix = {arXiv},
       eprint = {1006.3392},
 primaryClass = {astro-ph.CO},
       adsurl = {https://ui.adsabs.harvard.edu/abs/2010ApJ...723L.195I}
}

@ARTICLE{1999MNRAS.310.1147M,
       author = {{Moore}, B. and {Quinn}, T. and {Governato}, F. and {Stadel}, J. and {Lake}, G.},
        title = "{Cold collapse and the core catastrophe}",
      journal = {\mnras},
         year = 1999,
        month = dec,
       volume = {310},
       number = {4},
        pages = {1147-1152},
          doi = {10.1046/j.1365-8711.1999.03039.x},
archivePrefix = {arXiv},
       eprint = {astro-ph/9903164},
 primaryClass = {astro-ph},
       adsurl = {https://ui.adsabs.harvard.edu/abs/1999MNRAS.310.1147M}
}

@ARTICLE{2004MNRAS.353..624D,
       author = {{Diemand}, J{\"u}rg and {Moore}, Ben and {Stadel}, Joachim},
        title = "{Convergence and scatter of cluster density profiles}",
      journal = {\mnras},
         year = 2004,
        month = sep,
       volume = {353},
       number = {2},
        pages = {624-632},
          doi = {10.1111/j.1365-2966.2004.08094.x},
archivePrefix = {arXiv},
       eprint = {astro-ph/0402267},
 primaryClass = {astro-ph},
       adsurl = {https://ui.adsabs.harvard.edu/abs/2004MNRAS.353..624D}
}

@ARTICLE{2021MNRAS.505...18E,
       author = {{Errani}, Rapha{\"e}l and {Navarro}, Julio F.},
        title = "{The asymptotic tidal remnants of cold dark matter subhaloes}",
      journal = {\mnras},
         year = 2021,
        month = jul,
       volume = {505},
       number = {1},
        pages = {18-32},
          doi = {10.1093/mnras/stab1215},
archivePrefix = {arXiv},
       eprint = {2011.07077},
 primaryClass = {astro-ph.GA},
       adsurl = {https://ui.adsabs.harvard.edu/abs/2021MNRAS.505...18E}
}

@ARTICLE{2022MNRAS.517.1398B,
       author = {{Benson}, Andrew J. and {Du}, Xiaolong},
        title = "{Tidal tracks and artificial disruption of cold dark matter haloes}",
      journal = {\mnras},
         year = 2022,
        month = nov,
       volume = {517},
       number = {1},
        pages = {1398-1406},
          doi = {10.1093/mnras/stac2750},
archivePrefix = {arXiv},
       eprint = {2206.01842},
 primaryClass = {astro-ph.GA},
       adsurl = {https://ui.adsabs.harvard.edu/abs/2022MNRAS.517.1398B}
}

@ARTICLE{2024PhRvD.110b3019D,
       author = {{Du}, Xiaolong and {Benson}, Andrew and {Zeng}, Zhichao Carton and {Treu}, Tommaso and {Peter}, Annika H.~G. and {Mace}, Charlie and {Jiang}, Fangzhou and {Yang}, Shengqi and {Gannon}, Charles and {Gilman}, Daniel and {Nierenberg}, Anna. M. and {Nadler}, Ethan O.},
        title = "{Tidal evolution of cored and cuspy dark matter halos}",
      journal = {\prd},
         year = 2024,
        month = jul,
       volume = {110},
       number = {2},
          eid = {023019},
        pages = {023019},
          doi = {10.1103/PhysRevD.110.023019},
archivePrefix = {arXiv},
       eprint = {2403.09597},
 primaryClass = {astro-ph.GA},
       adsurl = {https://ui.adsabs.harvard.edu/abs/2024PhRvD.110b3019D}
}

@ARTICLE{2006ApJ...641..647K,
       author = {{Kazantzidis}, Stelios and {Zentner}, Andrew R. and {Kravtsov}, Andrey V.},
        title = "{The Robustness of Dark Matter Density Profiles in Dissipationless Mergers}",
      journal = {\apj},
         year = 2006,
        month = apr,
       volume = {641},
       number = {2},
        pages = {647-664},
          doi = {10.1086/500579},
archivePrefix = {arXiv},
       eprint = {astro-ph/0510583},
 primaryClass = {astro-ph},
       adsurl = {https://ui.adsabs.harvard.edu/abs/2006ApJ...641..647K}
}

@ARTICLE{2014MNRAS.439..300L,
       author = {{Lovell}, Mark R. and {Frenk}, Carlos S. and {Eke}, Vincent R. and {Jenkins}, Adrian and {Gao}, Liang and {Theuns}, Tom},
        title = "{The properties of warm dark matter haloes}",
      journal = {\mnras},
         year = 2014,
        month = mar,
       volume = {439},
       number = {1},
        pages = {300-317},
          doi = {10.1093/mnras/stt2431},
archivePrefix = {arXiv},
       eprint = {1308.1399},
 primaryClass = {astro-ph.CO},
       adsurl = {https://ui.adsabs.harvard.edu/abs/2014MNRAS.439..300L}
}

@ARTICLE{2001MNRAS.321..559B,
       author = {{Bullock}, J.~S. and {Kolatt}, T.~S. and {Sigad}, Y. and {Somerville}, R.~S. and {Kravtsov}, A.~V. and {Klypin}, A.~A. and {Primack}, J.~R. and {Dekel}, A.},
        title = "{Profiles of dark haloes: evolution, scatter and environment}",
      journal = {\mnras},
         year = 2001,
        month = mar,
       volume = {321},
       number = {3},
        pages = {559-575},
          doi = {10.1046/j.1365-8711.2001.04068.x},
archivePrefix = {arXiv},
       eprint = {astro-ph/9908159},
 primaryClass = {astro-ph},
       adsurl = {https://ui.adsabs.harvard.edu/abs/2001MNRAS.321..559B}
}

@ARTICLE{1998MNRAS.300..146G,
       author = {{Ghigna}, Sebastiano and {Moore}, Ben and {Governato}, Fabio and {Lake}, George and {Quinn}, Thomas and {Stadel}, Joachim},
        title = "{Dark matter haloes within clusters}",
      journal = {\mnras},
         year = 1998,
        month = oct,
       volume = {300},
       number = {1},
        pages = {146-162},
          doi = {10.1046/j.1365-8711.1998.01918.x},
archivePrefix = {arXiv},
       eprint = {astro-ph/9801192},
 primaryClass = {astro-ph},
       adsurl = {https://ui.adsabs.harvard.edu/abs/1998MNRAS.300..146G}
}

@ARTICLE{1992ApJ...396L...1S,
       author = {{Smoot}, G.~F. and {Bennett}, C.~L. and {Kogut}, A. and {Wright}, E.~L. and {Aymon}, J. and {Boggess}, N.~W. and {Cheng}, E.~S. and {de Amici}, G. and {Gulkis}, S. and {Hauser}, M.~G. and {Hinshaw}, G. and {Jackson}, P.~D. and {Janssen}, M. and {Kaita}, E. and {Kelsall}, T. and {Keegstra}, P. and {Lineweaver}, C. and {Loewenstein}, K. and {Lubin}, P. and {Mather}, J. and {Meyer}, S.~S. and {Moseley}, S.~H. and {Murdock}, T. and {Rokke}, L. and {Silverberg}, R.~F. and {Tenorio}, L. and {Weiss}, R. and {Wilkinson}, D.~T.},
        title = "{Structure in the COBE Differential Microwave Radiometer First-Year Maps}",
      journal = {\apjl},
         year = 1992,
        month = sep,
       volume = {396},
        pages = {L1},
          doi = {10.1086/186504},
       adsurl = {https://ui.adsabs.harvard.edu/abs/1992ApJ...396L...1S}
}

@ARTICLE{2018PhR...730....1T,
       author = {{Tulin}, Sean and {Yu}, Hai-Bo},
        title = "{Dark matter self-interactions and small scale structure}",
      journal = {\physrep},
         year = 2018,
        month = feb,
       volume = {730},
        pages = {1-57},
          doi = {10.1016/j.physrep.2017.11.004},
archivePrefix = {arXiv},
       eprint = {1705.02358},
 primaryClass = {hep-ph},
       adsurl = {https://ui.adsabs.harvard.edu/abs/2018PhR...730....1T}
}

@BOOK{1990eaun.book.....K,
       author = {{Kolb}, Edward W. and {Turner}, Michael S.},
        title = "{The early universe}",
         year = 1990,
       volume = {69},
       series = {},
       adsurl = {https://ui.adsabs.harvard.edu/abs/1990eaun.book.....K}
}

@ARTICLE{2024ApJ...964..123J,
       author = {{Jung}, Minyong and {Roca-F{\`a}brega}, Santi and {Kim}, Ji-Hoon and {Genina}, Anna and {Hausammann}, Loic and {Kim}, Hyeonyong and {Lupi}, Alessandro and {Nagamine}, Kentaro and {Powell}, Johnny W. and {Revaz}, Yves and {Shimizu}, Ikkoh and {Vel{\'a}zquez}, H{\'e}ctor and {Ceverino}, Daniel and {Primack}, Joel R. and {Quinn}, Thomas R. and {Strawn}, Clayton and {Abel}, Tom and {Dekel}, Avishai and {Dong}, Bili and {Oh}, Boon Kiat and {Teyssier}, Romain and {AGORA Collaboration}},
        title = "{The AGORA High-resolution Galaxy Simulations Comparison Project. V. Satellite Galaxy Populations in a Cosmological Zoom-in Simulation of a Milky Way{\textendash}Mass Halo}",
      journal = {\apj},
         year = 2024,
        month = apr,
       volume = {964},
       number = {2},
          eid = {123},
        pages = {123},
          doi = {10.3847/1538-4357/ad245b},
archivePrefix = {arXiv},
       eprint = {2402.05392},
 primaryClass = {astro-ph.GA},
       adsurl = {https://ui.adsabs.harvard.edu/abs/2024ApJ...964..123J}
}

@ARTICLE{2014ApJ...786...87B,
       author = {{Brooks}, Alyson M. and {Zolotov}, Adi},
        title = "{Why Baryons Matter: The Kinematics of Dwarf Spheroidal Satellites}",
      journal = {\apj},
         year = 2014,
        month = may,
       volume = {786},
       number = {2},
          eid = {87},
        pages = {87},
          doi = {10.1088/0004-637X/786/2/87},
archivePrefix = {arXiv},
       eprint = {1207.2468},
 primaryClass = {astro-ph.CO},
       adsurl = {https://ui.adsabs.harvard.edu/abs/2014ApJ...786...87B}
}

@ARTICLE{2016MNRAS.458.1559Z,
       author = {{Zhu}, Qirong and {Marinacci}, Federico and {Maji}, Moupiya and {Li}, Yuexing and {Springel}, Volker and {Hernquist}, Lars},
        title = "{Baryonic impact on the dark matter distribution in Milky Way-sized galaxies and their satellites}",
      journal = {\mnras},
         year = 2016,
        month = may,
       volume = {458},
       number = {2},
        pages = {1559-1580},
          doi = {10.1093/mnras/stw374},
archivePrefix = {arXiv},
       eprint = {1506.05537},
 primaryClass = {astro-ph.CO},
       adsurl = {https://ui.adsabs.harvard.edu/abs/2016MNRAS.458.1559Z}
}




\appendix
\section{Survival criteria and region}\label{sec:apsurvive}

\begin{figure}
\begin{center}
\includegraphics[width=\columnwidth]{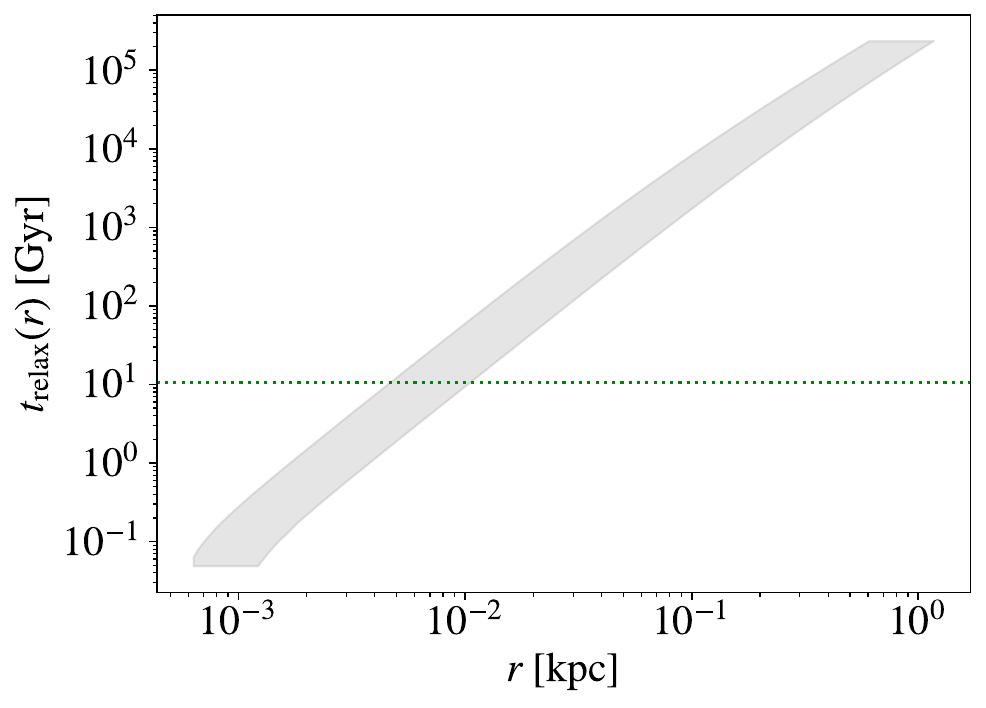}
\caption{Relaxation times for our runs, as a shadowed band which encompasses both initial NFW and prompt cusps. The green dotted line corresponds to $z=2$ and indicates from which radius the simulations can be trusted outwards.} 
\label{fig:relax}
\end{center}
\end{figure}

As noted in the main text, simulations start to misbehave when they suffer from lack of resolution. Because of this, we need to set some convergence criteria below which we are not able to trust the simulations anymore \citep[][\citetalias{2023MNRAS.518...93A}]{vandenBosch2018_num_criteria}.  In our case, 
we consider every data point with $\log_{10}f_\mathrm{b} > -3.5$ for NFW profiles and $\log_{10}f_\mathrm{b} > -2.5$ %
for prompt cusps. This corresponds to the respective values when $\log_{10} (r_\mathrm{max}/r_\mathrm{max,i}) = -1.5$, where two-body relaxation generates artificial effects; see Fig \ref{fig:relax}. This means that, for the runs whose $f_\mathrm{b}$ at $z=0$ falls below these thresholds, we still retain part of the data for analysis.   We have checked that the value for initial NFW subhaloes is in tune with the findings in \citetalias{2021MNRAS.505...18E}, where the convergence criterium is having at least 3000 particles inside $r_\mathrm{max}$. 

\begin{figure}
\begin{center}
\includegraphics[width=\columnwidth]{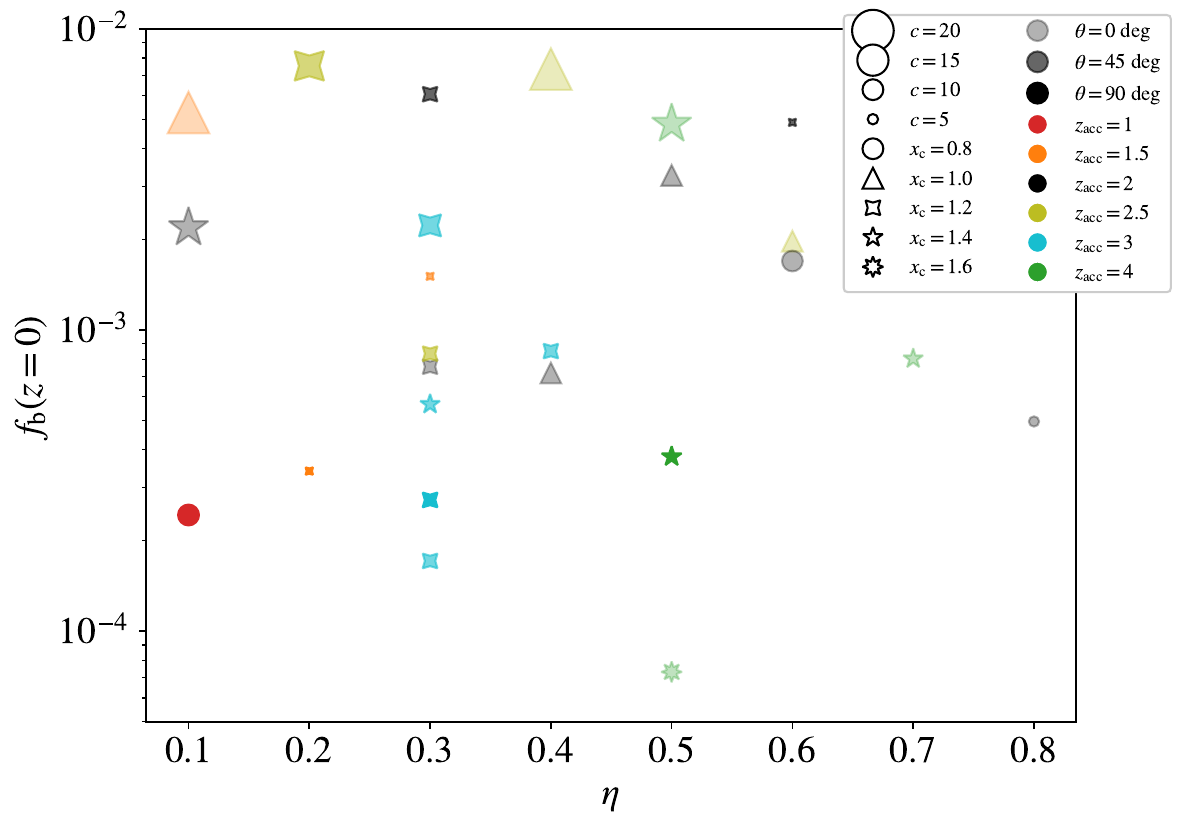} 
\caption{Bound mass fraction at present time for different simulation runs. Size represents NFW concentration at infall; shape represents orbital energy $x_\mathrm{c}$, which is higher for shapes with more spikes; colour represents accretion redshift $z_\mathrm{acc}$; transparency represents orbital inclination angle $\theta$, where 0 means parallel to the disc; $x$-axis is the circularity $\eta$. See \citetalias{2023MNRAS.518...93A} for further details on each of these parameters. An interactive version of this plot is available as an ancillary file.
} 
\label{fig:figsurvr}
\end{center}
\end{figure}

We have a multidimensional parameter space and it is not feasible to explore every possible combination of initial parameters that give $f_\mathrm{b}$ values below $0.01$ at late times, since the number of particles needed for many of those simulations to converge is at least several millions, which implies large computational execution times. Nonetheless, we can get a rough idea about the minimum value of a parameter needed to trust the simulation until $z=0$, 
according to our criterion, when the rest are fixed, by looking at Fig.~\ref{fig:figsurvr}. For instance, we can extract 
that if the subhalo initial concentration is 10, the accretion redshift is 2 and the orbit is parallel to the baryonic disc (i.e., $\theta = 0 $ deg), we need $\eta \gtrsim 0.3$ if $x_\mathrm{c} = 1.2$, or $\eta \gtrsim 0.4$ if  $x_\mathrm{c} = 1.0$.

On the other hand, for a subhalo on a parallel orbit with \( c = 20 \) and \( z_\mathrm{acc} = 3 \), we find that for \( \eta = 0.4 \), the simulation remains reliable for \( x_\mathrm{c} \leq 1.0 \). Conversely, for \( x_\mathrm{c} = 1.0 \), the entire run can be trusted for orbits with eccentricities greater than \( \eta = 0.4 \).

\section{Additional tidal tracks}\label{sec:aptt}

In this appendix we provide tidal tracks for  combinations of parameters other than the ones in the main text. For the sake of clarity, we also include plots to show explicitly how the tidal track is affected by the different parameters involved in the evolution of the subhalo, e.g. orbit circularity, accretion redshift, initial concentration...

\begin{figure}
\begin{center}
\includegraphics[width=\columnwidth]{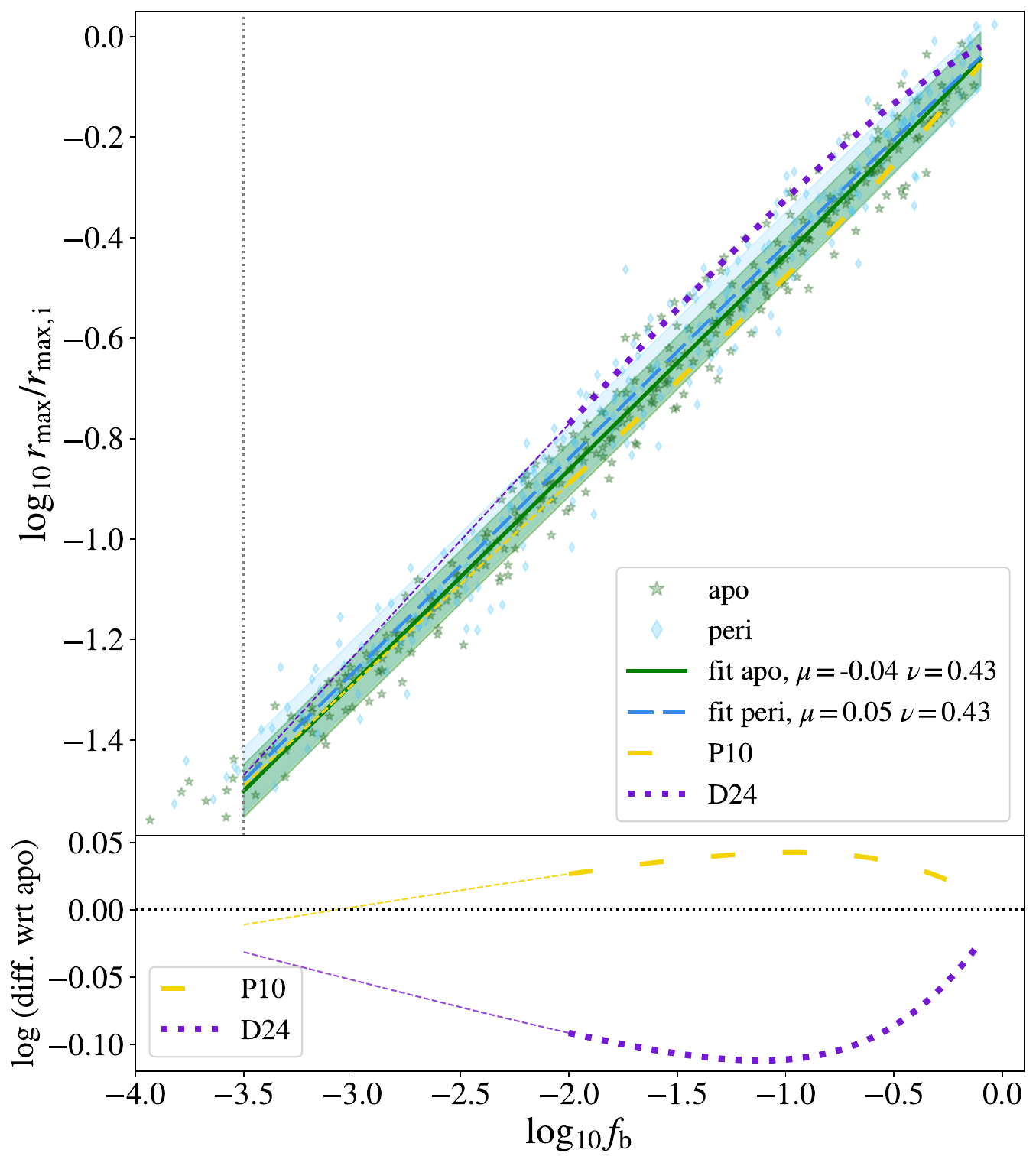}  
\caption{Similar to Fig.~\ref{fig:ttvmaxfb} but for $r_\mathrm{max}$. 
Relation between $f_\mathrm{b}$ and $r_\mathrm{max}$ divided by their initial values for the 
apocentres (green stars) and the pericentres (sky blue diamonds). Subhaloes exhibit an NFW density profile at accretion. The tidal track found for each subset is drawn as a solid green line in the former case and a dashed sky blue line in the latter, with shadowed bands for their respective scatter. Data can be trusted to the right of the vertical dotted line. The fits from \citetalias{Penarrubia2010} (loosely dashed yellow line) and \citetalias{2024PhRvD.110b3019D} (purple dotted line) are included, and thinned when they are extrapolated. We have included a lower panel with the difference between our best fit for the apocentres and the literature fits.
} 
\label{fig:figttrmax}
\end{center}
\end{figure}

We start with the tidal tracks involving $r_\mathrm{max}$. Fig.~\ref{fig:figttrmax} corresponds to the tidal track considering the bound mass fraction and $r_\mathrm{max}$ for NFW initial density profiles, as in Section~\ref{sec:tidaltrackNFW}. The figure also includes our fits as well as the ones from the literature. The best-fit parameters are included in Table~\ref{tab:ttbestfit}. 
Our $r_\mathrm{max} - f_\mathrm{b}$ fit for the apocentres agrees within the scatter with \citetalias{Penarrubia2010}, although the latter is consistently below. The \citetalias{2024PhRvD.110b3019D} result, on the other hand, is consistently above, although $\nu$  is similar for the apocentre case and $\mu$ is also negative; $\mu$ is closer to 0 in our fit. 

\begin{figure}
\begin{center}
\includegraphics[width=\columnwidth]{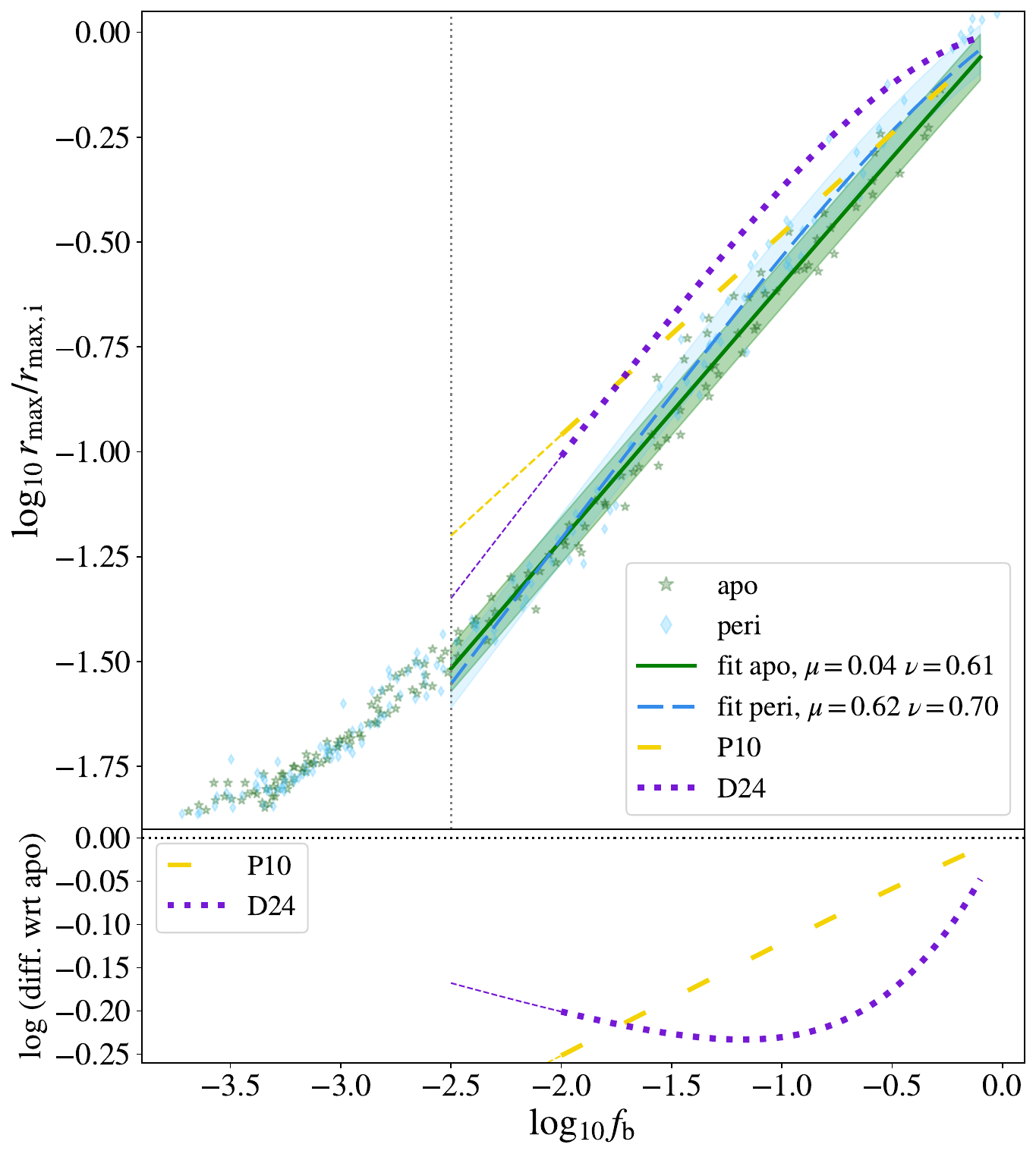}  
\caption{Similar to Fig.~\ref{fig:figpc1} but for $r_\mathrm{max}$. %
Relation between $f_\mathrm{b}$ and $r_\mathrm{max}$ divided by their initial values for the 
apocentres (green stars) and the pericentres (sky blue diamonds). Subhaloes exhibit an inner prompt cusp and an NFW tail at accretion. The tidal track found for each subset is drawn as a solid green line in the former case and a dashed sky blue line in the latter, with shadowed bands for their respective scatter. Data can be trusted to the right of the vertical dotted line. The fits from \citetalias{Penarrubia2010} (loosely dashed yellow line) and \citetalias{2024PhRvD.110b3019D} (purple dotted line) are included, and thinned when they are extrapolated. 
The lower panel displays the difference between our best fit for the apocentres and the literature fits. 
} 
\label{fig:figpc2}
\end{center}
\end{figure}

Fig.~\ref{fig:figpc2} corresponds to the tidal track between the bound mass fraction and $r_\mathrm{max}$ of initial density profiles following Eq.~\ref{eq:nfw} with an inner prompt cusp of slope $\gamma = -1.5$, as in Section~\ref{sec:tidaltrackpc}. For these subhaloes, $r_\mathrm{max}$ is much smaller in our simulations compared to the fits from previous works. The value at the pericentres is larger than the one at the apocentres for the same $f_\mathrm{b}$ during the first orbits, yet both become similar at later times. The best-fit parameters are included in Table~\ref{tab:ttbestfit}. Here, $\mu$ is close to 0 again in our fit using the apocentres.

Now, we explore the impact of different initial parameters on the tidal tracks.

\begin{figure*}
\begin{center}
\includegraphics[width=.48\textwidth]{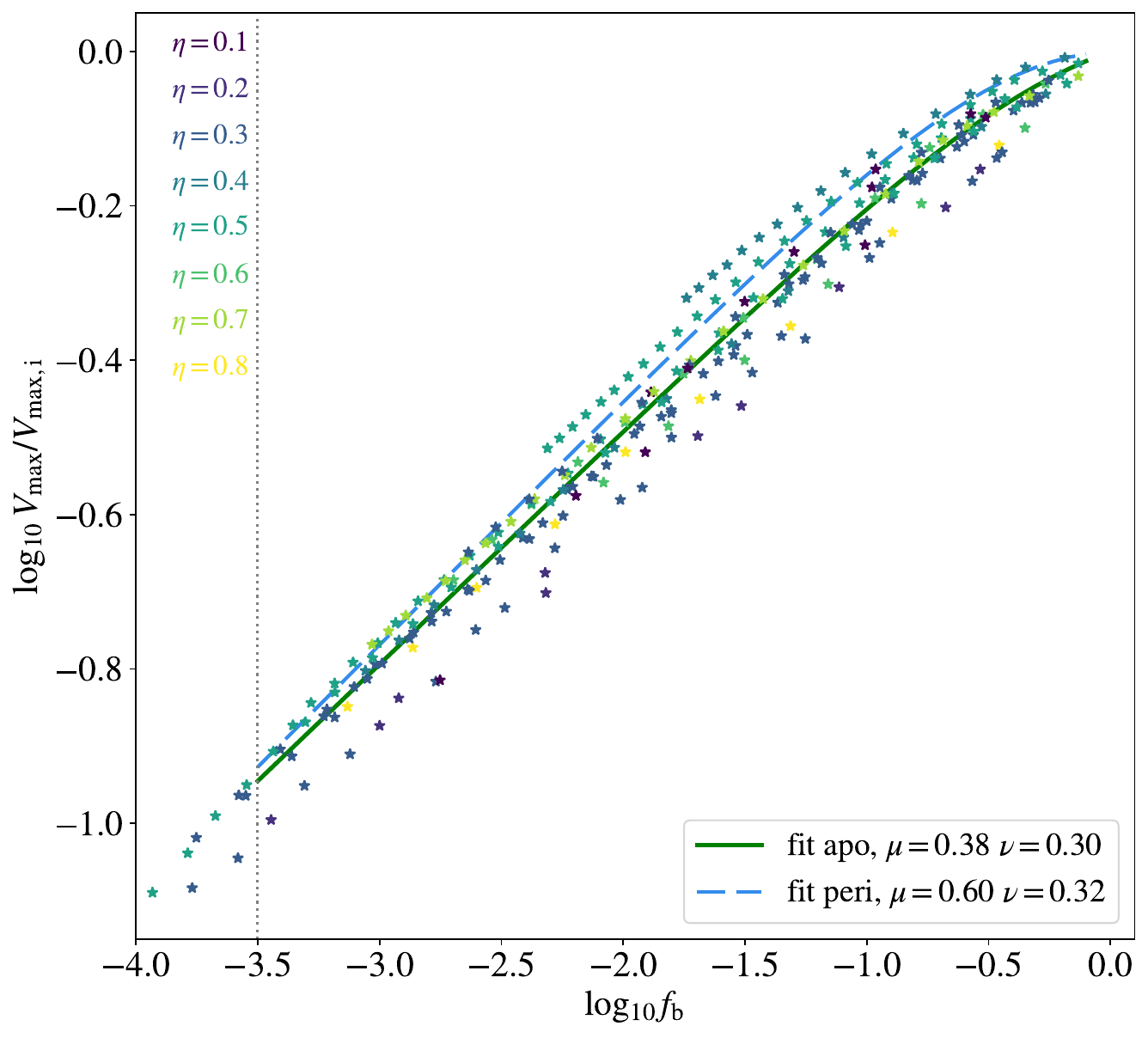} 
\includegraphics[width=.48\textwidth]{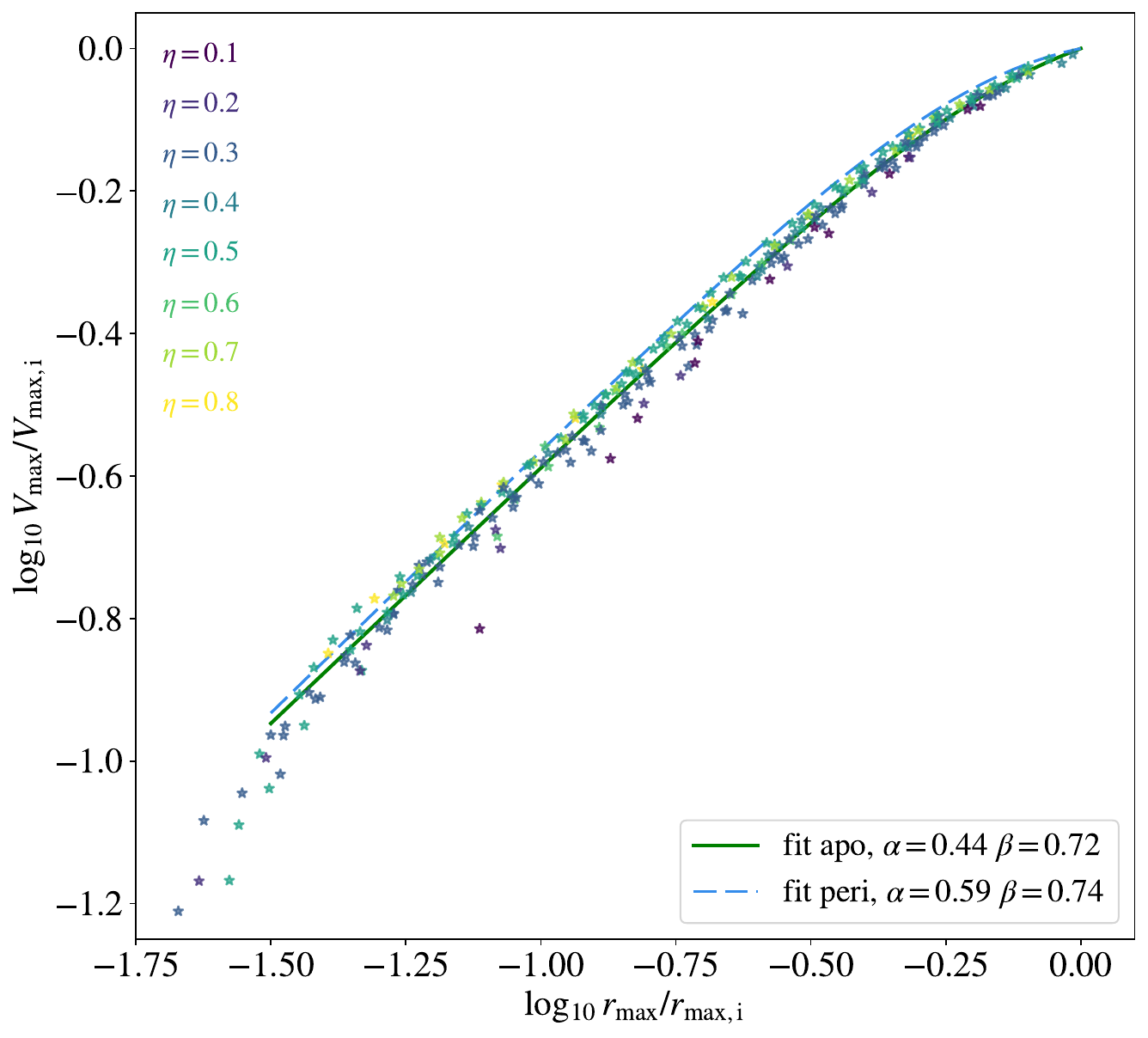} 
\caption{Same as Figs.~\ref{fig:ttvmaxfb} and~\ref{fig:ttvmaxrmax} but only for apocentres and colouring the points using the circularity of the respective run.%
} 
\label{fig:figttcirc}
\end{center}
\end{figure*}

In Fig.~\ref{fig:figttcirc} we can observe that the circularity is not relevant for the tidal track between $f_\mathrm{b}$ and $V_\mathrm{max}$, while it plays a role when plotting $V_\mathrm{max}$ against $r_\mathrm{max}$: subhaloes in more eccentric orbits, i.e. with lower circularities, reach a lower $V_\mathrm{max}$ for the same $r_\mathrm{max}$ after significant disruption. This is in agreement with Fig. 6 from \citetalias{2021MNRAS.505...18E}. Their pericentre-to-apocentre ratios span from 1:1 to 1:20, while our orbits are more elliptic in general, with ratios ranging from nearly 1:4 to 1:45. 

\begin{figure*}
\begin{center}
\includegraphics[width=.48\textwidth]{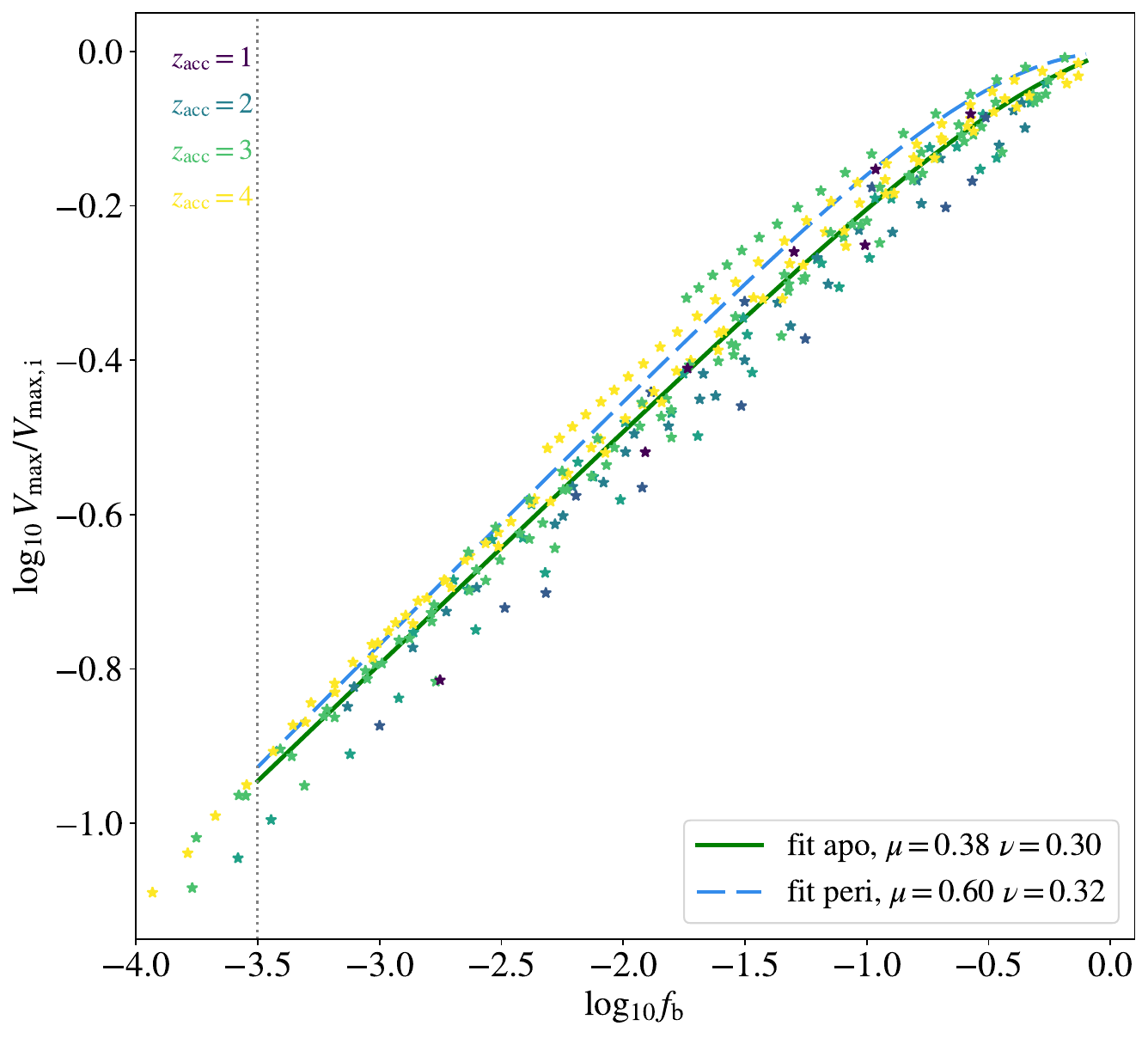} 
\includegraphics[width=.48\textwidth]{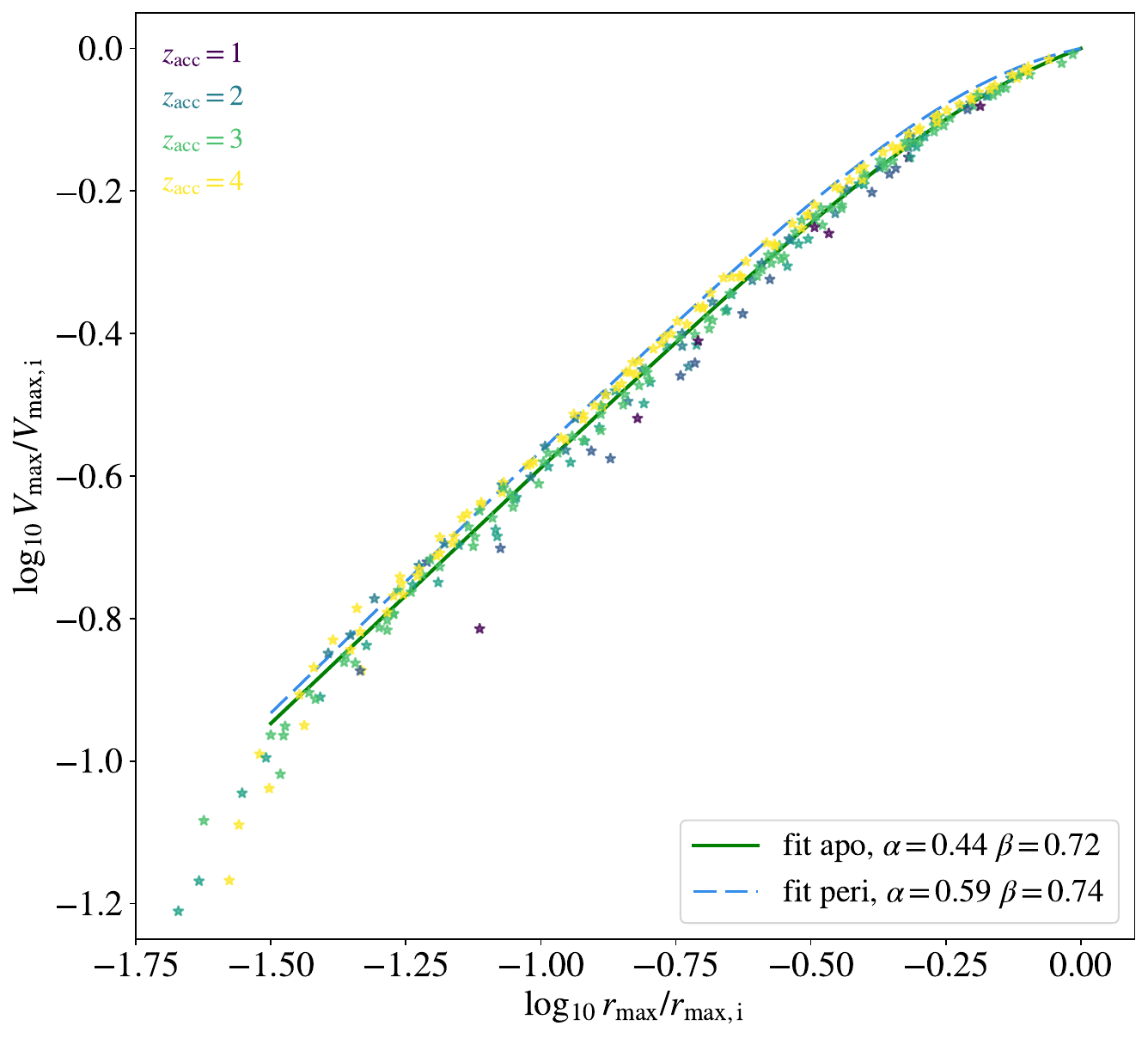} 
\caption{Same as Figs.~\ref{fig:ttvmaxfb} and~\ref{fig:ttvmaxrmax} but only for apocentres and colouring the points using the accretion redshift of the respective run.} %
\label{fig:figttzacc}
\end{center}
\end{figure*}

In Fig.~\ref{fig:figttzacc} we find the opposite for $z_\mathrm{acc}$: its relevance strengthens when considering the tidal track between $f_\mathrm{b}$ and $V_\mathrm{max}$, where subhaloes accreted earlier do experience smaller changes in $V_\mathrm{max}$ for the same $f_\mathrm{b}$ compared to subhaloes accreted later. This can be explained since the former subhaloes would be more concentrated.

\begin{figure*}
\begin{center}
\includegraphics[width=.48\textwidth]{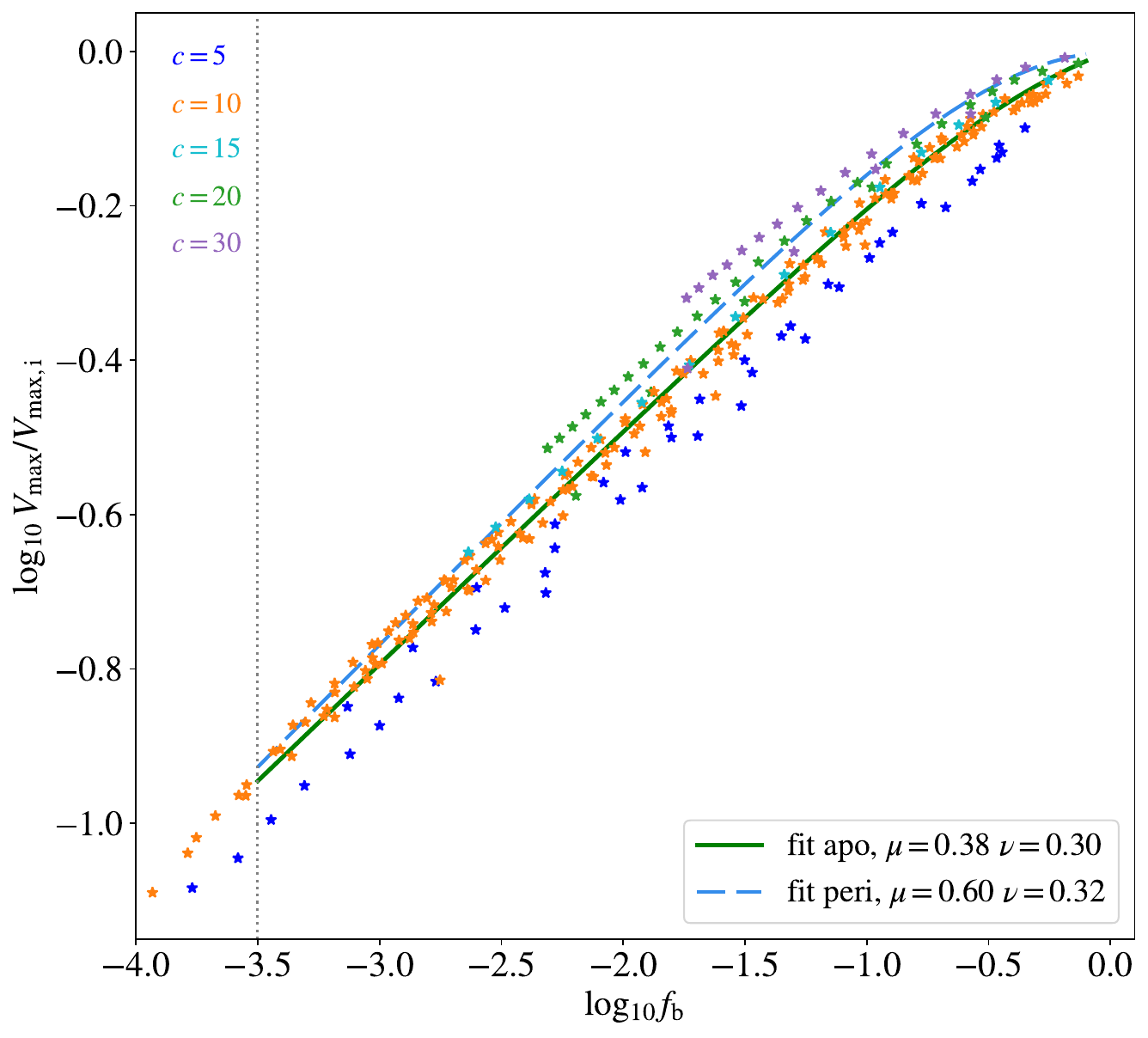} %
\includegraphics[width=.48\textwidth]{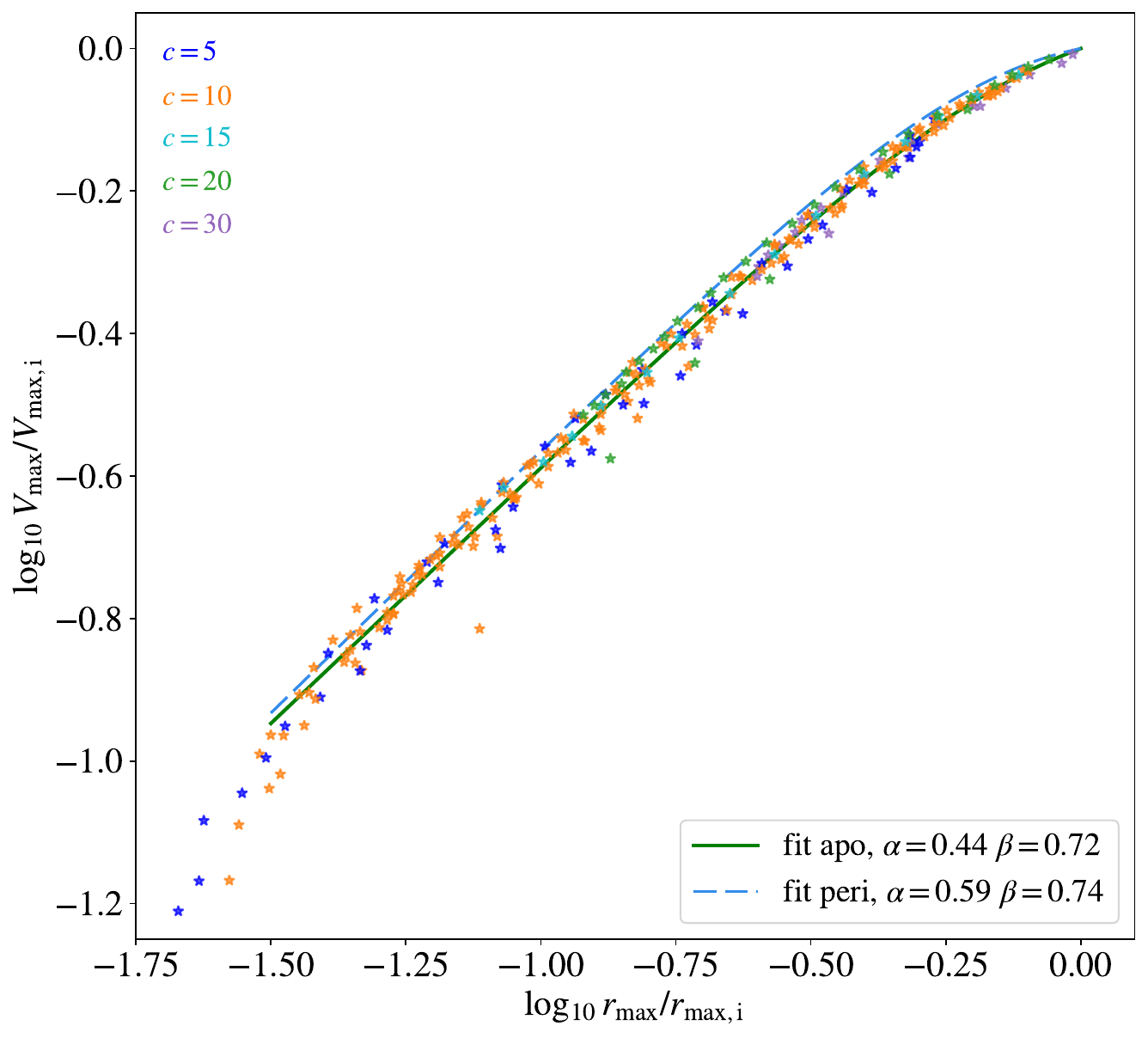}
\caption{Same as Figs.~\ref{fig:ttvmaxfb} and~\ref{fig:ttvmaxrmax} but only for apocentres and colouring the points using the initial concentration of the respective run.} %
\label{fig:figttc}
\end{center}
\end{figure*}

In Fig.~\ref{fig:figttc} we agree with \citet{2019MNRAS.490.2091G}, observing a dependence of the tidal track between $f_\mathrm{b}$ and $V_\mathrm{max}$ on the subhalo concentration: $V_\mathrm{max}$ decreases more for the same $f_\mathrm{b}$ when the initial concentration is smaller. Conversely, no such dependence is found when plotting $V_\mathrm{max}$ against $r_\mathrm{max}$.

\begin{figure*}
\begin{center}
\includegraphics[width=.48\textwidth]{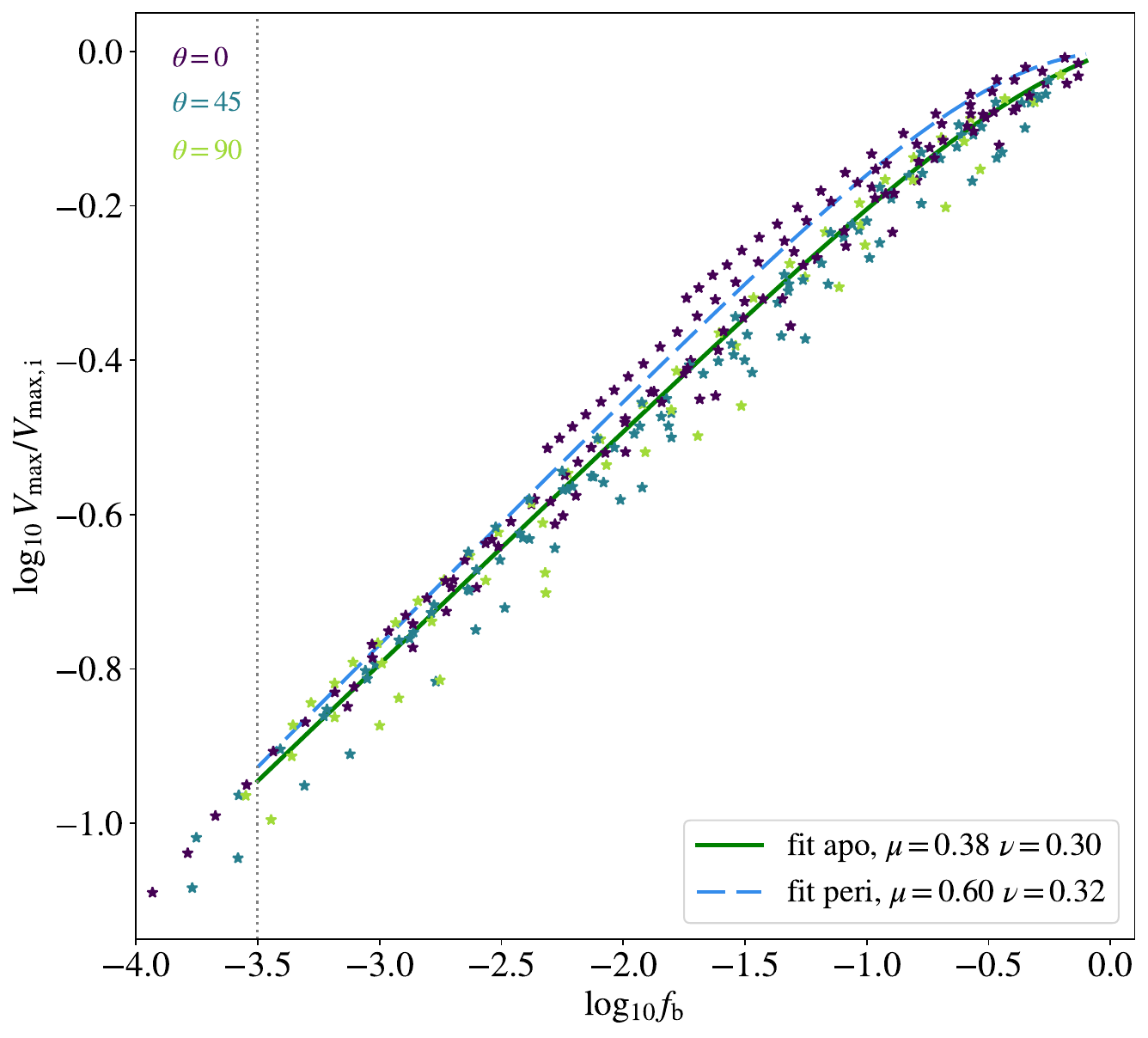} 
\includegraphics[width=.48\textwidth]{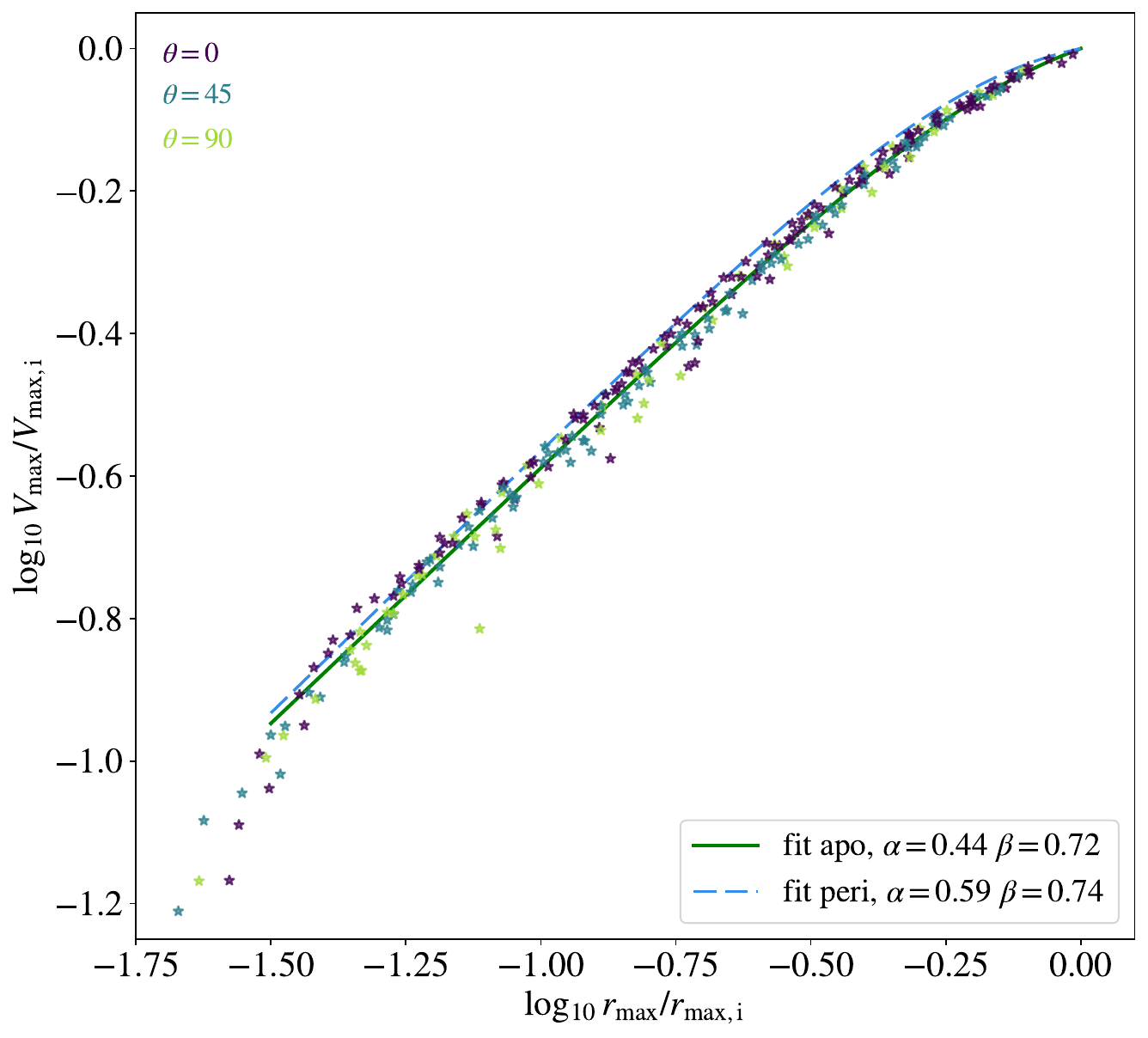} 
\caption{Same as Figs.~\ref{fig:ttvmaxfb} and~\ref{fig:ttvmaxrmax} but only for apocentres and colouring the points using the inclination angle of the respective run.} %
\label{fig:figttinc}
\end{center}
\end{figure*}

Finally, in Fig.~\ref{fig:figttinc} we have not found any remarkable dependence of the tidal tracks on the inclination angle.

\begin{figure}
\begin{center}
\includegraphics[width=\columnwidth]{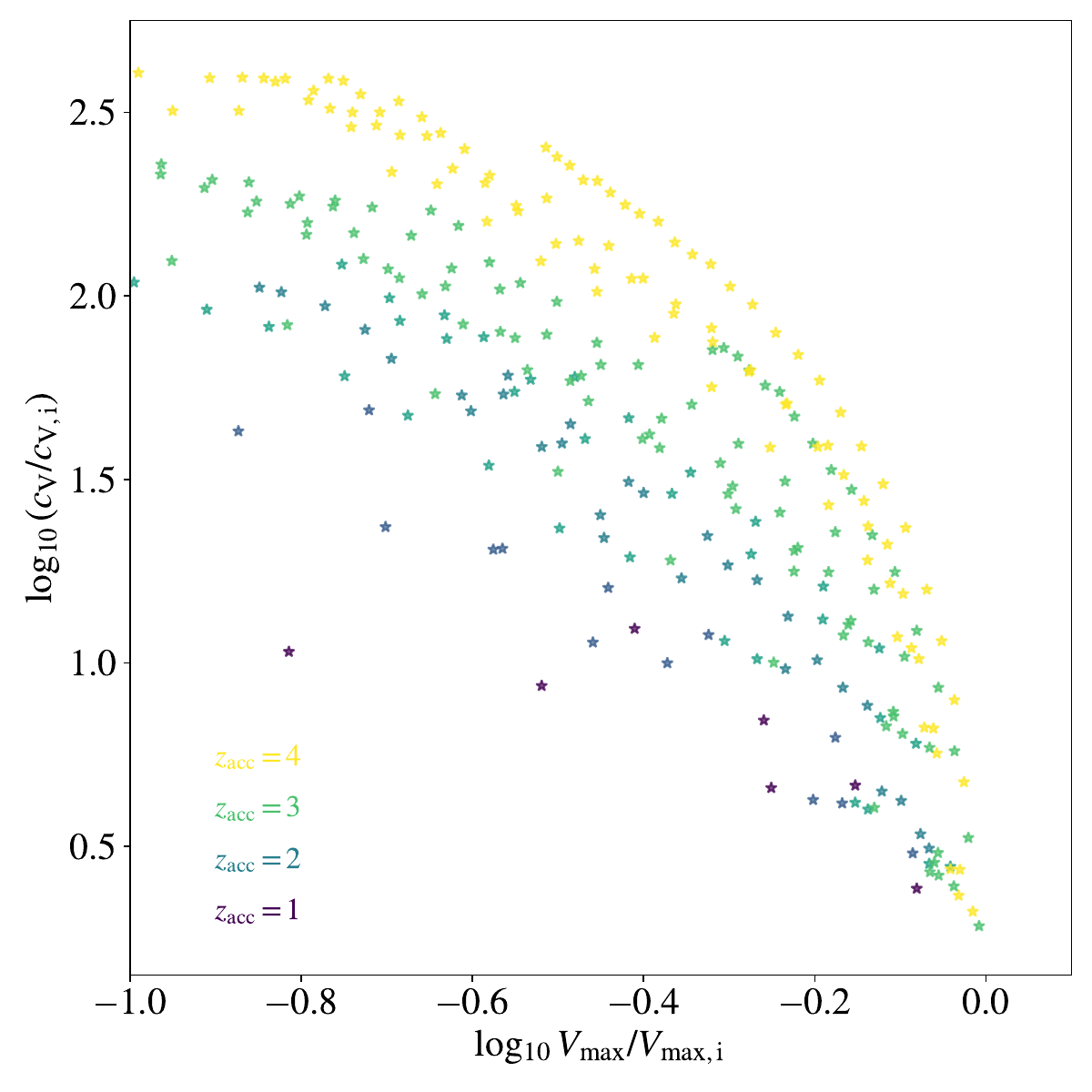 } 
\caption{Similar to Fig.~\ref{fig:figvelc2} but without the Hubble parameter normalisation. Evolution of the ratio $c_\mathrm{V} / c_\mathrm{V,i}$ in log for different simulations with an initial NFW profile. The $x$ axis is the ratio $V_\mathrm{max} / V_\mathrm{max,i}$ in log, which becomes more negative with time.  Apocentre values are coloured depending on $z_\mathrm{acc}$, and a clear dependence on this parameter appears.
}
\label{fig:figvelc3}
\end{center}
\end{figure}

The concentration tidal track without the redshift normalisation via the Hubble parameter is shown in Fig.~\ref{fig:figvelc3}. 
 In this case, we observe a significantly larger scatter compared to Fig.~\ref{fig:figvelc2}.

\bsp	
\label{lastpage}
\end{document}